\documentclass[review,11pt]{elsarticle}
\usepackage{epstopdf} 
\usepackage[pagewise,mathlines]{lineno}

\usepackage[top=2.75cm,bottom=2.75cm,left=3.5cm,right=3.5cm]{geometry}

\usepackage{amsmath}
\usepackage{bm}
\usepackage{threeparttable}
\usepackage{amssymb}
\usepackage{array}

\usepackage{graphicx}
\usepackage{subfigure}

\usepackage{color}

\begin{document}

\begin{frontmatter}
\title{Design of a double-breast gradient coil with controlled anterior–posterior gradient variation for diffusion-weighted imaging}

\author[rvt]{Feng Jia\corref{cor1}}
\ead{feng.jia@uniklinik-freiburg.de}
\author[rvb]{Gerrit Cornelis Arends}
\author[rvt]{Philipp Amrein}
\author[rvb]{Edwin Versteeg}
\author[rvb]{Dennis W. J. Klomp}
\author[rvt]{Maxim Zaitsev}
\author[rvb,rvc]{Chantal M. W. Tax}
\author[rvt]{Sebastian Littin}

\cortext[cor1]{Corresponding author}

\address[rvt]{Division of Medical Physics, Department of Diagnostic and Interventional Radiology, Medical Center - University of Freiburg, Faculty of Medicine - University of Freiburg, Freiburg, Germany}
\address[rvb]{Center for Image Sciences, University Medical Center Utrecht, Utrecht, The Netherlands}
\address[rvc]{Cardiff University Brain Research Imaging Centre (CUBRIC), School ofPhysics and Astronomy, Cardiff University, Cardiff, United Kingdom}


\begin{abstract}

Introduction
High-performance gradients poses a promise for breast diffusion-weighted imaging (DWI) but are restricted by physiological limits in whole-body scanners. While local nonlinear coils offer higher amplitudes, they often suffer from severe gradient reduction near the chest wall.

Methods
We introduced an optimization framework incorporating a constraint to control anterior–posterior gradient variation. A width-based figure of merit was defined to evaluate performance regarding coil efficiency and minimum wire width. A prototype was constructed to validate the design methodology.

Results
The optimized coil achieved a 2.35-fold efficiency increase over standard linear coils. Compared to previous nonlinear designs, the new constraint reduced spatial variation by 35.7\% and improved minimum efficiency near the chest wall by 2.6-fold. Experimental field maps matched simulations with errors under 8\%.

Discussion
The proposed method effectively mitigates the trade-off between gradient strength and spatial uniformity along anterior–posterior direction. By enhancing performance in the posterior breast region, the design addresses a critical limitation of previous local coils.

Conclusion
This framework enables the development of high-performance, robust local gradient coils, facilitating the clinical implementation of advanced DWI protocols for breast cancer screening. 

\end{abstract}

\begin{keyword}
magnetic resonance imaging \sep gradient coil design \sep diffusion weighted imaging \sep non-linear spatial encoding magnetic fields

\end{keyword}

\end{frontmatter}


\section{Introduction}
\label{sec:Introduction} 

Breast cancer is the most common malignancy in women globally and a leading cause of cancer-related mortality, highlighting the necessity for effective screening and early detection strategies. Magnetic resonance imaging (MRI) has become an indispensable modality in breast cancer diagnostics, particularly for high-risk populations and women with dense breast tissue, where mammography sensitivity is limited \cite{iima_road_2023}. Screening‑compatible MR protocols must be short, robust, and quantitative. In a clinical MRI scanner, two pulse‑sequence pillars are dynamic contrast‑enhanced (DCE) MRI and diffusion‑weighted imaging (DWI). Preliminary evidence suggests that DWI could be an alternative to DCE MRI for breast cancer screening while avoiding gadolinium risks \cite{partridge_diffusion-weighted_2017, baxter_meta-analysis_2019}.

For DWI, high b‑value diffusion encoding and short echo time (TE) are important to improve lesion conspicuity \cite{molendowska_diffusion_2024} and quantitative biomarkers such as the apparent diffusion coefficient (ADC) \cite{baltzer_diffusion-weighted_2020}. However, shorter TE for the same b-value are strongly constrained by gradient performance: the maximum gradient strength and slew rate \cite{laun_nmr-based_2012,setsompop_pushing_2013,lasic_apparent_2016,jones_microstructural_2018,ludwig_diffusion_2022}. At present the highest performance whole-body gradient system with 300 mT/m gradient strength is a part of the experimental Connectome whole-body sized scanner \cite{setsompop_pushing_2013}. However, the main limitation of such gradient systems is posed by physiology, namely peripheral nerve stimulation (PNS) and cardiac stimulation \cite{setsompop_pushing_2013}. PNS limitations inversely correlate to the size of the gradient coil. Therefore smaller and more localized gradients allow circumventing PNS limitations \cite{zhang_peripheral_2003,tan_peripheral_2020}. 

Nonlinear gradient coils allow for generating even higher gradient amplitudes locally than conventional linear local gradient coils. We have introduced a nonlinear local single-breast gradient coil capable of producing a gradient strength of over 1 T/m \cite{jia_design_2021, littin_single_2020} for diffusion weighting. However, a large variation of gradient amplitudes along the posterior direction limits the diagnostic value of this setup close to the chest wall. To address this, our current work introduces a novel constraint in the coil design to control y-axis gradient variation. 

In addition to the need for higher gradient amplitudes, several engineering constraints, such as sufficiently wide wire tracks, need be considered in the design of a local breast coil. Typically, the minimum width of the coil wire tracks limits the maximum current that can flow through the wire, thereby restricting the achievable maximum encoding gradient for imaging. In this study, we propose a novel figure of merit (FOM) that relates to both minimum wire width and gradient coil efficiency. This figure of merit is used to evaluate coil performance and guide the formulation of design optimization problems. Using this optimization framework, a double-breast nonlinear gradient coil was designed for cancer screening, and a prototype was constructed to demonstrate the validity of the design methodology.

\section{Methods}
\label{sec:Methods} 
\subsection{A width-based performance measure}
To establish a figure of merit related to the minimum wire width for a nonlinear coil, it is important to first examine the corresponding performance parameter used to evaluate a conventional linear coil. In the context of a linear gradient coil, the width-based figure of merit $\beta_J$ is defined as $\beta_J := \eta w_{\min}$, as outlined in \cite{PooleCrozier2012a}. Here, $w_{\min}$ represents the minimum wire width, and $\eta$ signifies the efficiency of the linear graident coil. The coil efficiency $\eta$ (T/(m$\cdot$A)) is defined as the ratio of the gradient amplitude $G$ of the magnetic encoding field generated by the coil to the coil current $I$, expressed as $\eta := G/I$ \cite{turner_minimum_1988}. $\beta_J$ has the units of T/A when $w_{\min}$ is given in meters. Typically, the value of $G$ is calculated at the origin \cite{turner_minimum_1988} or at the center of a region of interest (ROI) \cite{sanchez_simple_2007} for a linear coil. Therefore, we can deduce that $\beta_J = Gw_{\min}/I = G/{|\vec{J}|}_{\max}$, where ${|\vec{J}|}_{\max}$ indicates the maximum norm of the electrical current density vector $\vec{J}$ flowing through a current-carrying surface.

For nonlinear coils, the gradient strength of encoding fields typically exhibits spatial variation within the ROI. To capture a meaningful nonlinear encoding field within the ROI, it is logical to aggregate the gradient strength over all test points in the ROI. This aggregate is then considered in the definition of a width-based figure of merit for a nonlinear coil. Based on these considerations, the width-based FOM is defined as follows:
\begin{linenomath}
	\begin{align}
		\beta_J&:=\frac{\sum_{\vec{x}_i\in{\text{ROI}}}\lvert\nabla B_z(\vec{x}_i)\rvert}{k I}w_{\min} = \frac{\sum_{\vec{x}_i\in{\text{ROI}}}\lvert\nabla B_z(\vec{x}_i)\rvert}{k{|\vec{J}|}_{\max}}
		= \bar{\eta}w_{\min}. \label{equ:NBCD_betaJ}  
	\end{align}
\end{linenomath}  
Here, $k$ denotes the total number of test points $\vec{x}_i$ in the ROI, and $\bar{\eta}$ is the average coil efficiency, given by $\bar{\eta}:=\sum_{\vec{x}_i \in \text{ROI}} \lvert\nabla B_z(\vec{x}_i)\rvert/(kI)$. The $B_z$ is the z-component of the magnetic field $\vec{B}$ generated by current density $\vec{J}$, which is calculated using the Biot-Savart law. Note that the formula (\ref{equ:NBCD_betaJ}) for $\beta_J$ can also be utilized for calculations of the width-based FOM for a linear coil in subsequent examples, enabling fair comparisons between the performance of linear and nonlinear coils.

According to equation (\ref{equ:NBCD_betaJ}), a larger $\beta_J$ for a coil implies a higher average coil efficiency over the ROI for a given minimum wire width of the coil. That also means that the coil generates higher average gradient strength when a fixed current was applied on the coil with a given minimum wire width. 

\subsection{Formulation of the optimizaiton problem}
To propose a suitable optimization problem in the design of a nonlinear breast coil for diffusion weighting, several requirements were considered. The first requirement was that the coil generated an encoding field with strong gradients within the ROI. Given that the minimum wire width of a coil was typically determined by a decision maker based on the target maximum current and other engineering constraints, one approach to achieving stronger gradients was to increase the width-based FOM $\beta_J$ during the coil design process.

To enhance the width-based figure of merit $\beta_J$ for the nonlinear coil design, the maximum norm of the current density $\vec{J}$ needed to be reduced while maintaining the average gradient strength of the encoding fields across the ROI, as described by equation (\ref{equ:NBCD_betaJ}). A successful approach \cite{jia_design_2017, JiaKorvink2011a} to reducing the maximum norm of the electric current density $\vec{J}$ on a current-carrying surface $\Gamma$ involved minimizing the p-norm ($\lVert J \rVert_p:=(\int_{\Gamma}|\vec{J}|^p d\Gamma)^{1/p} (p>2)$) of the surface current density, which converges to the maximum norm of $\vec{J}$ when $p$ approaches infinity. Moreover, coil power dissipation was incorporated as a regularization term in the optimization problem to make the resulting coil layout smooth \cite{PooleCrozier2012a}.

The second requirement was to control y-axis gradient variation, possibly enhancing the diagnostic value of the diffusion weighting close to the chest wall. The third requirement was that the encoding field generated by a non-linear breast coil have a nearly constant gradient magnitude within each coronal slice \cite{jia_design_2021}. This requirement will allow the resulting coil to be employed for diffusion weighting within each coronal slice in a manner similar to the currently established protocols that rely on linear gradients \cite{littin_Approaching_2021}. Finally, electromagnetic force and torque of the coil also needed to be controlled during the design process.

\begin{figure}[htbp]
	\centering \resizebox{0.9\textwidth}{!}{\includegraphics{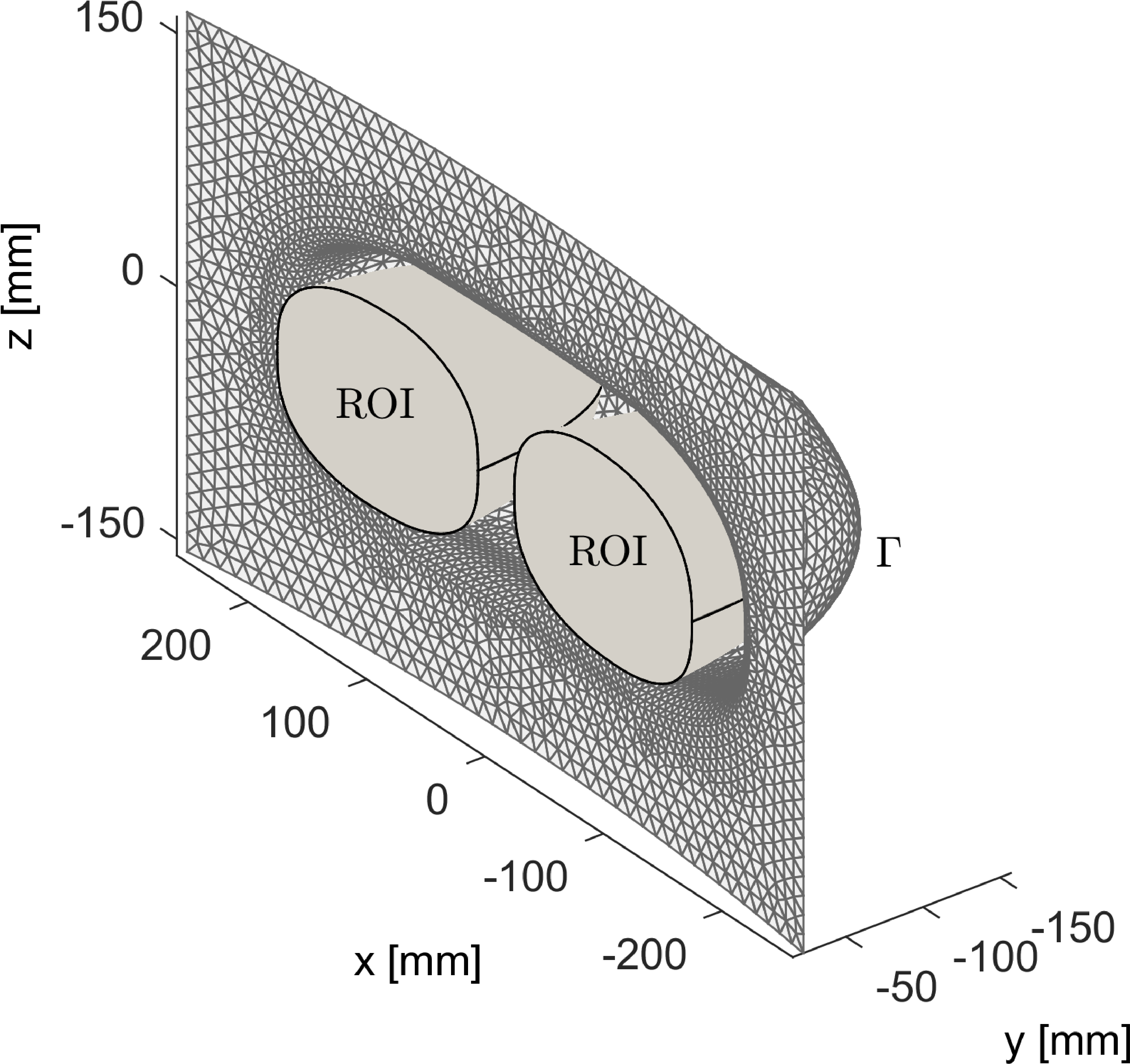}} 
	\caption{Geometries of the ROI and current-carrying surface $\Gamma$.}
	\label{fig:CCSMeshROI_fig1} 
\end{figure}

Based on the above requirements, the following optimization problem was proposed:
\begin{linenomath}
	\begin{align}
		\min_{\psi}\mbox{ }& \sqrt{P(\psi)}+ \alpha_J \lVert\vec{J}(\psi)\rVert_{p} ,\label{equ:NBCD_non-linearF}\\
		\mbox{subject to } &\mbox{  } \frac{1}{k}\sum_{\vec{x}_i\in{\text{ROI}}}\left|\nabla B_z(\psi,\vec{x}_i)\right|\geq C_g, \nonumber \\
		&\mbox{  }D_y:=\left( \sum_{\vec{x}_i\in{\text{ROI}}}\left(\frac{\partial\left|\nabla B_z(\psi,\vec{x}_i)\right|}{\partial y}\right)^2 \right)^\frac{1}{2}\leq k C_y C_g, \label{equ:NBCD_YAxisConstraints}\\
		&\mbox{  }D_s:=\left( \sum_{\vec{x}_i\in{\text{ROI}}}\left(\frac{\partial\left|\nabla B_z(\psi,\vec{x}_i)\right|}{\partial x}\right)^2+\left(\frac{\partial\left|\nabla B_z(\psi,\vec{x}_i)\right|}{\partial z}\right)^2 \right)^\frac{1}{2}\leq k C_s C_g, \label{equ:NBCD_CoronalSliceConstraints}\\
		&\mbox{ } |F_x|:=\left|\int_{\Gamma}B_0J_y(\psi)d\Gamma\right|\leq F_{\max}C_g,\nonumber \\
		&\mbox{ } |F_y|:=\left|-\int_{\Gamma}B_0J_x(\psi)d\Gamma\right|\leq F_{\max}C_g,\nonumber \\
		&\mbox{ } |M_x|:=\left|\int_{\Gamma}B_0J_x(\psi)zd\Gamma\right|\leq M_{\max}C_g,\nonumber \\
		&\mbox{ } |M_y|:=\left|\int_{\Gamma}B_0J_y(\psi)zd\Gamma\right|\leq M_{\max}C_g,\nonumber \\
		&\mbox{ } |M_z|:=\left|-\int_{\Gamma}B_0\left(J_x(\psi)x+J_y(\psi)y\right)d\Gamma\right|\leq M_{\max}C_g.\nonumber
	\end{align}
\end{linenomath}
Here, $\psi$ denotes the scalar piecewise-linear stream function \citep{peeren_stream_2003} of the electric current density vector $\vec{J}(\psi):=(J_x(\psi),J_y(\psi),J_z(\psi))^T$ and $\vec{J}(\psi) = \nabla\times(\psi\vec{n})$ on a current-carrying surface $\Gamma$ (Fig. \ref{fig:CCSMeshROI_fig1}) with a normal unit vector $\vec{n}$. $P(\psi):=\lVert\vec{J}(\psi)\rVert_{2}^2 /(\tau\sigma_e)$ is power dissipated by the coil, where $\tau$ and $\sigma_e$ indicate the thickness and the electrical conductivity of the surface $\Gamma$, respectively. The vectors $\vec{x}_i = (x_i, y_i, z_i)^T$, $i=1,\ldots, k$, denote the positions of the test points in the ROI. The z-component of the magnetic field, $B_z(\psi,\vec{x}_i$), is generated by  $\vec{J}(\psi)$ and evaluated at $\vec{x}_i$ using the Biot-Savart law.

In this problem, $D_s:=\left( \sum_{\vec{x}_i\in{\text{ROI}}}\left({\partial\left|\nabla B_z(\psi,\vec{x}_i)\right|}/{\partial x}\right)^2+\left({\partial\left|\nabla B_z(\psi,\vec{x}_i)\right|}/{\partial z}\right)^2 \right)^\frac{1}{2}$ indicates the per-slice constraint, which is used to homogenize the gradient strength of $B_z$ within each of coronal slices, as proposed in our previous work \cite{jia_design_2021}. Compared to the previous optimization formulation in \cite{jia_design_2021}, a new constraint for $D_y:=\left( \sum_{\vec{x}_i\in{\text{ROI}}}\left({\partial\left|\nabla B_z(\psi,\vec{x}_i)\right|}/{\partial y}\right)^2 \right)^\frac{1}{2}$ is introduced to control the gradient variation along the y-axis. 

In this problem, $F_x$ and $F_y$ indicate x- and y-components of the net force on the coil, respectively, in the presence of a main magnetic field $B_0$ along z-direction. In the examples of this manuscript, the main field $B_0$ is assumed to be homogeneous with the strength of 3 T since the built prototype of the designed coil will be used close to the isocenter of the main magnet generating field $B_0$. Therefore, the constraints of $F_x$ and $F_y$ are naturally satisfied and can be ignored during the optimization of Problem (\ref{equ:NBCD_non-linearF}). $M_x$, $M_y$ and $M_z$ denote x-, y- and z-components of the magnetic torque executed on the coil, respectively. $F_{\max}$ and $M_{\max}$ are two given positive constants with the unit of N and N$\cdot$m in order to set a limitation of the total force and the total magnetic torque, respectively.

The optimization problem also contains four positive tuning parameters, i.e. $\alpha_J$, $C_g$, $C_s$ and $Cy$. However, considering that the stream function $\psi$ in the current optimization problem is linearly dependent on $C_g$, we only need to consider three tuning parameters $\alpha_J$, $C_s$ and $C_y$.

In order to guarantee that resulting wire patterns are closed on the current-carrying surface $\Gamma$, all values of the stream function $\psi$ must remain constant on each closed boundary of $\Gamma$  \citep{peeren_stream_2003}. Moreover, $\psi$ is typically specified as a given constant, such as 0, at one point in $\Gamma$ or on one closed boundary of $\Gamma$ to avoid the existence of infinitely many solutions of Problem (\ref{equ:NBCD_non-linearF}). Without this requirement for any stream function $\psi$, which is a solution of Problem (\ref{equ:NBCD_non-linearF}), $\psi$ plus any constant is also a solution of Problem (\ref{equ:NBCD_non-linearF}). As shown in Figure \ref{fig:CCSMeshROI_fig1}, the current-carrying surface $\Gamma$ has only one boundary and $\psi$ on the boundary is set to 0 in all subsequent numerical examples.

\subsection{Two other performance measures}
In addition to the width-based figure of merit $\beta_J$, the resistive FOM $\beta_P$, inductive FOM $\beta_W$ and the coefficient of variation $CV_D$ of the coil efficiency $\eta$ over the domain $D$ are used to assess and compare different breast coil designs. The formulae for $\beta_P$, $\beta_W$ \cite{jia_design_2021} and $CV_D$ are defined as follows:  
\begin{linenomath}
	\begin{align}
	\beta_P&:=\frac{\sum_{\vec{x}_i\in{\text{ROI}}}\lvert\nabla B_z(\psi,\vec{x}_i)\rvert}{k\sqrt{P(\psi)}}, \label{equ:NBCD_betaP}   \\
	\beta_W&:=\frac{\sum_{\vec{x}_i\in{\text{ROI}}}\lvert\nabla B_z(\psi,\vec{x}_i)\rvert}{k\sqrt{W(\psi)}}, \label{equ:NBCD_betaW}  \\
	CV_D&:= \frac{{\text{std}}(\eta)_D}{{\text{mean}}(\eta)_D}. \label{equ:NBCD_CVD}
	\end{align}
\end{linenomath}                                         
Here, $W(\psi):=\frac{\mu_0}{8\pi}\int_{\Gamma}\int_{\Gamma'}\frac{\vec{J}(\psi(\vec{x}))\cdot\vec{J}(\psi(\vec{x}'))}{|\vec{x}-\vec{x}'|} d\Gamma'd\Gamma$ \cite{lemdiasov_stream_2005} indicates coil magnetic energy and $\mu_0$ denotes the magnetic constant of $4\pi\times10^{-7}$ H/m. The std($\eta$)$_D$ and mean($\eta$)$_D$ denote the standard deviation and mean value of the coil efficiency $\eta$ over the domain $D$, respectively.

As shown in Equation (\ref{equ:NBCD_betaP}), $\beta_P$, with the unit of T/(m$\cdot$A$\cdot\Omega^{1/2}$), is defined as the average gradient strength of the encoding field over the ROI divided by the square root of the power dissipation. For a given dissipated power, a larger $\beta_P$ indicates a higher average gradient strength. Similarly, as presented in Equation (\ref{equ:NBCD_betaW}), $\beta_W$, with the unit of T/(m$\cdot$A$\cdot$H$^{1/2}$), is described as the average gradient strength of the encoding field over the ROI divided by the square root of the stored magnetic energy. Thus, for given magnetic energy, a larger $\beta_W$ implies a higher average gradient strength and better coil performance. Finally, as described in Equation (\ref{equ:NBCD_CVD}), the coefficient of variation $CV_D$, with no unit, is employed to assess the spatial uniformity of the gradient strength over a specified domain $D$. In this work, the domain $D$ represents either an individual coronal slice ($D=s$) or the entire ROI ($D=ROI$). 

\subsection{Numerical optimization procedure}
In order to assist a decision maker to select a suitable breast coil, Problem (\ref{equ:NBCD_non-linearF}) with different values of $\alpha_J$, $C_s$ and $C_y$ was solved to obtain different non-linear coil layouts. In order to obtain suitable values of the tuning parameters for a non-linear breast coil design comparable to its corresponding linear one, the whole procedure of selecting the tuning parameters has been divided into three steps. In the first step, a linear breast coil is designed using the method described in \cite{jia_design_2021} when the target field $B_z^*$ is provided. Here, $B_z^*$ is specified as linear $G_z$ with the gradient strength of 50 mT/m, as recommended in \cite{jia_design_2021}. For the resulting linear coil, $D_y$ and $D_s$ indicated by $D_y^l$ and $D_s^l$ were calculated using equations (\ref{equ:NBCD_YAxisConstraints}) and (\ref{equ:NBCD_CoronalSliceConstraints}), respectively. In the examples, $C_g$ was specified as to be $\left( \sum_{\vec{x}_i\in{\text{ROI}}}\left({\left|\nabla B_z(\psi,\vec{x}_i)\right|}\right)^2 \right)^\frac{1}{2}/k$, where $B_z$ was generated by the resulting linear coil. The two constants $C_y^l$ and $C_s^l$ were defined by $D_y^l/k/C_g$ and $D_s^l/k/C_g$, respectively, which were used for next steps. 

Secondly, the parameter $C_s$ was selected from a list that decreases proportionally starting from $C_s^l$. For each given $C_s$, Problem (\ref{equ:NBCD_non-linearF}) with $\alpha_J$ = 0 and without the constraint (\ref{equ:NBCD_YAxisConstraints}) was solved to obtain a non-linear breast coil. For the resulting non-linear coil, $D_y$ indicated by $D_y^m$ was calculated using equation (\ref{equ:NBCD_YAxisConstraints}) and the constant $C_y^m$ was defined by  $D_y^m/k/C_g$. In the examples, $C_s$ was taken from [$C_s^l$ $C_s^l$/2.5 $C_s^l$/5] and the resulting stream function from the first step was used as an initial value of the current problem. Finally, the parameters $C_y$ and $\alpha_J$ were taken from the intervals [$C_y^l$ $C_y^m$] and [0 $\alpha_J^m$] with the steps $C_y^s$ and $\alpha_J^s$, respectively. 
In the examples, $\alpha_J^m$, $C_y^s$ and $\alpha_J^s$ were specified as to be 8e-3, $(C_y^m-C_y^l)$/10 and 2.5e-4, respectively.

\subsection{Coil geometry and physical parameters}
Volumes and surfaces were designed using the 3D design suite Inventor (Autodesk, San Rafael, CA, USA). For the examples shown, the current-carrying surface $\Gamma$ (Fig. \ref{fig:CCSMeshROI_fig1}) was designed to have one volume, large enough to house an additional radio frequency (RF) transceiver coil, with enough space to fitting the necessary electrical components. The RF coil has two separate independent cups for each breast, which are the ROIs of the coil design  (Fig. \ref{fig:CCSMeshROI_fig1}). The volume of each ROI was chosen to include at least $85\%$ of the female population. Each ROI extends 160mm x 125mm x 120mm (horizontal x sagittal x longitudinal), resulting in a net volume of 1.44L. The current carrying surface consists of a top part along the horizontal direction, which is slightly curved for patient comfort and has a rectangular base shape of 520mm x 320mm (horizontal x longitudinal). The main opening is 290mm x 152mm (horizontal x longitudinal) wide and has a depth between 126mm and 136mm (sagittal) from the top of the curved surface. In Autodesk Inventor the bridge curve and loft functions were used to design the target volumes, the current carrying surface, and the RF carrier structure. The thickness $\tau$ and the conductivity $\sigma_e$ were set as 3 mm and  $5.998\times10^4$ S/mm, respectively. $F_{\max}$ and $M_{\max}$ are specified as $1\times10^{-4}$ N and $1\times10^{-4}$ N$\cdot$m, respectively, to limit the total net magnetic force and torque of a designed coil.

\subsection{Prototype and field map measurement}
One optimal stream function was selected and transformed to a discretized wire pattern using the automatized algorithm \cite{amrein_coilgen_2022}. A prototype of the discretized coil was built in-house \cite{Arends_Littin_2025,Arends_Tax_2026}. In order to demonstrate the validity of the design method, an initial field map of the prototype was measured using a modified 2D multi-slice spoiled gradient echo sequence (GRE) similar to \cite{AnnaZaitsev2013a}. The modified sequence consists of two parts. The first part is a standard 2D multi-slice spoiled GRE sequence and the second part is the same sequence with an extra gradient pulse on the prototype (0.5 ms, 2 A). The gradient pulse parameters were chosen to avoid wrapping of the resulting phase map. The imaging parameters were: TE = 7 ms, TR = 10 ms, flip angle = 20 degree, in-plane resolution = 4 mm, field of view = 320 X 320 mm, slice thickness = 4 mm and 55 slices. All the measured data were acquired using a water phantom and a 3T Achieva MRI system (Philips, Best, The Netherlands).

\section{Results}
\label{sec:Results} 
This section is organized into four subsections. The first subsection presents a comparison between a linear gradient coil and two non-linear designs—one without and one with the proposed $y$-axis gradient variation constraint. The second subsection~\ref{subsec:NoneLinearCoils} investigates the influence of key tuning parameters ($C_s$, $C_y$, and $\alpha_J$) on multiple figures of merit, highlighting performance trade-offs. The third subsection~\ref{subsec:NoneLinearCoils3p5mm} evaluates a series of non-linear designs that satisfy a minimum wire width constraint of 3.5~mm, focusing on gradient uniformity and layout complexity. Finally, subsection~\ref{subsec:Experiments} reports on the experimental validation of a prototype coil, including comparisons between measured and simulated magnetic fields.

\subsection{Comparison of Linear and Non-linear Coils}
\label{subsec:LinearCoils}
Table~\ref{table:Comp_Linear_NonLinearCoils} presents a quantitative comparison between a linear $G_z$ coil, an old non-linear gradient coil, and a new non-linear gradient coil, all designed using the same $C_s = 0.1589$. The old non-linear coil was optimized without the $y$-axis gradient variation constraint~(\ref{equ:NBCD_YAxisConstraints}), as described in~\cite{jia_design_2021}, while the new non-linear coil incorporates this additional constraint to improve gradient uniformity along the anterior–posterior direction.

Compared to the linear coil, both non-linear designs achieve substantially higher encoding efficiency. Specifically, the average coil efficiency $\bar{\eta}$ of the old and new non-linear coils is approximately 2.79 and 2.35 times higher than that of the linear coil, respectively. The width-based figure of merit $\beta_J$ improves by factors of approximately 3.78 (old non-linear) and 4.25 (new non-linear) over the linear design. These increases demonstrate the advantage of nonlinear encoding in producing stronger local gradients per unit current.

However, this gain in efficiency comes with increased spatial non-uniformity. Compared to the linear coil, the coefficient of variation $CV_{{ROI}}$ for the old non-linear coil increases by a factor of approximately 5.5. By including the $C_y$ constraint in the new non-linear coil, $CV_{{ROI}}$ shows a 35.7\% decrease compared to the old design. Moreover, the minimum coil efficiency near the chest wall also improves significantly. Compared to the old non-linear coil, the minimum value of $\eta$ in the new non-linear coil increases by a factor of approximately 2.63.

Although the new non-linear coil exhibits slightly lower $\bar{\eta}$ and performance metrics $\beta_P$ and $\beta_W$ than the old non-linear version (decreased by factors of approximately 1.18 and 1.17, respectively), it achieves a markedly better trade-off between efficiency and gradient uniformity. Furthermore, the new coil features a larger minimum wire width $w_{\min}$ (3.233~mm vs.\ 2.431~mm), which is beneficial for fabrication and thermal management.

\subsection{Performance impact of different tuning parameters in non-linear coil designs}
\label{subsec:NoneLinearCoils}
\begin{figure}[htbp]
	\centering \subfigure[$\beta_J$ when $C_s$ = 0.1589]{ \label{fig:BetaJs0to8CcD1_fig3a} \includegraphics[height=0.174\textheight]{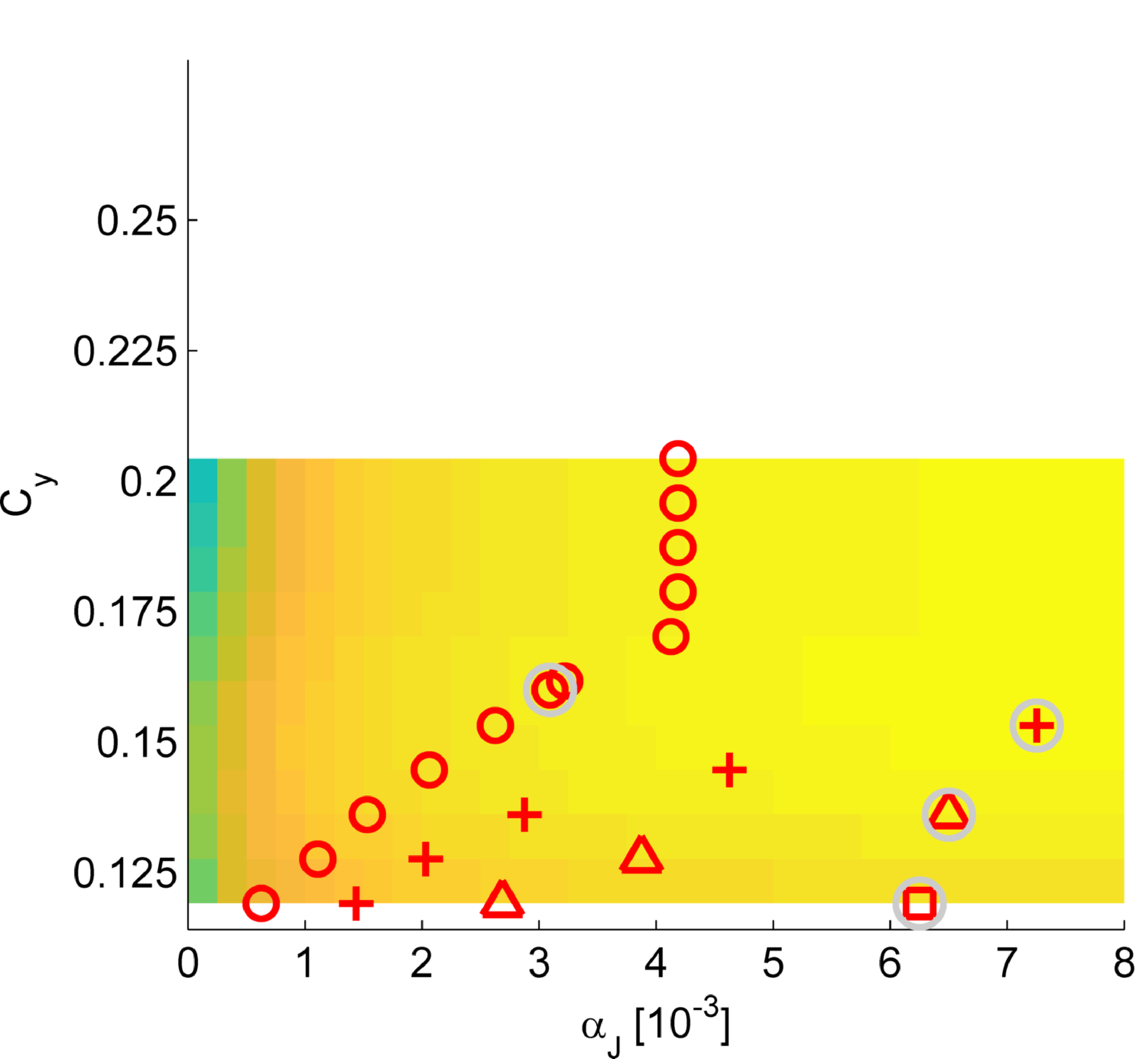}
	} \subfigure[$\beta_J$ when $C_s$ = 0.06357]{ \label{fig:BetaJs0to8CcD2p5_fig3d}  \includegraphics[height=0.174\textheight]{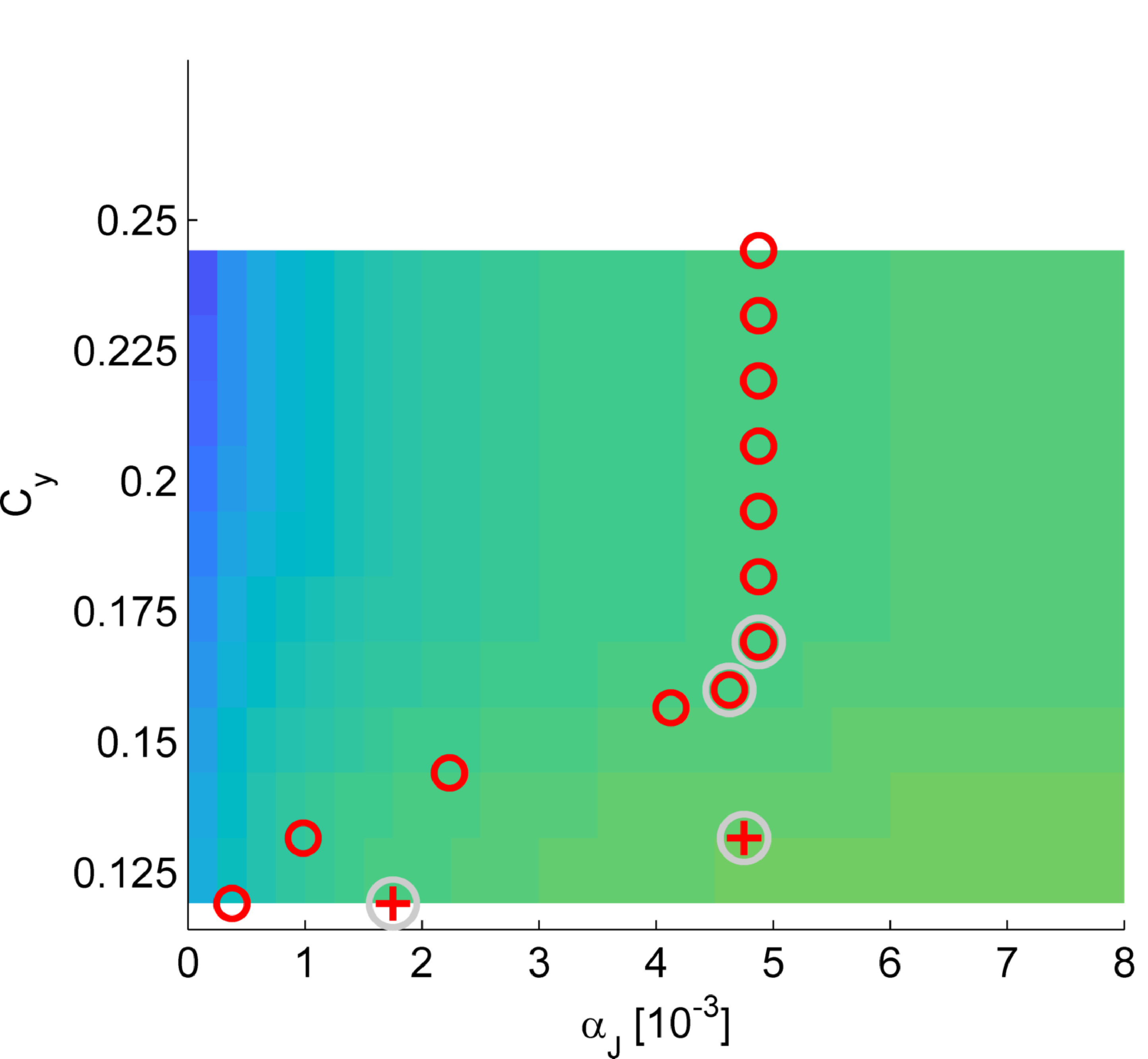}
	} \subfigure[$\beta_J$ when $C_s$ = 0.03178]{ \label{fig:BetaJs0to8CcD5_fig3g} \includegraphics[height=0.174\textheight]{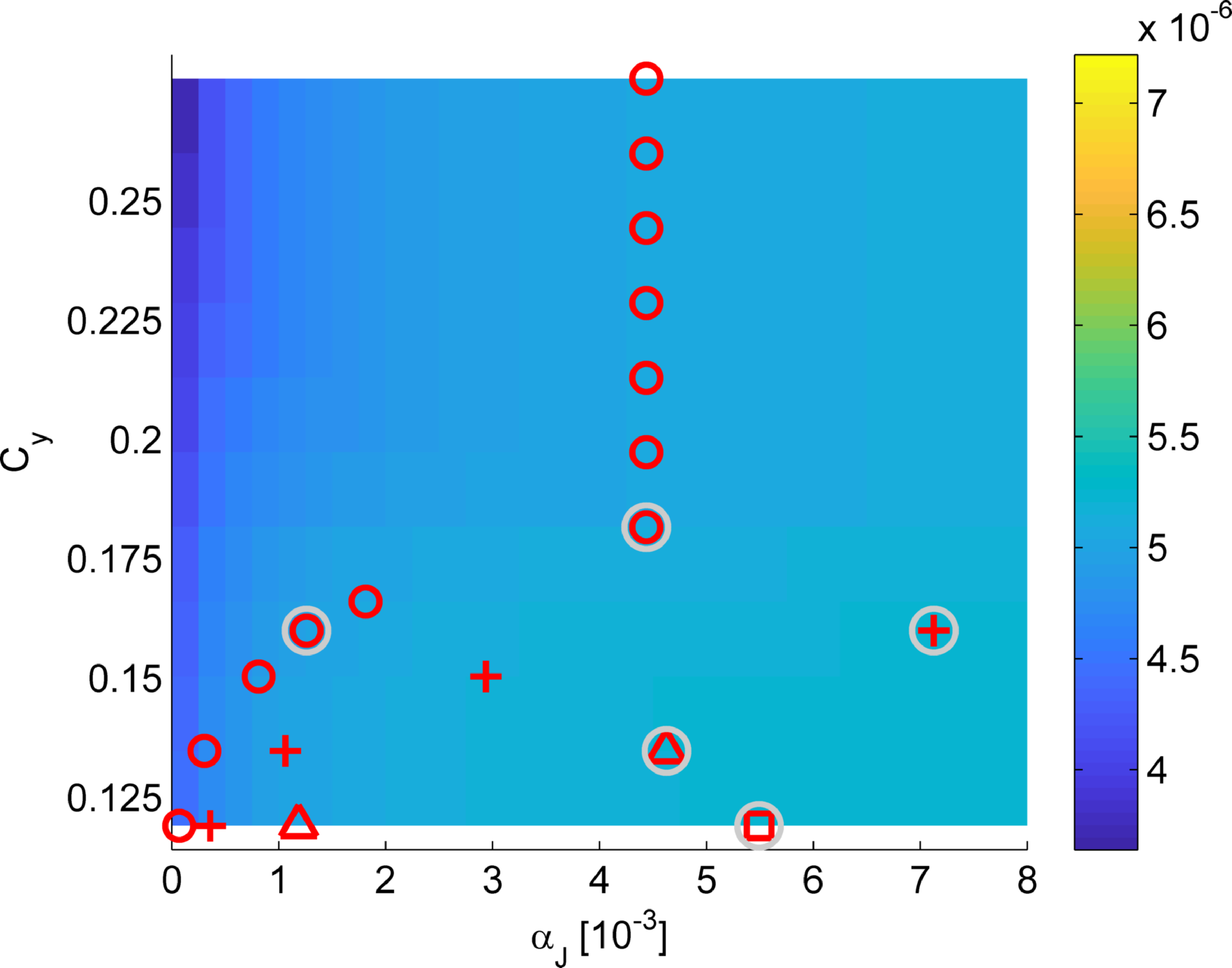}
	} \subfigure[$\beta_P$ when $C_s$ = 0.1589]{ \label{fig:BetaPs0to8CcD1_fig3b} \includegraphics[height=0.174\textheight]{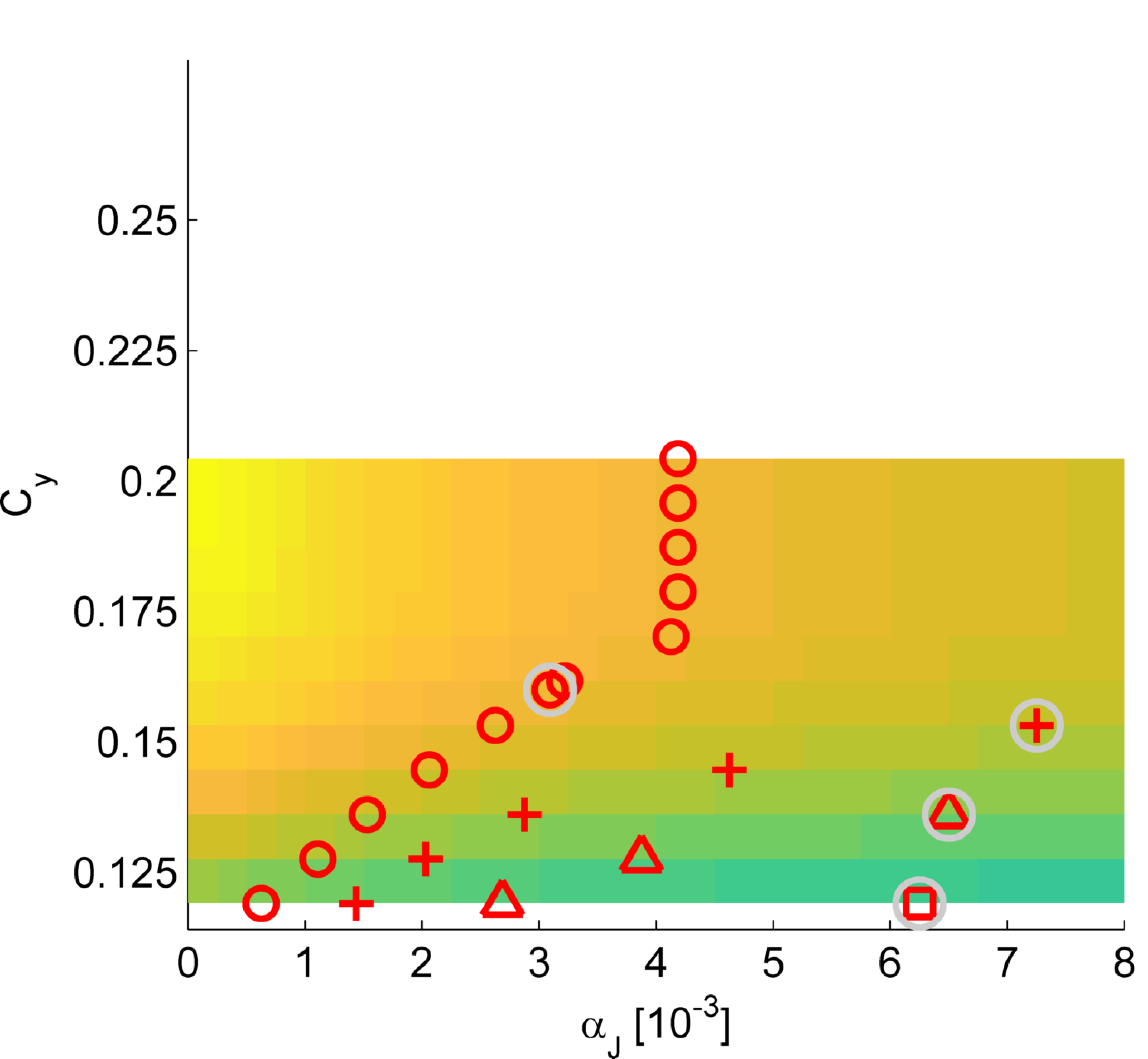}
	} \subfigure[$\beta_P$ when $C_s$ = 0.06357]{ \label{fig:BetaPs0to8CcD2p5_fig3e} \includegraphics[height=0.174\textheight]{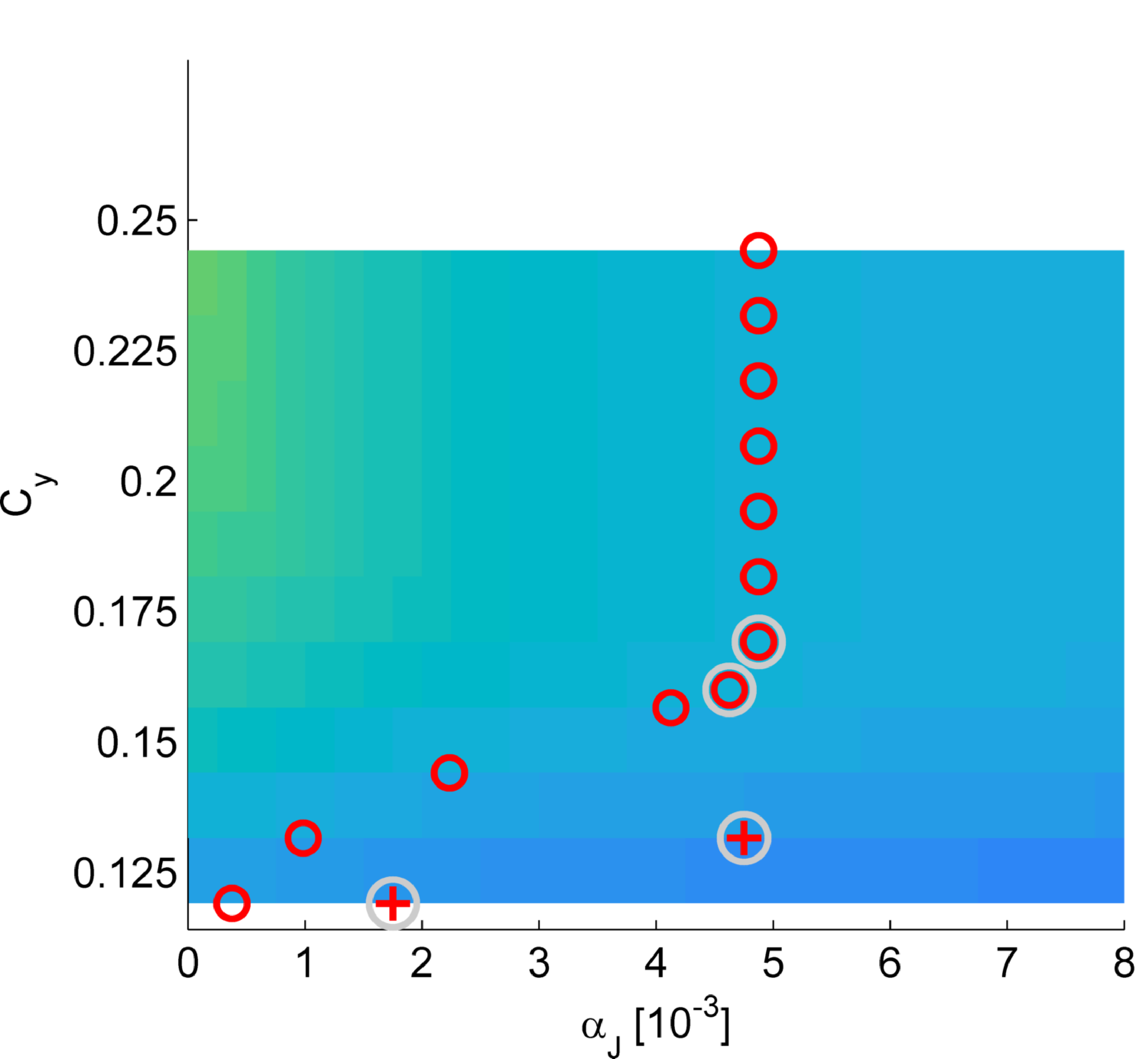}
	} \subfigure[$\beta_P$ when $C_s$ = 0.03178]{ \label{fig:BetaPs0to8CcD5_fig3h} \includegraphics[height=0.174\textheight]{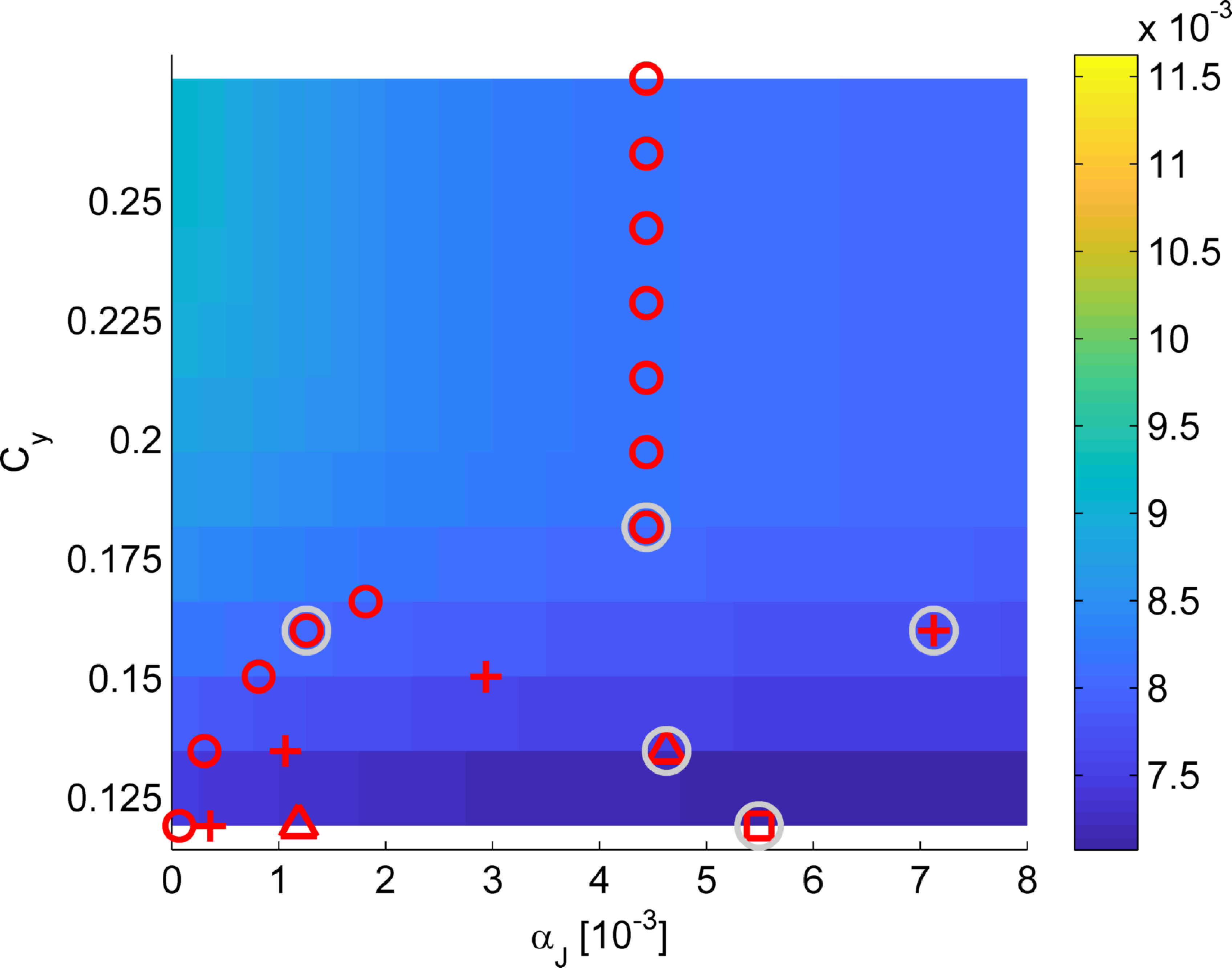}
	} \subfigure[$\beta_W$ when $C_s$ = 0.1589]{ \label{fig:BetaWs0to8CcD1_fig3c} \includegraphics[height=0.165\textheight]{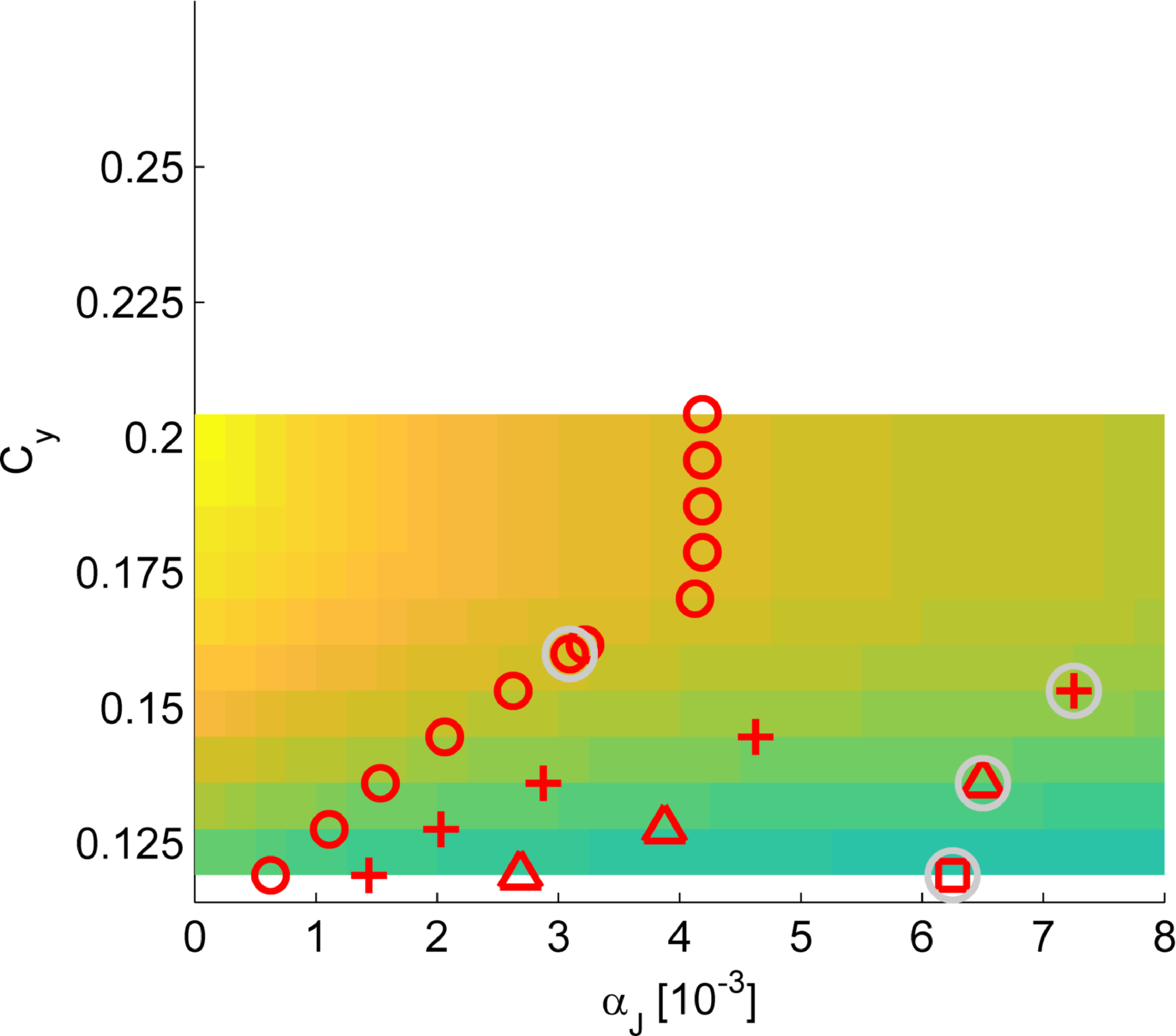}
	}  \subfigure[$\beta_W$ when $C_s$ = 0.06357]{ \label{fig:BetaWs0to8CcD2p5_fig3f} \includegraphics[height=0.165\textheight]{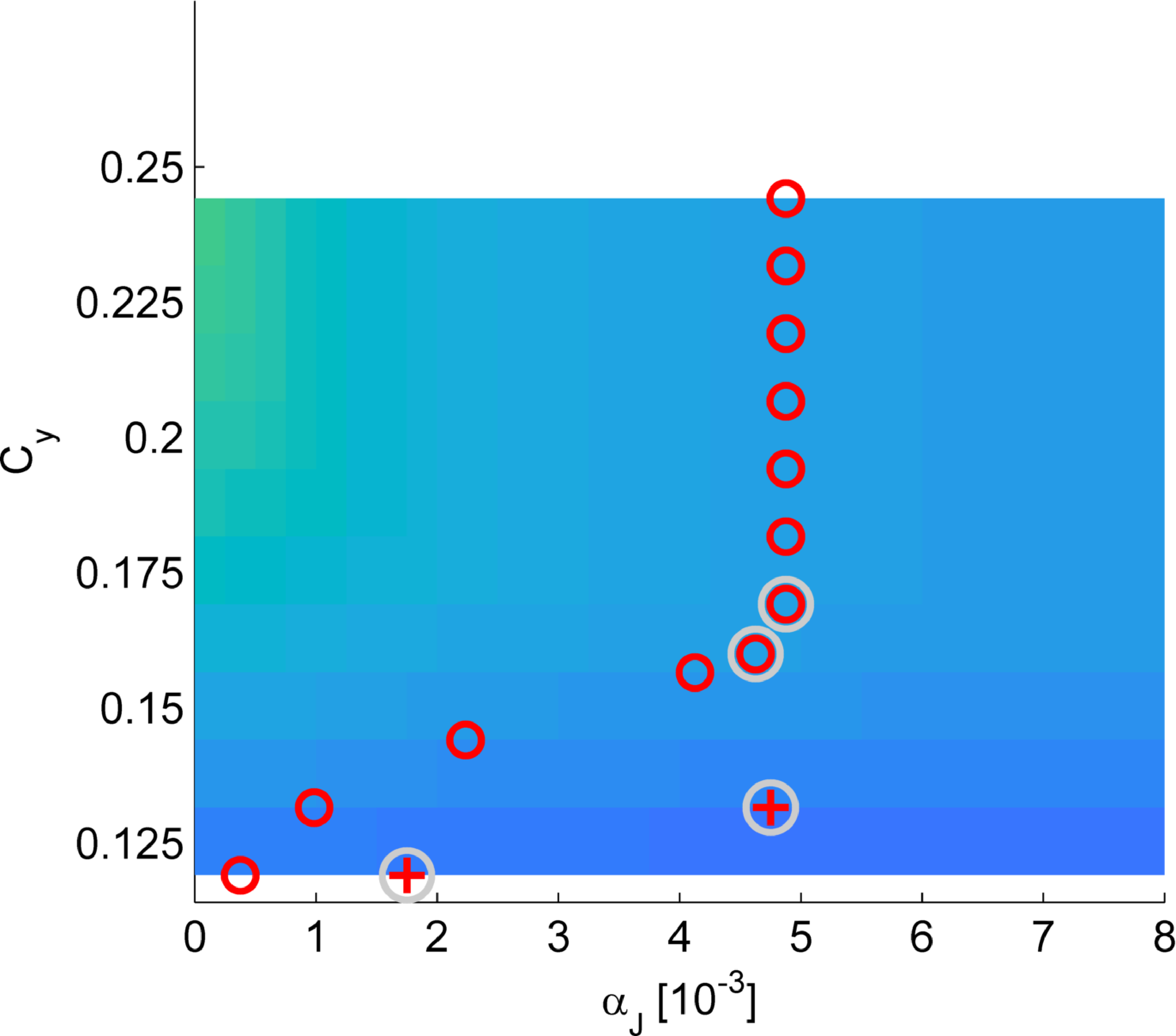}
	}  \subfigure[$\beta_W$ when $C_s$ = 0.03178]{ \label{fig:BetaWs0to8CcD5_fig3i} \includegraphics[height=0.165\textheight]{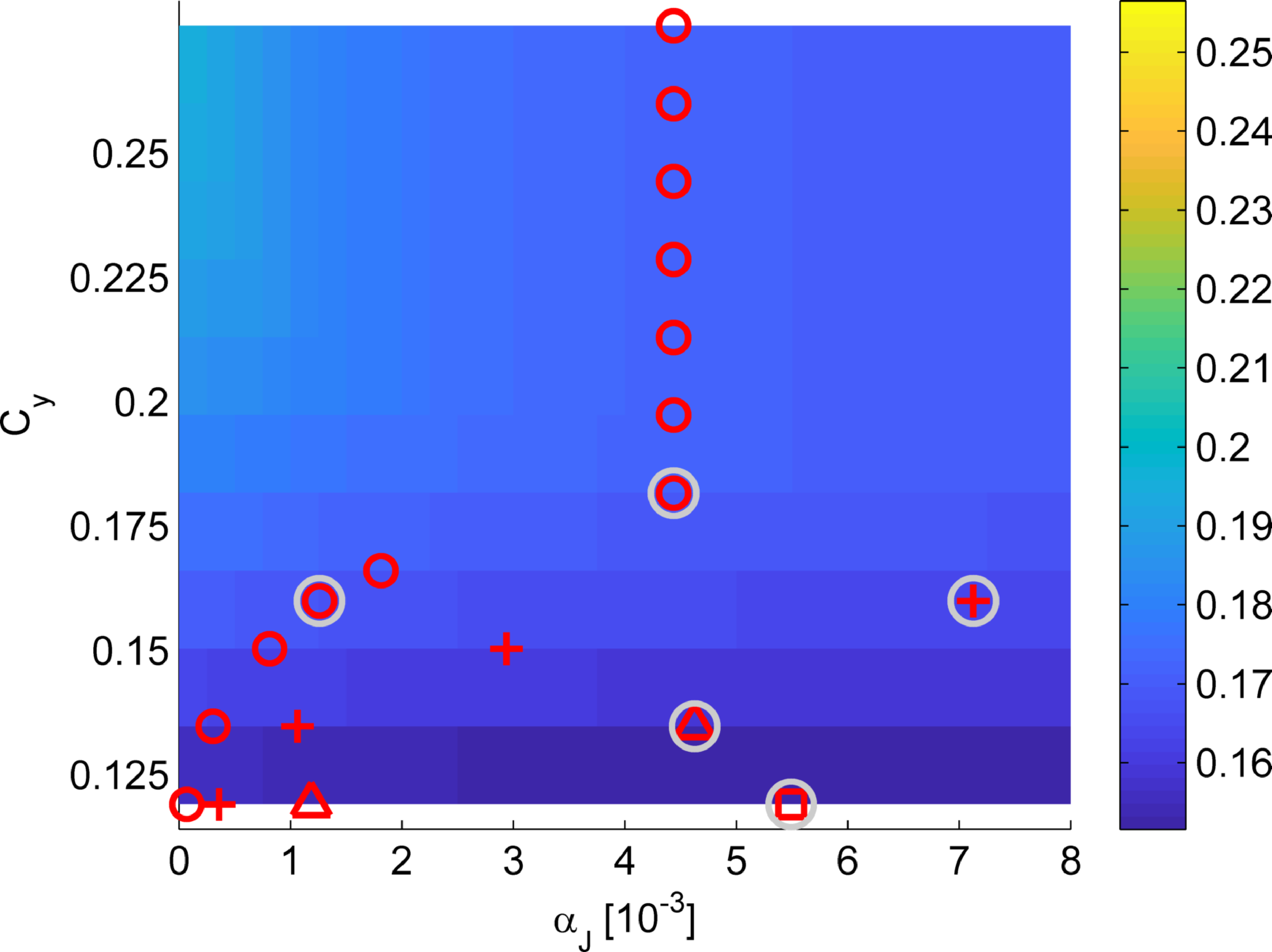}
	}\caption{Figures of merit $\beta_J$, $\beta_P$ and $\beta_W$ for $C_s$ = 0.1589 (a, d, g), 0.06357 (b, e, h) and 0.03178 (c, f, i). Red markers (circles, crosses, triangles, and squares) denote 26, 28, 30, and 32 wire turns, respectively, where the number of turns in each case is determined by the minimum wire width constraint of 3.5 mm. Performance parameters corresponding to the points marked by gray circles are detailed in Tables \ref{table:NBCD_Non-LinearCoils_CCD1}, \ref{table:NBCD_Non-LinearCoils_CCD2p5}, and \ref{table:NBCD_Non-LinearCoils_CCD5}. Subfigures (a)–(c), (d)–(f), and (g)–(i) each share a common color bar.}
	\label{fig:FOMs_fig2} 
\end{figure}

Figure \ref{fig:FOMs_fig2}.a, \ref{fig:FOMs_fig2}.b, and \ref{fig:FOMs_fig2}.c illustrate the variation of $\beta_J$ with respect to different values of $C_y$ and $\alpha_J$ for $C_s = 0.1589$, $0.06357$, and $0.03178$, respectively. As observed, $\beta_J$ decreases as $C_s$ decreases, given fixed values of $C_y$ and $\alpha_J$. Specifically, compared to $C_s = 0.1589$, the mean values of $\beta_J$ for $C_s = 0.06357$ and $C_s = 0.03178$ are reduced by factors of approximately 1.224 and 1.396, respectively.

For a given $C_s$, $\beta_J$ increases rapidly at lower values of $\alpha_J$, followed by a more gradual rise as $\alpha_J$ increases, for fixed $C_y$. When $C_s = 0.1589$, $\beta_J$ initially increases with $C_y$ for a fixed $\alpha_J$, but then slightly decreases or levels off. This behavior suggests that optimal average coil efficiency $\bar{\eta}$ is achieved at higher $\alpha_J$ combined with a moderately increased $C_y$, for a given minimum wire width. In contrast, when $C_s = 0.06357$ and $0.03178$, $\beta_J$ tends to decrease as $C_y$ increases at a fixed $\alpha_J$, and subsequently either slightly declines further or remains stable. This indicates that, in these cases, optimal coil efficiency $\bar{\eta}$ is achieved at larger $\alpha_J$ and lower $C_y$, highlighting a shift in the optimal parameter regime as $C_s$ decreases.

The second and third rows of Figure \ref{fig:FOMs_fig2} present the trends of $\beta_P$ and $\beta_W$, respectively. Similar to $\beta_J$, both $\beta_P$ and $\beta_W$ decrease with decreasing $C_s$, for fixed $C_y$ and $\alpha_J$. Specifically, the mean values of $\beta_P$ for $C_s = 0.06357$ and $0.03178$ are reduced by factors of approximately 1.17 and 1.315, respectively, compared to $C_s = 0.1589$. Likewise, the mean values of $\beta_W$ decrease by factors of 1.2 and 1.343, respectively. For a given $C_s$, both $\beta_P$ and $\beta_W$ exhibit a decreasing trend with increasing $\alpha_J$ at fixed $C_y$. Additionally, as $C_y$ increases for fixed $\alpha_J$, both $\beta_P$ and $\beta_W$ initially decrease and then either slightly decline further or stabilize. These trends suggest that optimal values of $\beta_P$ and $\beta_W$ are generally achieved at higher $\alpha_J$ and moderately elevated $C_y$.

\subsection{Non-linear coil designs with the minimum width of 3.5 mm}
\label{subsec:NoneLinearCoils3p5mm}

Figure \ref{fig:FOMs_fig2} also highlights the points with a minimum width of 3.5 mm. The red points delineated by circles, crosses, triangles and rectangles represent wire turn numbers of 26, 28, 30 and 32, respectively. Some points with relatively high $\beta_J$ values are marked with gray circles. Their specific performance metrics for $C_s = 0.1589$, $0.06357$, and $0.03178$ are presented in Table \ref{table:NBCD_Non-LinearCoils_CCD1}, \ref{table:NBCD_Non-LinearCoils_CCD2p5} and \ref{table:NBCD_Non-LinearCoils_CCD5} respectively.

Table \ref{table:NBCD_Non-LinearCoils_CCD1} compares the performance of four non-linear gradient coil designs optimized with $C_s = 0.1589$, while varying the y-axis gradient variation constraint $C_y$. For each design, the parameter $\alpha_J$ was individually adjusted to ensure that the minimum wire width constraint of 3.5 mm was satisfied. As $C_y$ increases from 0.1191 to 0.16, a clear deterioration in gradient uniformity is observed. The standard deviation of coil efficiency $\sigma_\eta$ and the coefficient of variation $CV_{ROI}$ over the ROI increase by factors of 1.27 and 1.25, respectively. In contrast, the average coil efficiency $\bar{\eta}$ remains relatively stable across the range of $C_y$. An improvement in $\bar{\eta}$ by a factor of approximately 1.035 is observed as $C_y$ increases from 0.1191 to 0.1532. However, when $C_y$ increases further from 0.1532 to 0.16, $\bar{\eta}$ decreases by a factor of 1.02. These observations suggest that a moderate increase in $C_y$ can slightly enhance average efficiency, but at the cost of reduced gradient uniformity across coronal slices.

Table \ref{table:NBCD_Non-LinearCoils_CCD2p5} compares the performance of four non-linear gradient coil designs optimized with $C_s = 0.06357$, while varying the y-axis gradient variation constraint $C_y$. As $C_y$ increases from 0.1191 to 0.1692, a progressive deterioration in gradient uniformity is again observed. The standard deviation $\sigma_\eta$ and the coefficient of variation $CV_{ROI}$ over the ROI increase by factors of approximately 1.341 and 1.344, respectively. Meanwhile, the average coil efficiency $\bar{\eta}$ remains relatively stable. A slight improvement is seen as $C_y$ increases from 0.1191 to 0.1316, followed by a marginal decrease as $C_y$ increases further to 0.1692. The overall variation in $\bar{\eta}$ across this range remains within a factor of approximately 1.02, indicating that relaxing $C_y$ has minimal impact on efficiency in this regime.

Table \ref{table:NBCD_Non-LinearCoils_CCD5} summarizes the performance of five non-linear gradient coil designs optimized with $C_s = 0.03178$, while varying the constraint $C_y$. As $C_y$ increases from 0.1191 to 0.1817, a notable decline in gradient uniformity is observed. The standard deviation $\sigma_\eta$ and coefficient of variation $CV_{ROI}$ over the ROI increase by factors of approximately 1.44 and 1.49, respectively. These trends confirm that relaxing the $C_y$ constraint leads to increasing heterogeneity in the encoding gradient along the y-axis. Unlike the cases with higher $C_s$, the average coil efficiency $\bar{\eta}$ does not exhibit a clear improvement with increasing $C_y$. Across the full range tested, the variation in $\bar{\eta}$ remains within a factor of approximately 1.05, indicating limited sensitivity of average efficiency to changes in $C_y$ when strong $C_y$ is applied.

Table \ref{table:NBCD_Non-LinearCoils_CCD5} also includes two designs with $C_y = 0.16$, which differ in their values of $\alpha_J$ (1.26e-3 and 7.125e-3). Compared to the lower $\alpha_J$, both $\bar{\eta}$ and $\sigma_\eta$ increase by a factor of approximately 1.04 in the higher $\alpha_J$ case. Interestingly, $CV_{ROI}$ over the ROI slightly decreases by a factor of 1.0032, suggesting that $CV_{ROI}$ may serve as a more stable and robust indicator of gradient uniformity than $\sigma_\eta$. Additionally, the design with higher $\alpha_J$ requires an increased number of contours to maintain the minimum wire width of 3.5 mm. This increase in coil-layout complexity may present more challenges in coil fabrication and winding.

Tables \ref{table:NBCD_Non-LinearCoils_CCD1}, \ref{table:NBCD_Non-LinearCoils_CCD2p5}, and \ref{table:NBCD_Non-LinearCoils_CCD5} also show that these optimized coils all possess an inductance of less than 200 $\mu$H, representing more than a 5-fold reduction compared to a conventional whole-body gradient coil \cite{setsompop_pushing_2013}. These results suggest that a higher slew rate can be achieved. Moreover, the maximum components of the torque vector are exceptionally small (on the order of 1e-4 $\cdot$m), indicating that the coil layouts are nearly mechanically balanced within a constant main magnetic field.

\begin{figure}[htbp]
	\centering \resizebox{0.9\textwidth}{!}{\includegraphics{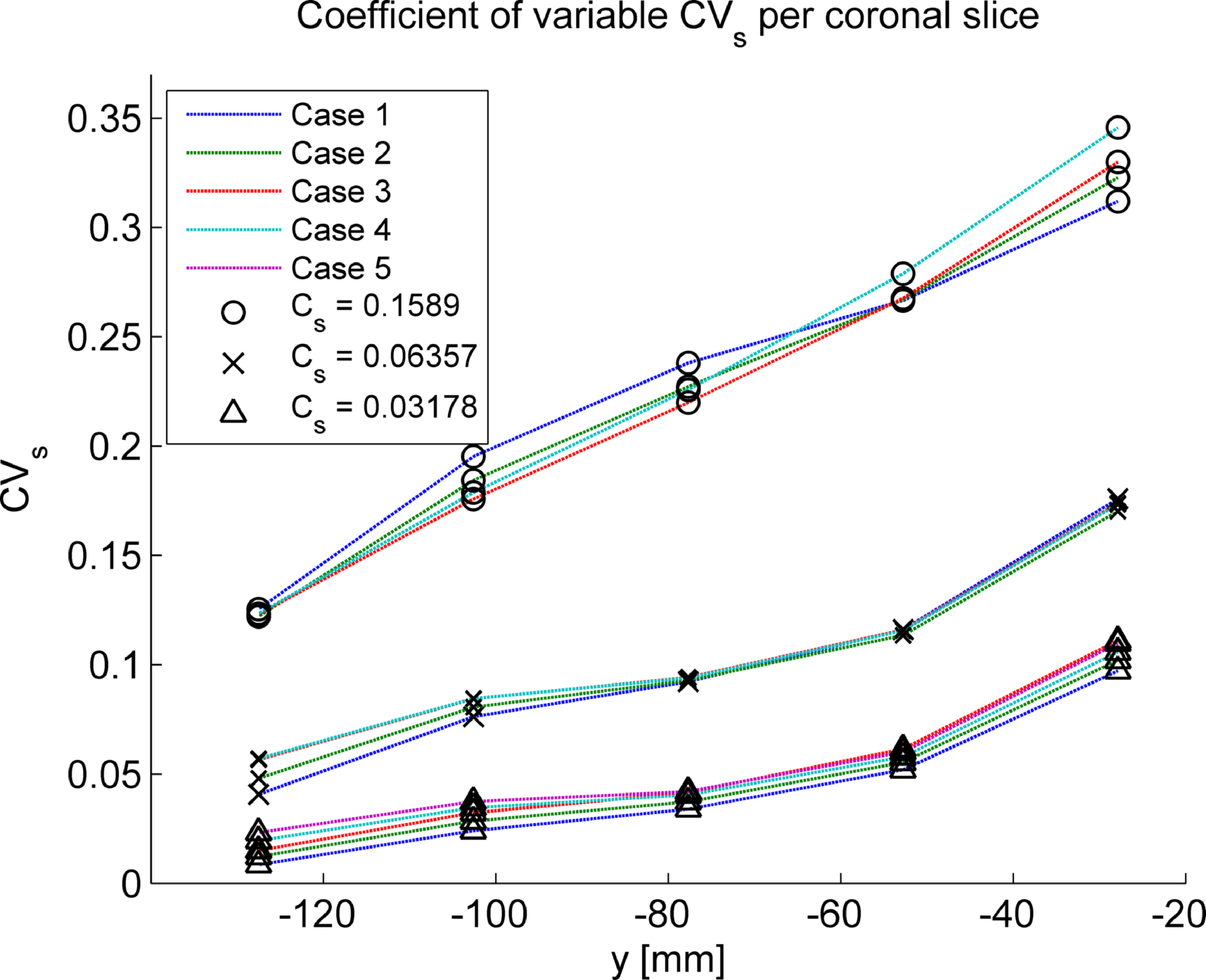}} 
	\caption{Coefficients of variable per coronal slice. For $C_s$ = 0.1589, 0.06357 and 0.03178, the corresponding tuning and performance parameters for different cases (different $\alpha_J$ or $C_y$) are presented in Tables \ref{table:NBCD_Non-LinearCoils_CCD1}, \ref{table:NBCD_Non-LinearCoils_CCD2p5} and \ref{table:NBCD_Non-LinearCoils_CCD5}, respectively. $C_s$ = 0.1589 and $C_s$ =  0.06357 have four different cases and $C_s$ = 0.03178 has five cases.}
	\label{fig:CVAllCurvesSlices_fig3} 
\end{figure}

Figure \ref{fig:CVAllCurvesSlices_fig3} shows the coefficient of variation per coronal slice, denoted as $CV_s$, as a function of the slice position $y$, spanning from anterior ($y = -127.5$ mm) to posterior region ($y = -27.9$ mm) of the ROI. Several different nonlinear coil design cases are presented for $C_s = 0.1589$, $0.06357$, and $0.03178$, represented by circular, cross and triangular markers, respectively. From case 1 to case 5, the corresponding value of $C_y$ increases or remains constant. 

For all values of $C_s$, the average value of $CV_s$ increases progressively with slice position, from anterior ($y = -127.5$~mm) to posterior ($y = -27.9$~mm). This trend reflects increasing intra-slice gradient non-uniformity in slices closer to the chest wall. Specifically, for $C_s = 0.1589$, the average $CV_s$ rises from 0.1236 in the anterior slice to 0.3276 in the most posterior slice. For $C_s = 0.06357$, the corresponding values increase from 0.0506 to 0.1734, and for $C_s = 0.03178$, from 0.0157 to 0.1052. These results confirm that stronger per-slice constraints (i.e., lower $C_s$) result in reduced intra-slice variation over the ROI. 

Considering that the designed linear coil exhibits a coefficient of variation of 0.1022 over the ROI, as shown in Table~\ref{table:Comp_Linear_NonLinearCoils}, the nonlinear designs with $C_s = 0.03178$ may offer comparable or even superior gradient uniformity within each coronal slice, making them a favorable choice when aiming to match the gradient homogeneity achieved by linear gradient coils.

For a given $C_s$, Figure~\ref{fig:CVAllCurvesSlices_fig3} also illustrates the effect of $C_y$ on the coefficient of variation $CV_s$. The maximum standard deviation of $CV_s$ across all coronal slices is 0.0142 for $C_s = 0.1589$, 0.0078 for $C_s = 0.06357$, and 0.0058 for $C_s = 0.03178$. However, in contrast to the trend observed for $CV_{\text{ROI}}$ (as shown in Tables~\ref{table:NBCD_Non-LinearCoils_CCD1}, \ref{table:NBCD_Non-LinearCoils_CCD2p5}, and \ref{table:NBCD_Non-LinearCoils_CCD5}), increasing $C_y$ does not consistently lead to an increase in $CV_s$ at each individual slice.

\begin{figure}[htbp]
	\centering \subfigure[$\alpha_J$ = 0.006375 and $C_y$ = 0.1191]{ \label{fig:EtaD1_fig4a} \includegraphics[width=0.475\textwidth]{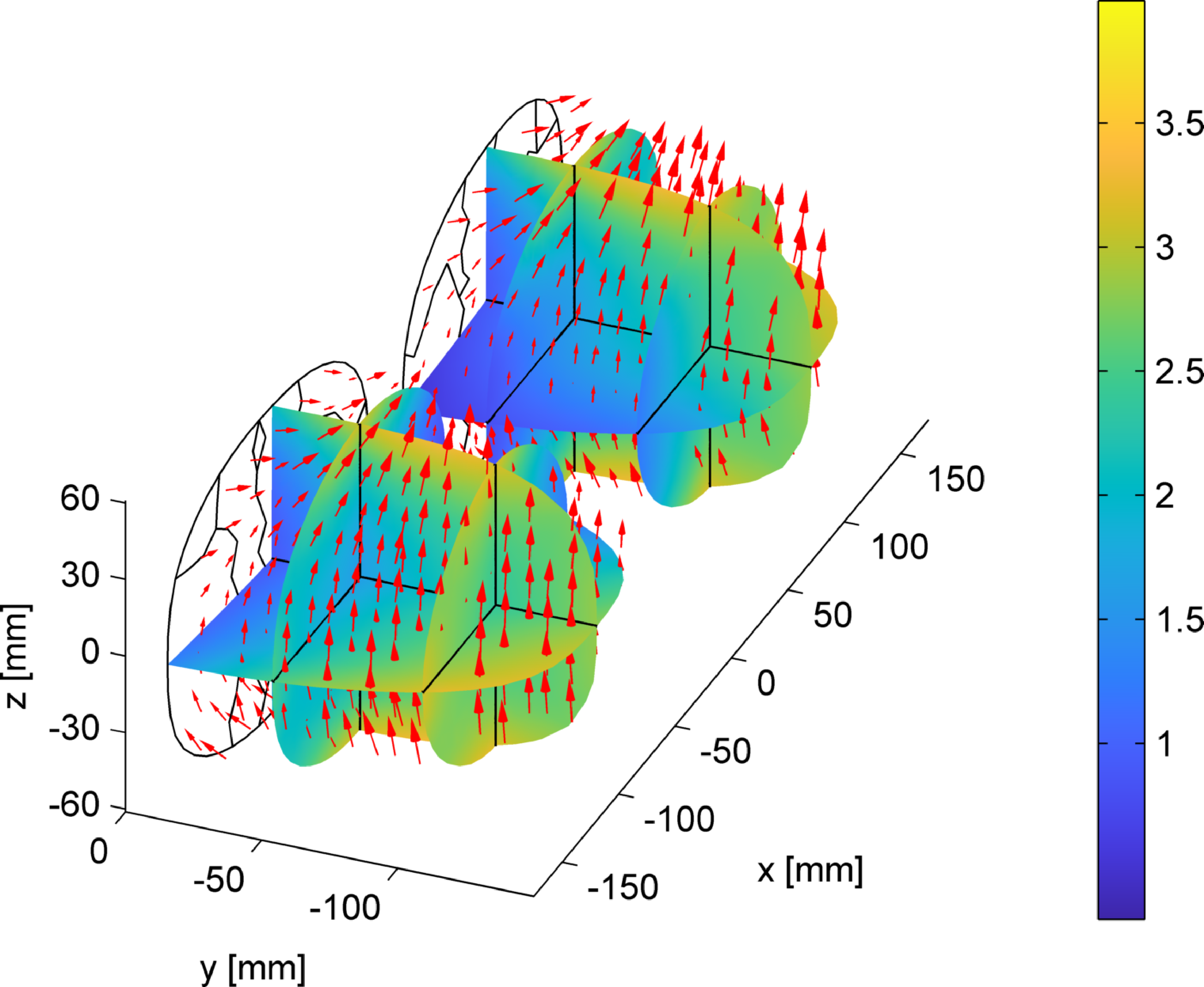}
	} \subfigure[$\alpha_J$ = 0.0066875 and $C_y$ = 0.1361]{ \label{fig:EtaD1_fig4b} \includegraphics[width=0.475\textwidth]{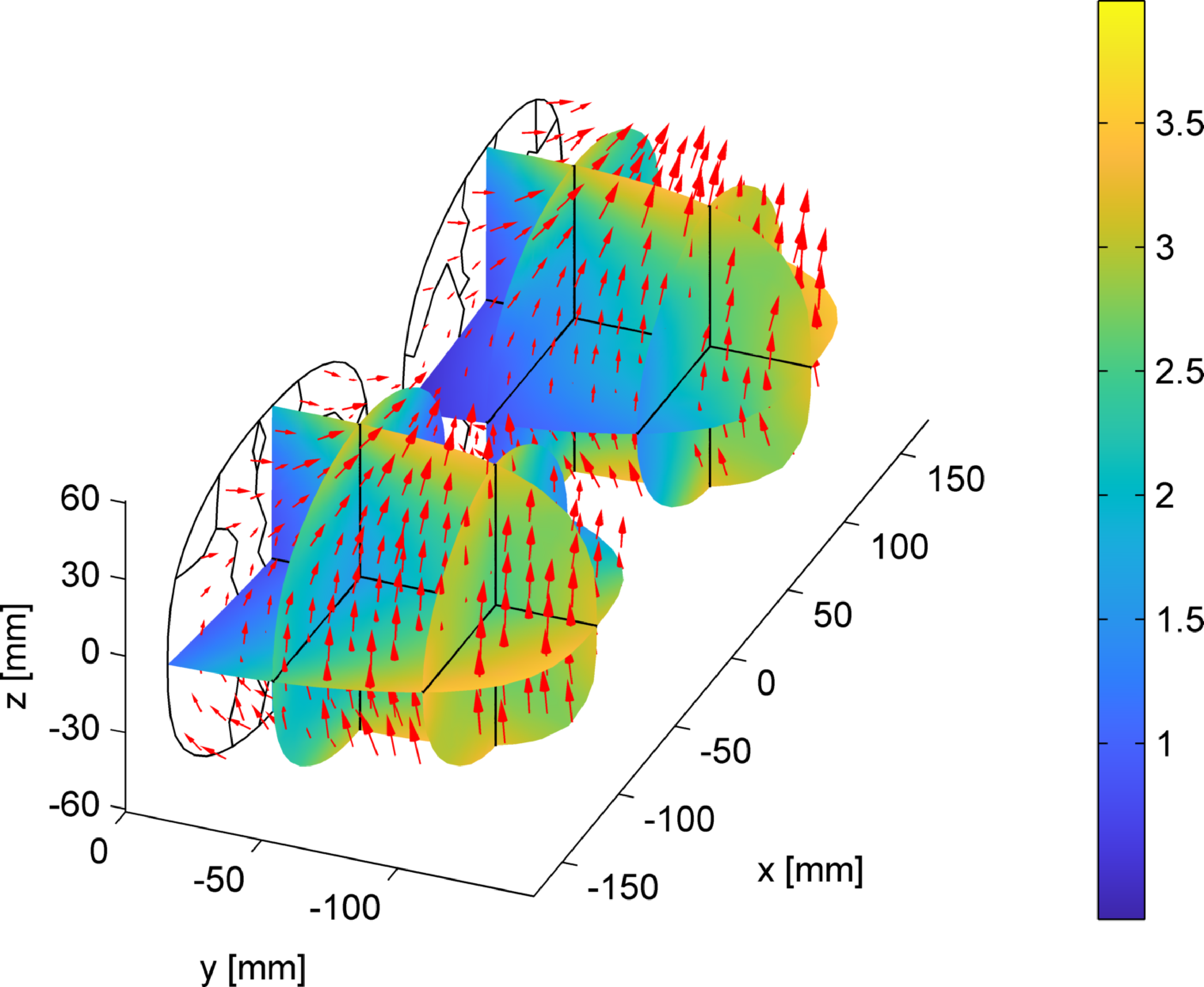}
	} \subfigure[$\alpha_J$ = 0.0074375 and $C_y$ = 0.1532]{ \label{fig:EtaD1_fig4c} \includegraphics[width=0.475\textwidth]{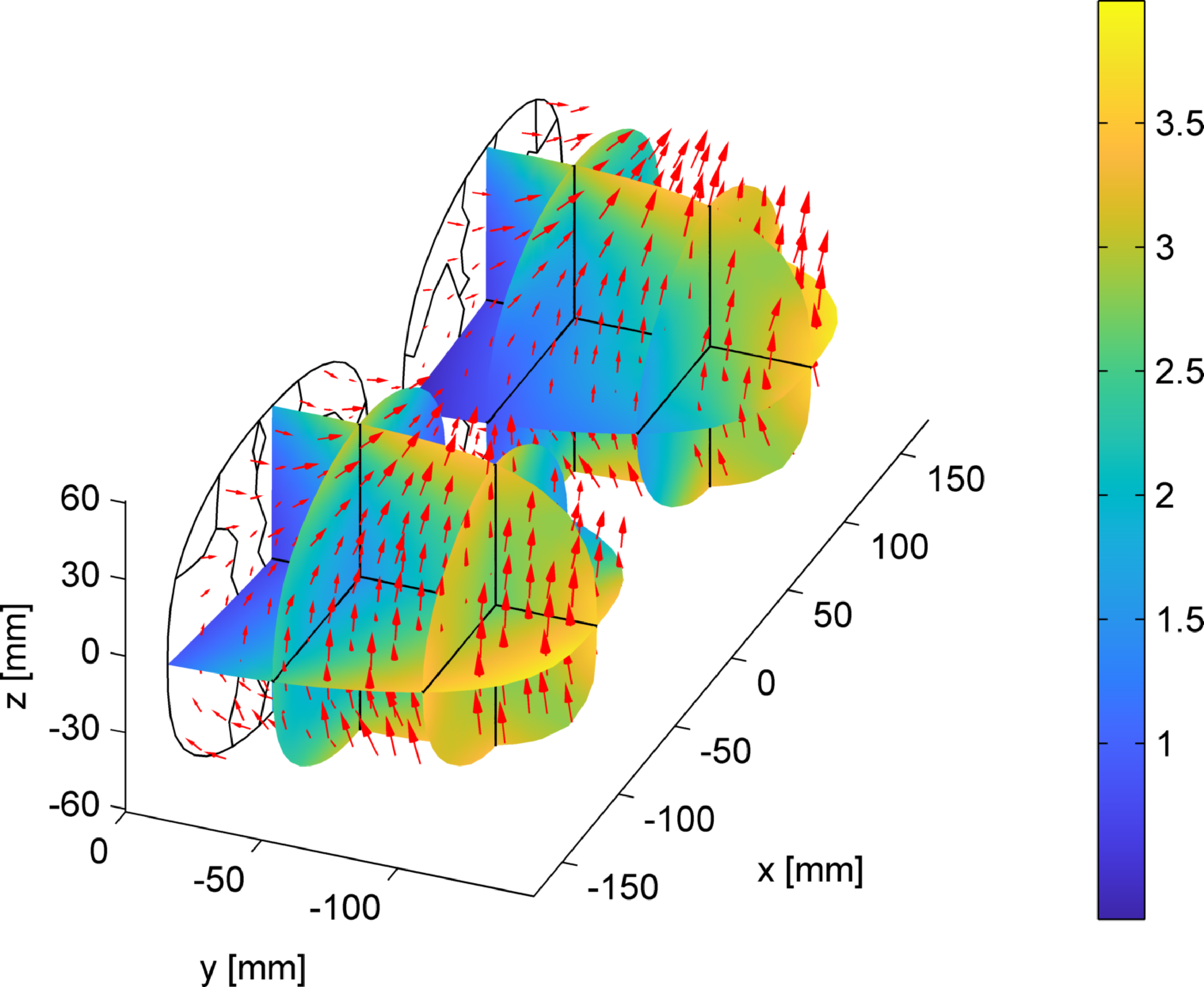}
	} \subfigure[$\alpha_J$ = 0.003094 and $C_y$ = 0.16]{ \label{fig:EtaD1_fig4d} \includegraphics[width=0.475\textwidth]{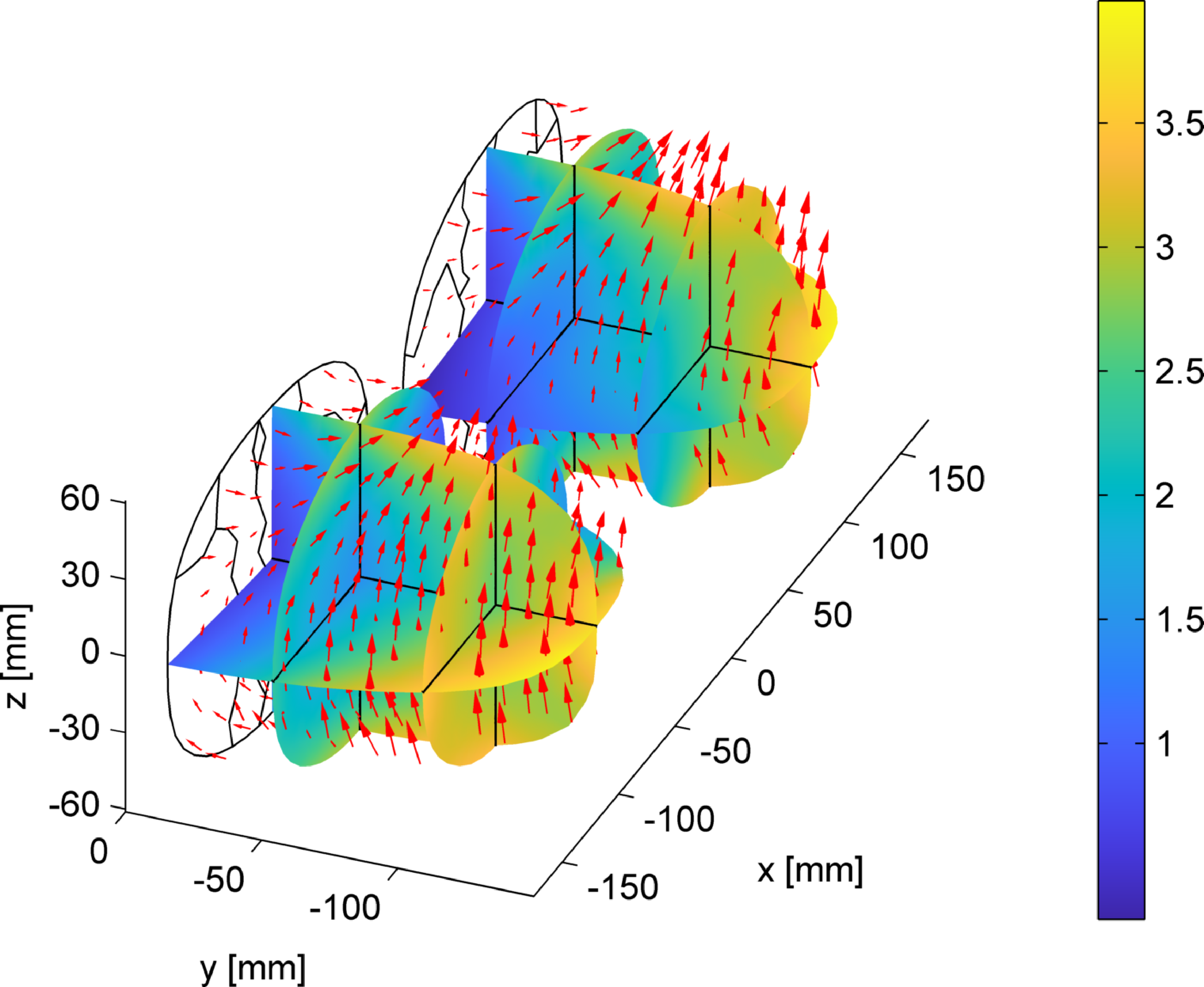}
	}\caption{Spatial distribution of coil efficiency $\eta$ [mT/m/A] (color map) for different values of $\alpha_J$ and $C_y$ at $C_s = $ 0.1589. The gradient vectors of the magnetic field $B_z$ per unit current are indicated by red arrows.}
	\label{fig:EtaD1_fig4} 
\end{figure}

\begin{figure}[htbp]
	\centering \subfigure[$\alpha_J$ = 0.001875 and $C_y$ = 0.1191]{ \label{fig:EtaD2p5_fig5a} \includegraphics[width=0.475\textwidth]{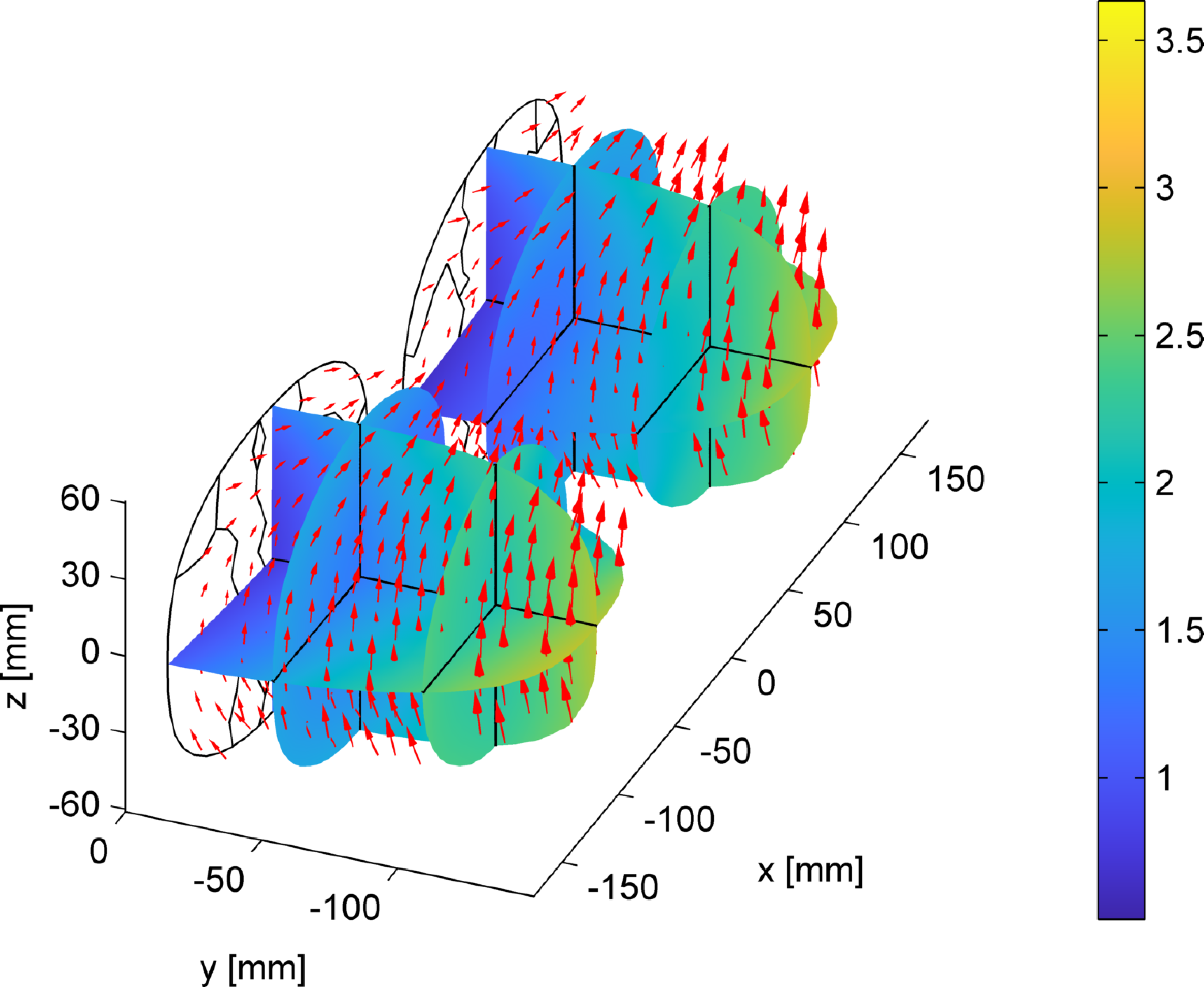}
	} \subfigure[$\alpha_J$ = 0.004875 and $C_y$ = 0.1361]{ \label{fig:EtaD2p5_fig5b} \includegraphics[width=0.475\textwidth]{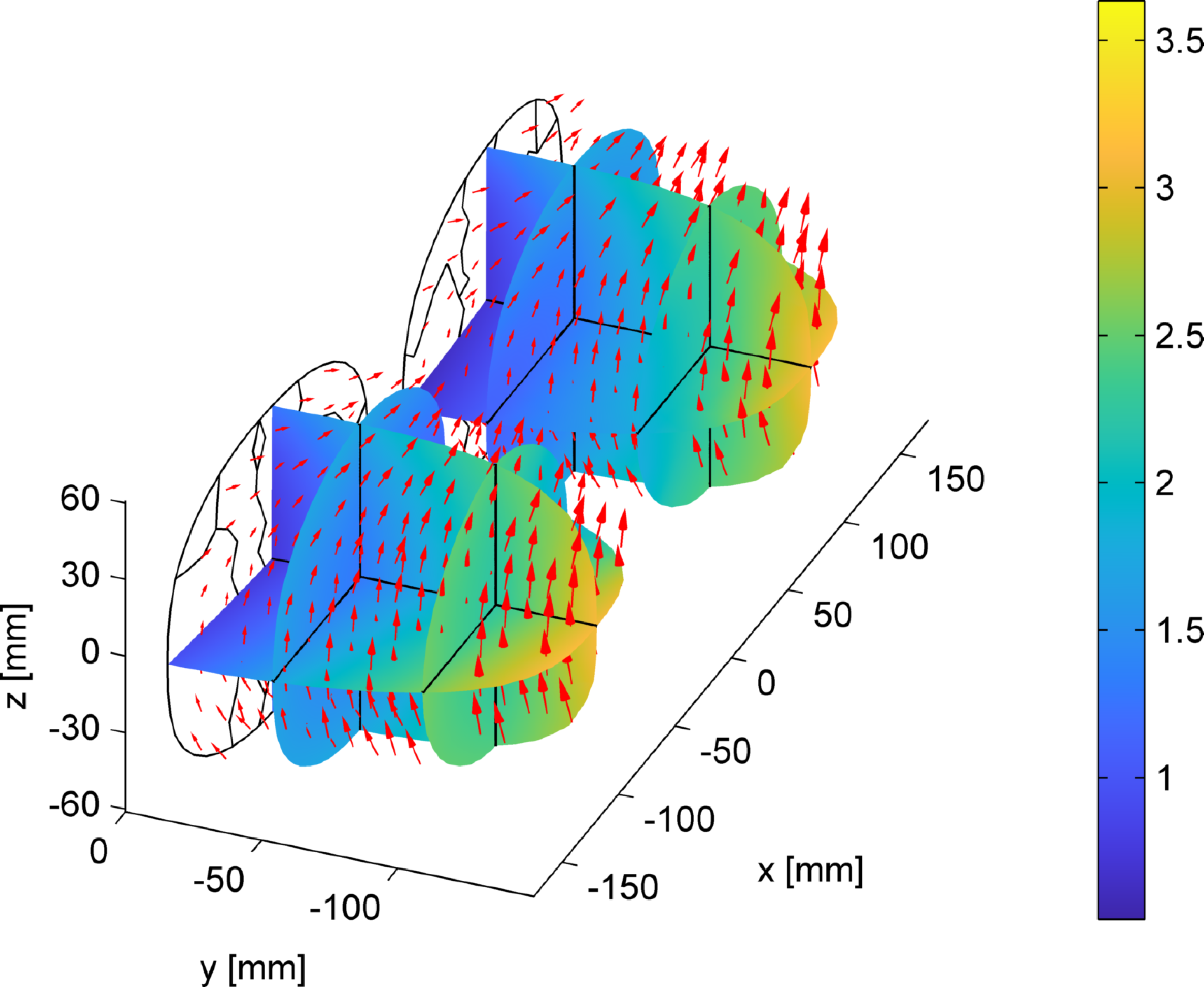}
	} \subfigure[$\alpha_J$ = 0.004625 and $C_y$ = 0.16]{ \label{fig:EtaD2p5_fig5c} \includegraphics[width=0.475\textwidth]{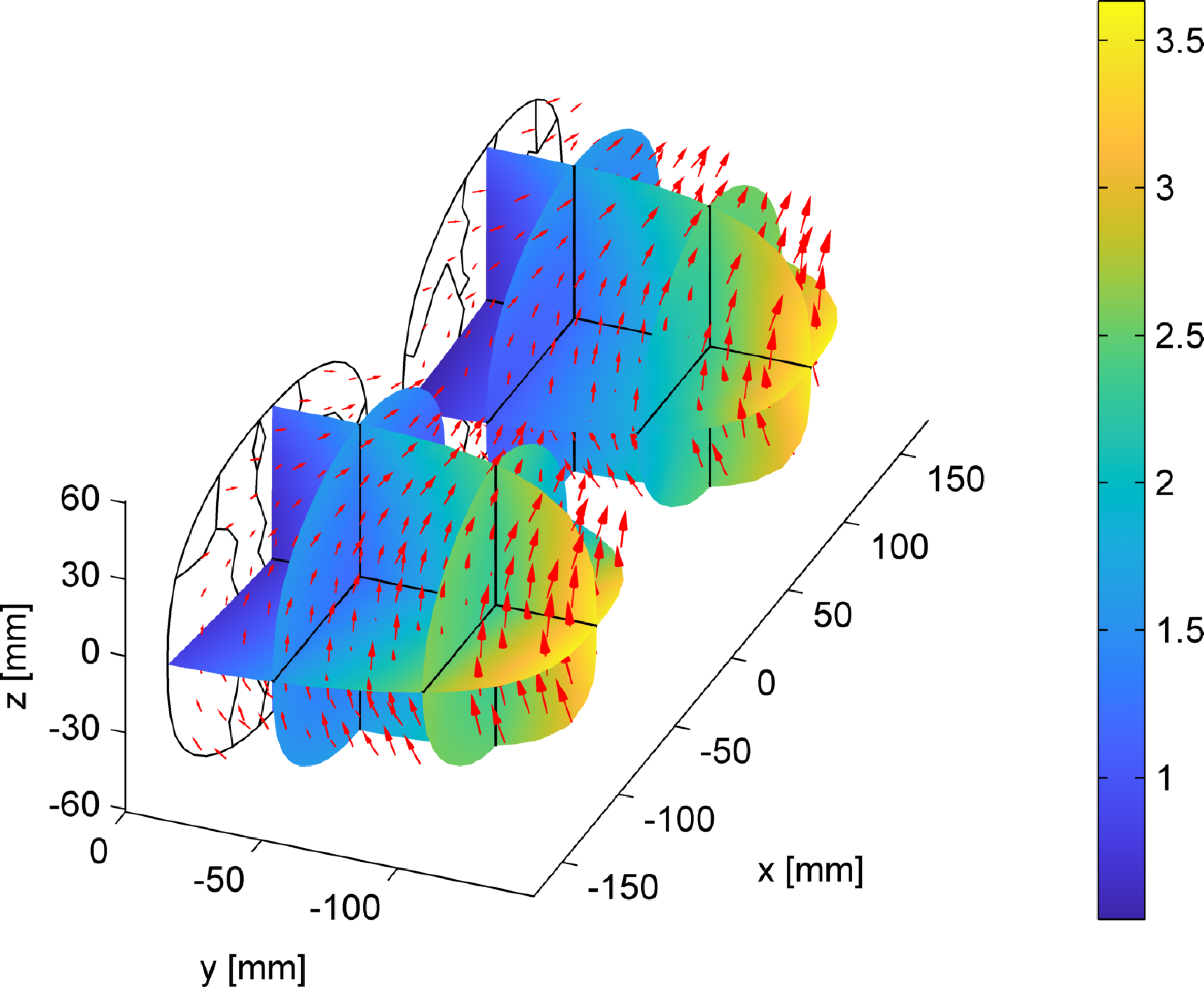}
	} \subfigure[$\alpha_J$ = 0.004875 and $C_y$ = 0.1692]{ \label{fig:EtaD2p5_fig5d} \includegraphics[width=0.475\textwidth]{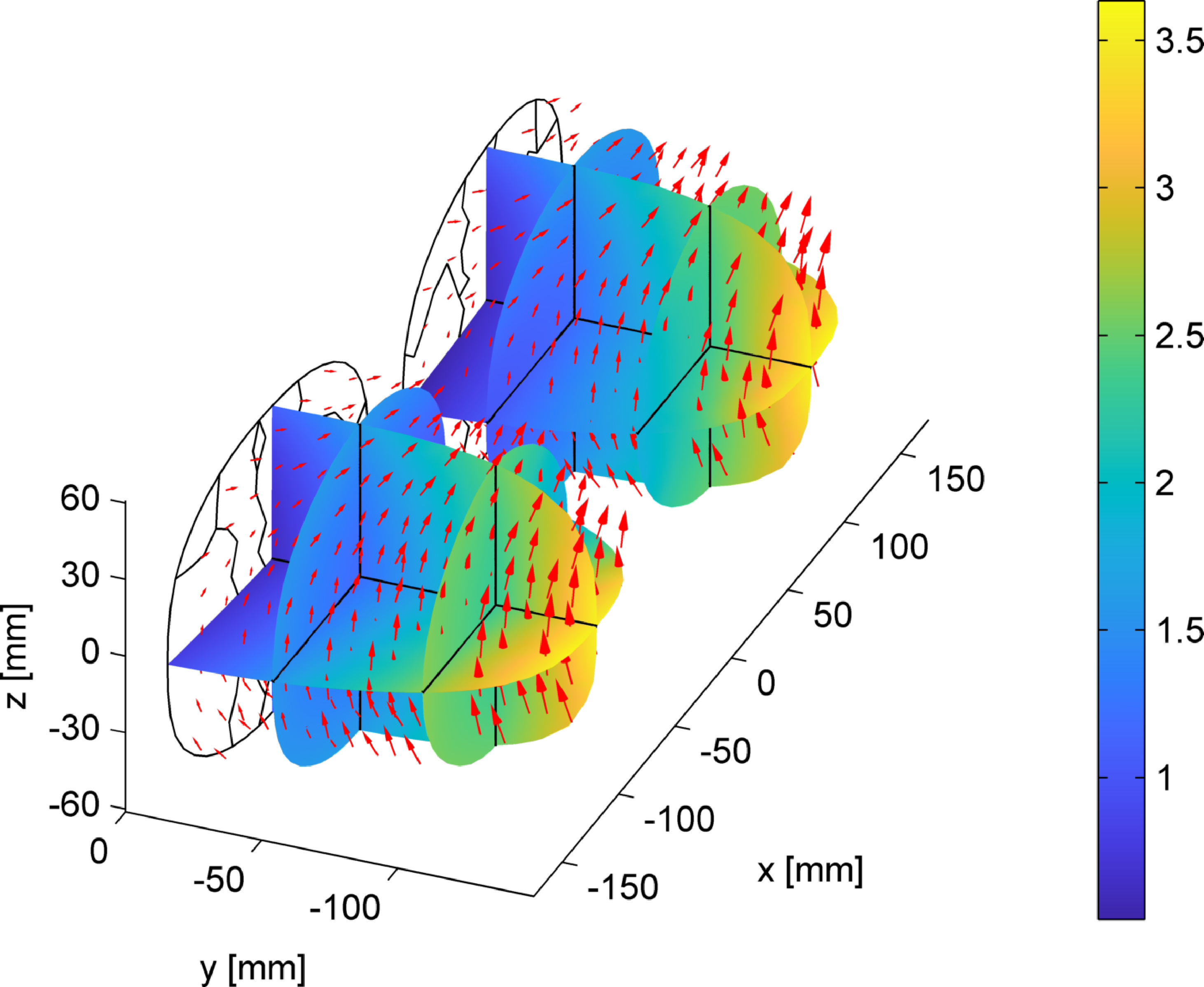}
	}\caption{Spatial distribution of coil efficiency $\eta$ [mT/m/A] (color map) for different values of $\alpha_J$ and $C_y$ at $C_s = $ 0.06357. The gradient vectors of the magnetic field $B_z$ per unit current are indicated by red arrows.}
	\label{fig:EtaD2p5_fig5} 
\end{figure}

\begin{figure}[htbp]
	\centering \subfigure[$\alpha_J$ = 0.00549 and $C_y$ = 0.1191]{ \label{fig:EtaD5_fig6a} \includegraphics[width=0.475\textwidth]{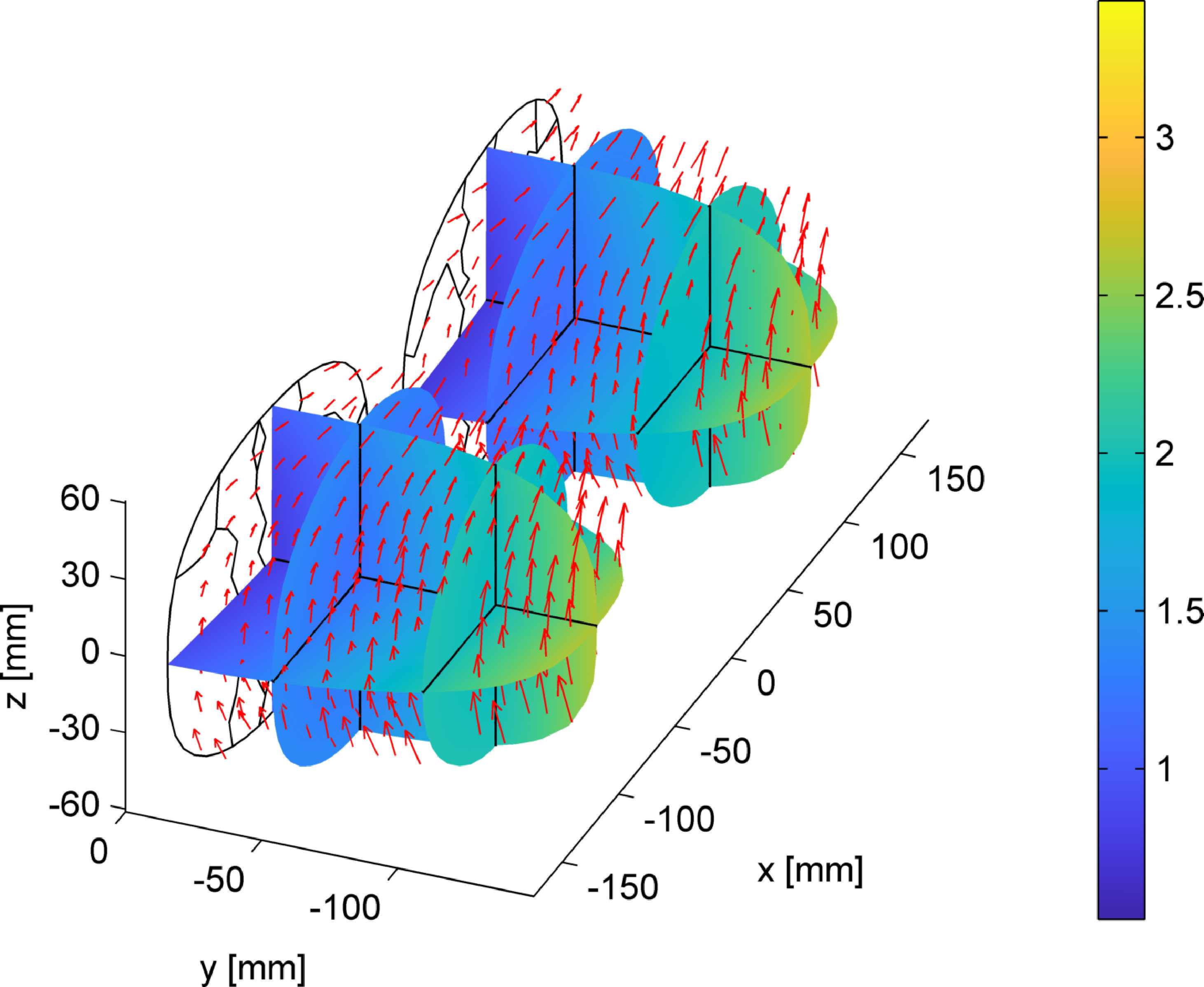}
	} \subfigure[$\alpha_J$ = 0.00126 and $C_y$ = 0.16]{ \label{fig:EtaD5_fig6b} \includegraphics[width=0.475\textwidth]{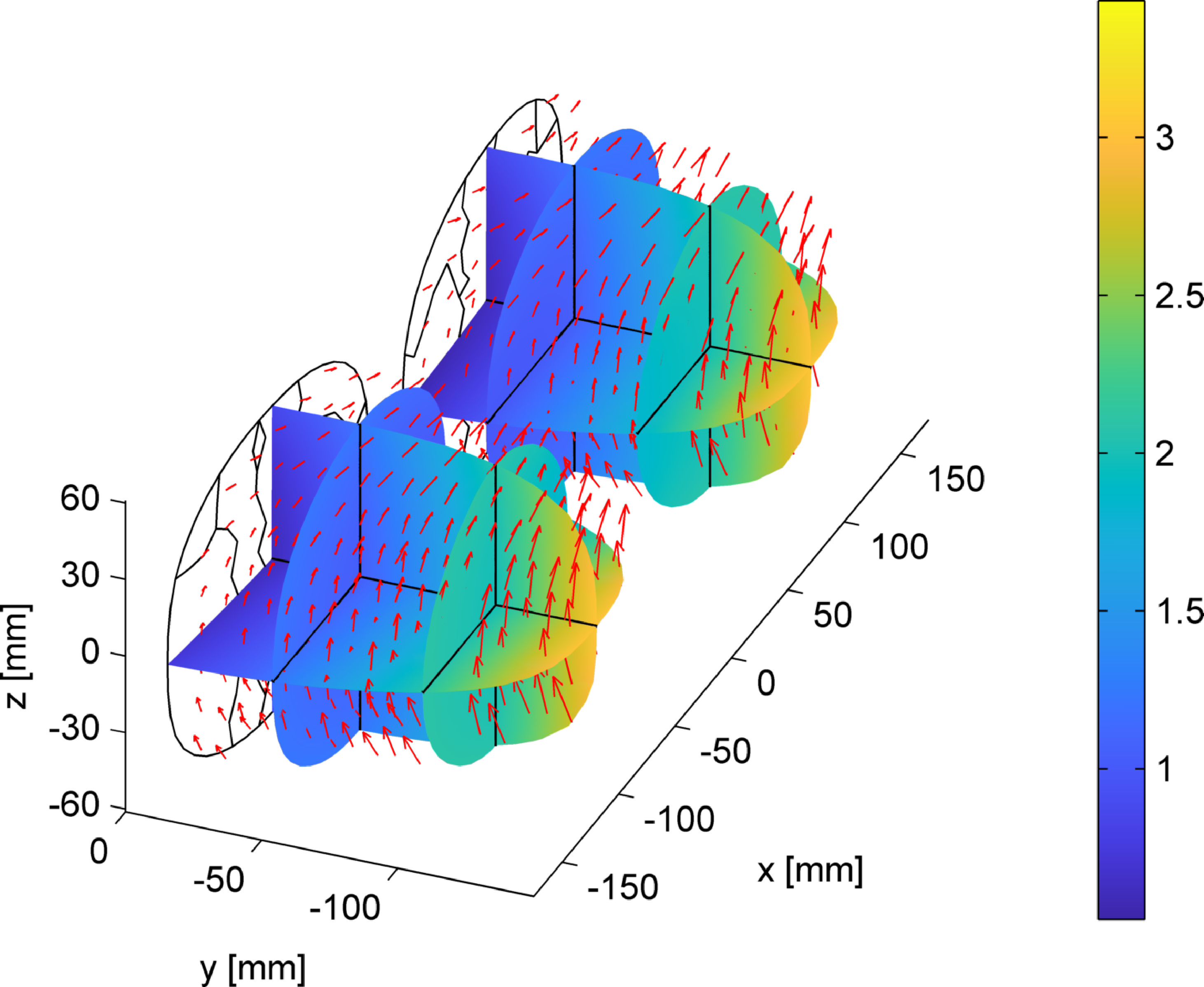}
	} \subfigure[$\alpha_J$ = 0.007125 and $C_y$ = 0.16]{ \label{fig:EtaD5_fig6c} \includegraphics[width=0.475\textwidth]{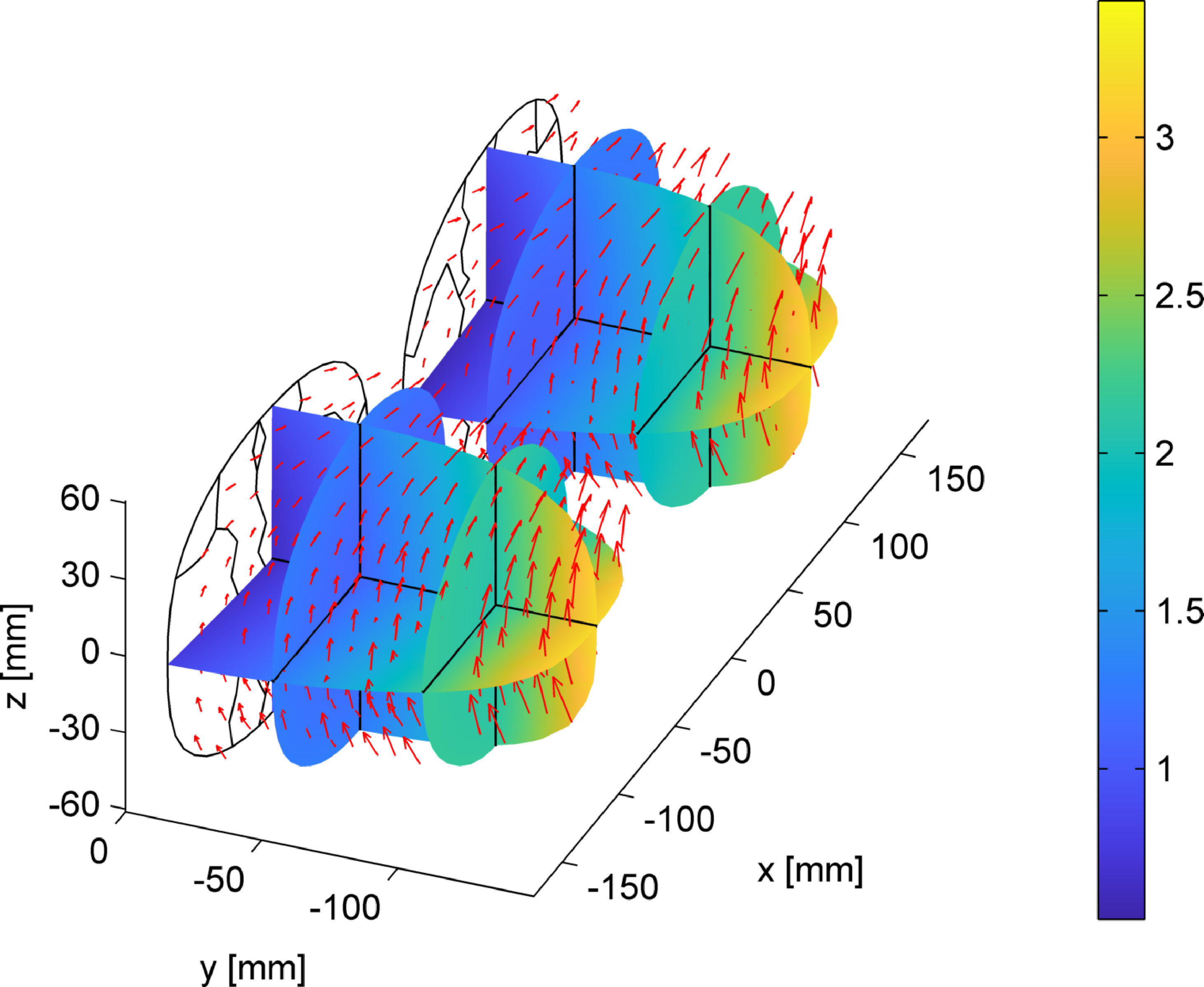}
	} \subfigure[$\alpha_J$ = 0.0044375 and $C_y$ = 0.1817]{ \label{fig:EtaD5_fig6d} \includegraphics[width=0.475\textwidth]{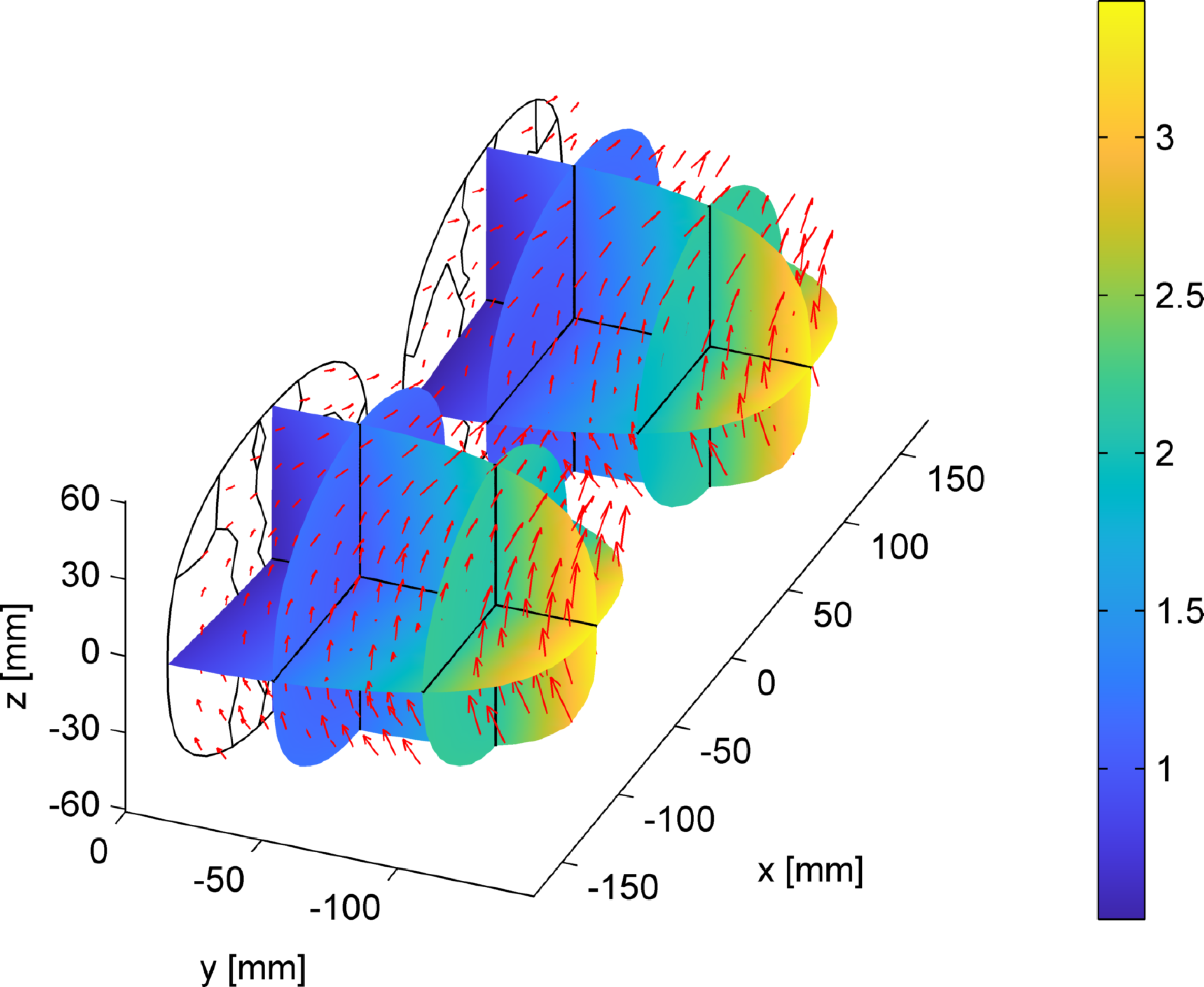}
	}\caption{Spatial distribution of coil efficiency $\eta$ [mT/m/A] (color map) for different values of $\alpha_J$ and $C_y$ at $C_s = $ 0.03178. The gradient vectors of the magnetic field $B_z$ per unit current are indicated by red arrows.}
	\label{fig:EtaD5_fig6} 
\end{figure}

\begin{figure}[htbp]
	\centering \subfigure[$\alpha_J$ = 0.006375 and $C_y$ = 0.1191]{ \label{fig:SFMD1_fig7a} \includegraphics[width=0.475\textwidth]{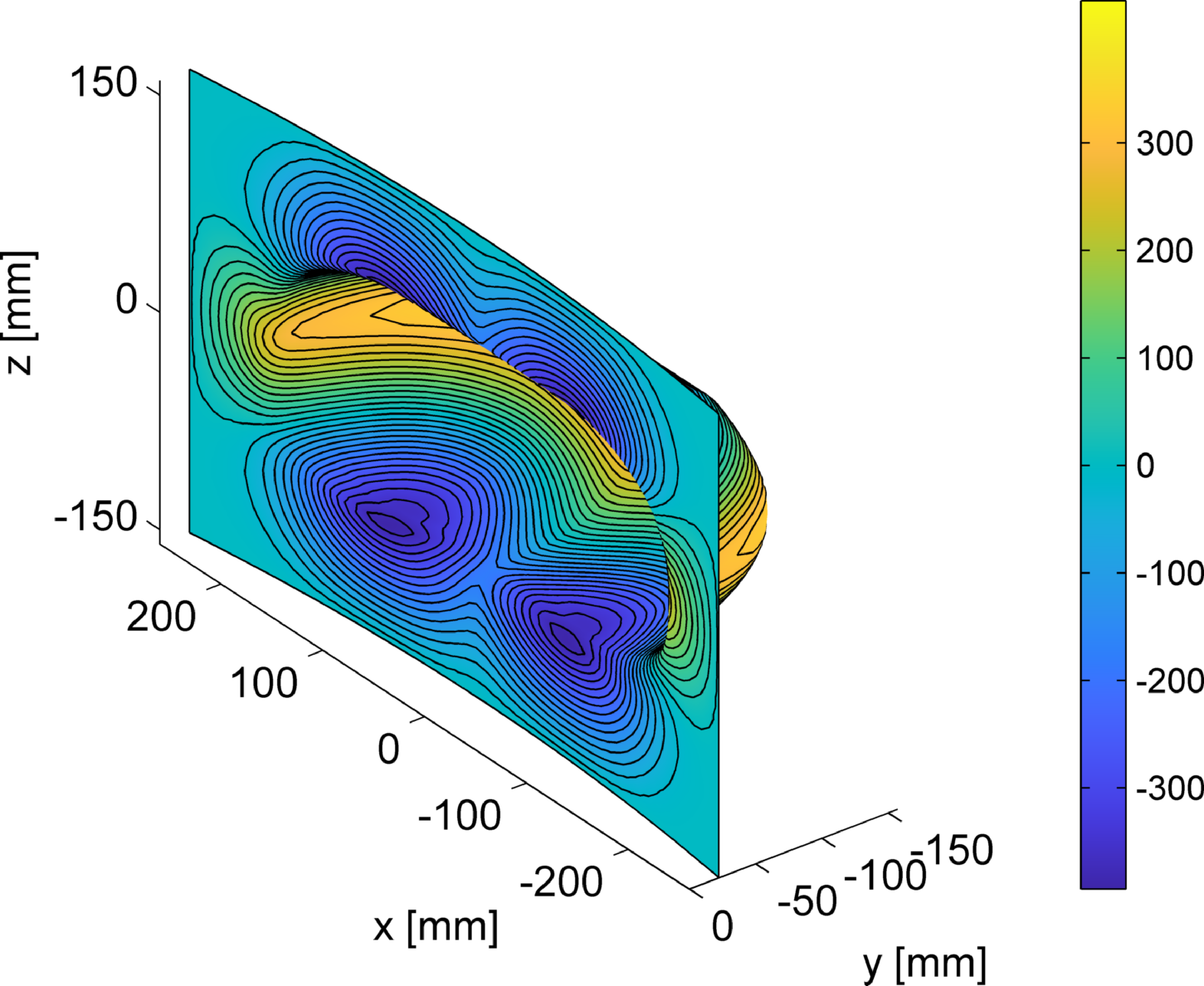}
	} \subfigure[$\alpha_J$ = 0.0066875 and $C_y$ = 0.1361]{ \label{fig:SFMD1_fig7b} \includegraphics[width=0.475\textwidth]{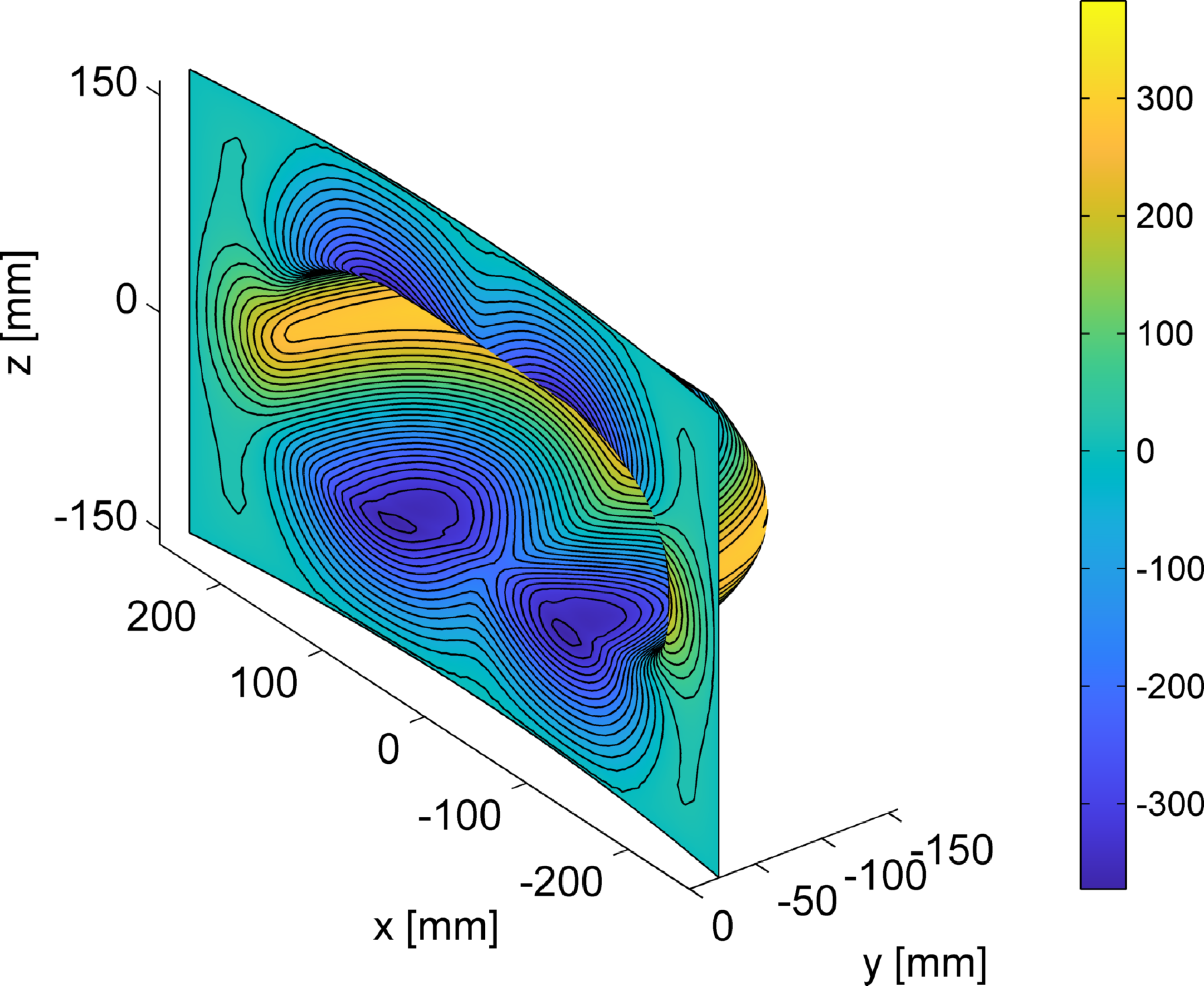}
	} \subfigure[$\alpha_J$ = 0.0074375 and $C_y$ = 0.1532]{ \label{fig:SFMD1_fig7c} \includegraphics[width=0.475\textwidth]{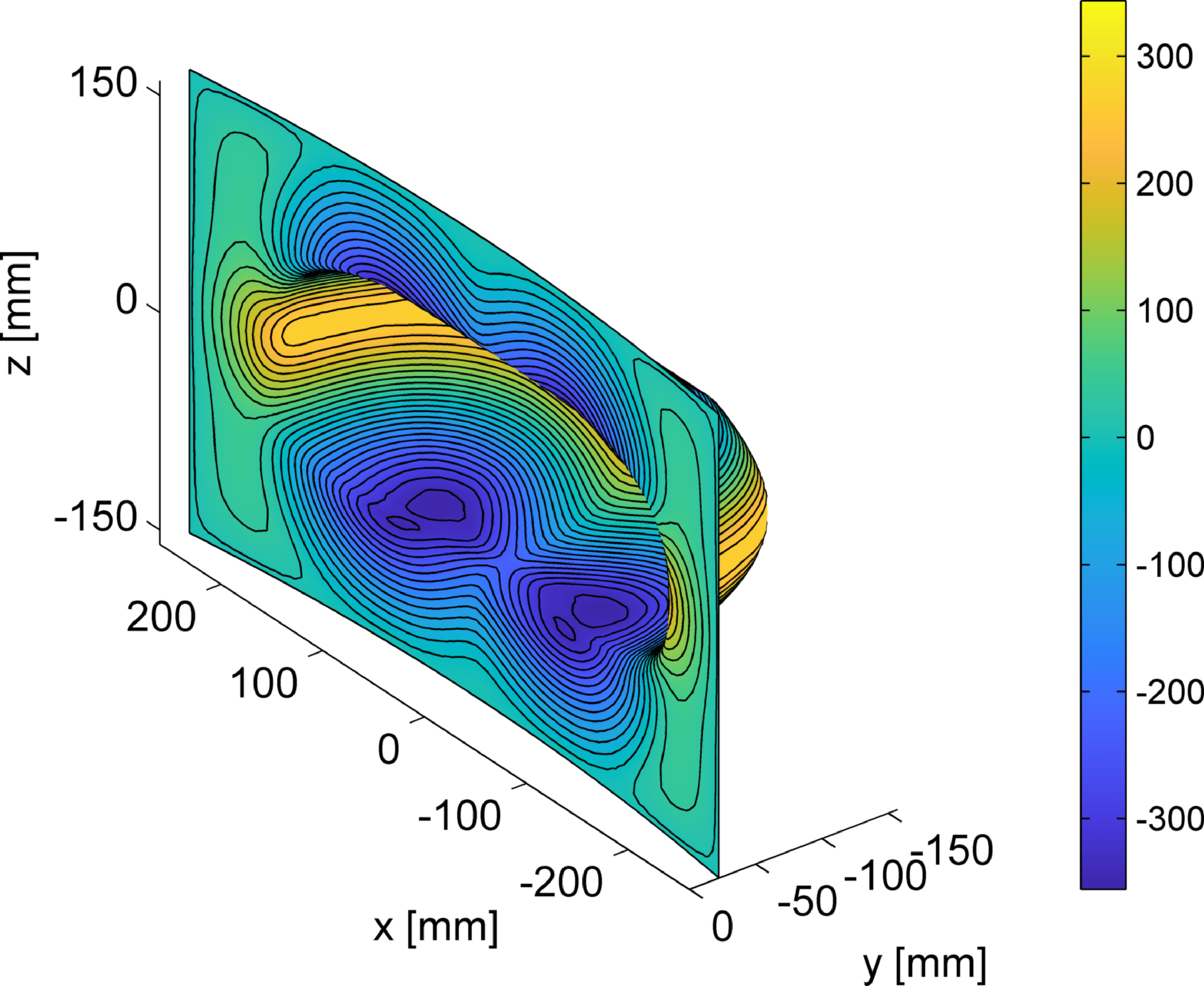}
	} \subfigure[$\alpha_J$ = 0.003094 and $C_y$ = 0.16]{ \label{fig:SFMD1_fig7d} \includegraphics[width=0.475\textwidth]{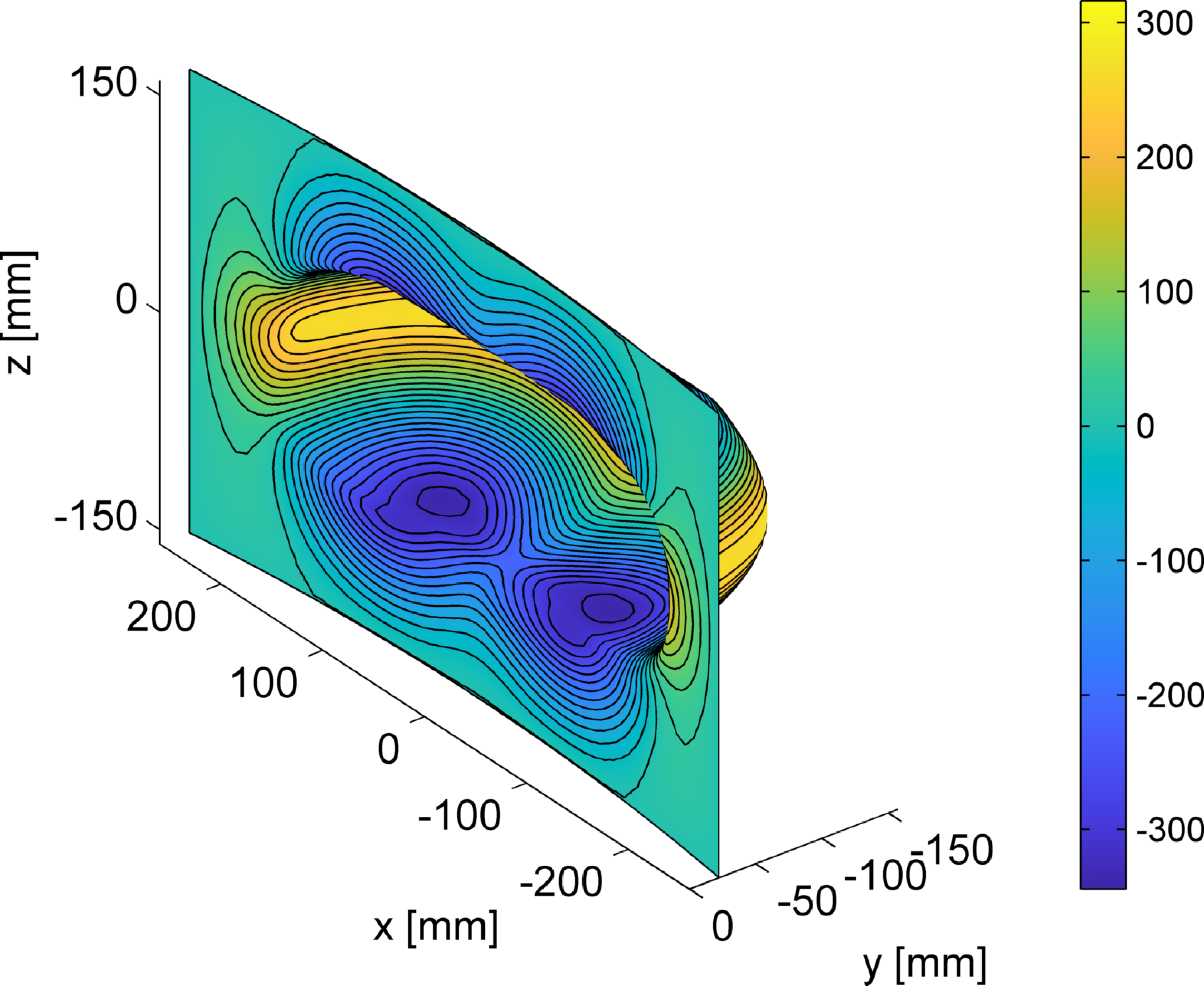}
	}\caption{Spatial distribution of resulting stream function $\psi$ [A] for varying parameters $\alpha_J$ and $C_y$ at $C_s = $ 0.1589. The continuous black contour lines (isolines) define the optimized coil winding layouts mapped onto the 3D surface geometry.}
	\label{fig:SFMD1_fig7} 
\end{figure}

\begin{figure}[htbp]
	\centering \subfigure[$\alpha_J$ = 0.001875 and $C_y$ = 0.1191]{ \label{fig:SFMD2p5_fig8a} \includegraphics[width=0.475\textwidth]{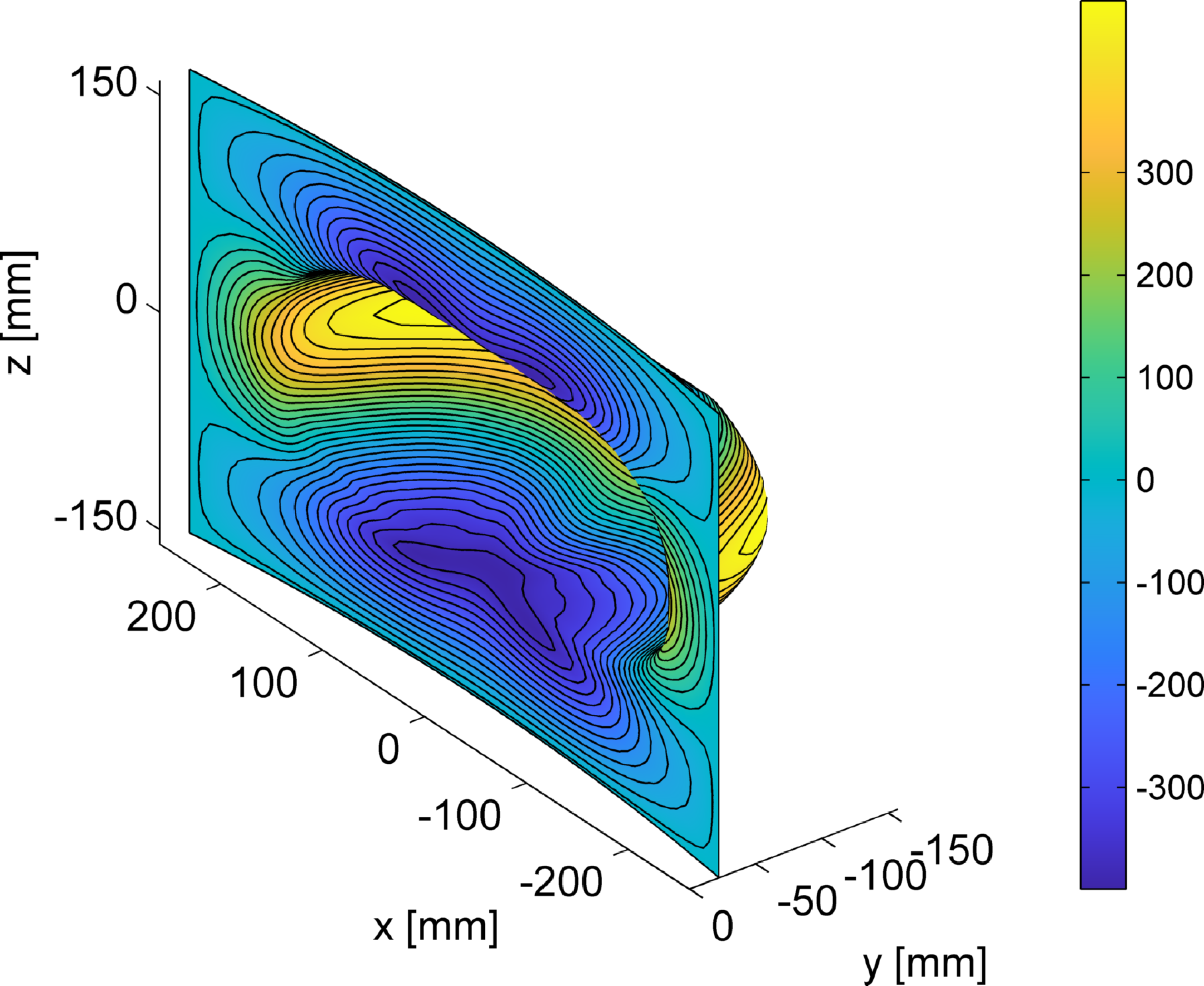}
	} \subfigure[$\alpha_J$ = 0.004875 and $C_y$ = 0.1361]{ \label{fig:SFMD2p5_fig8b} \includegraphics[width=0.475\textwidth]{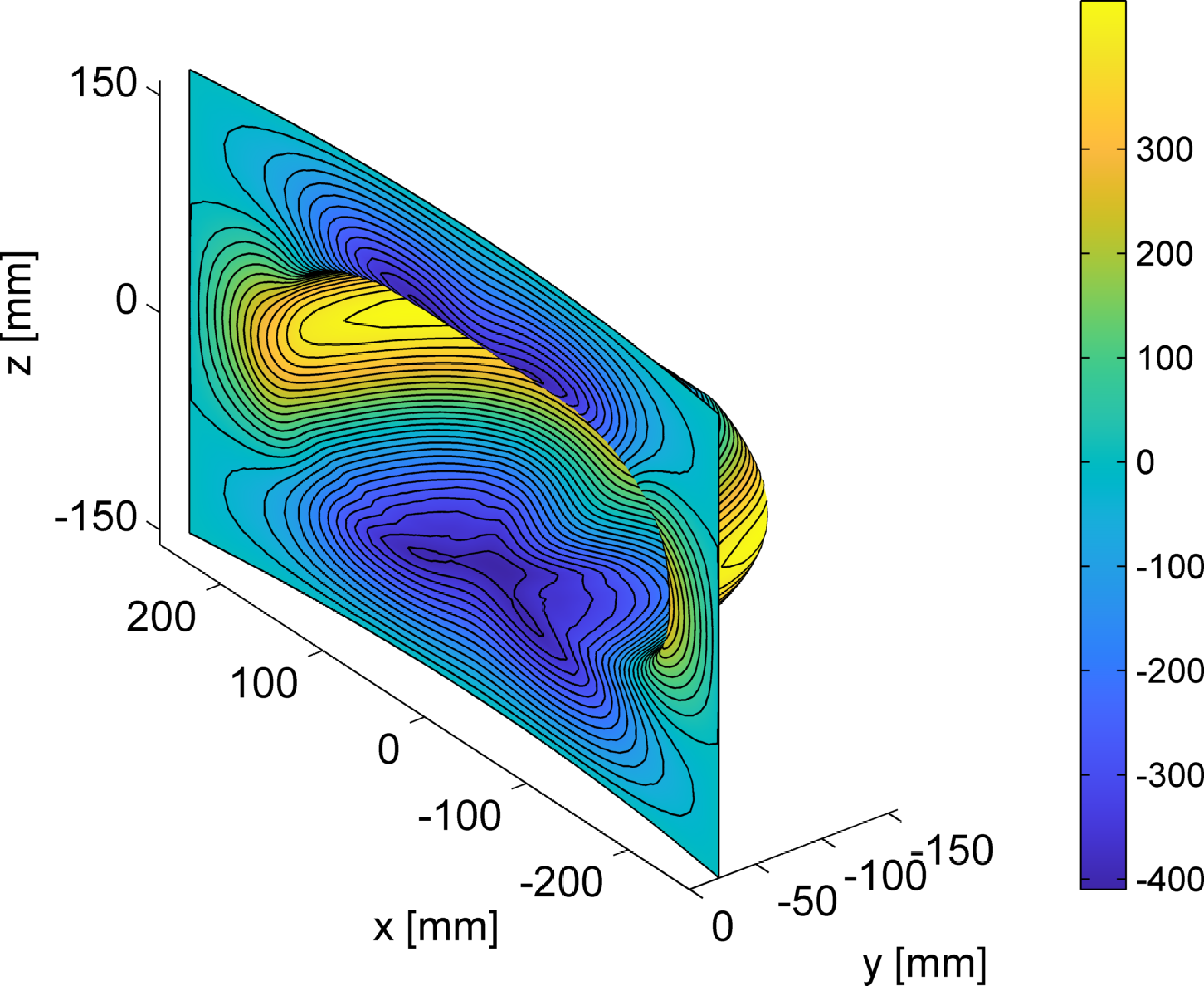}
	} \subfigure[$\alpha_J$ = 0.004625 and $C_y$ = 0.16]{ \label{fig:SFMD2p5_fig8c} \includegraphics[width=0.475\textwidth]{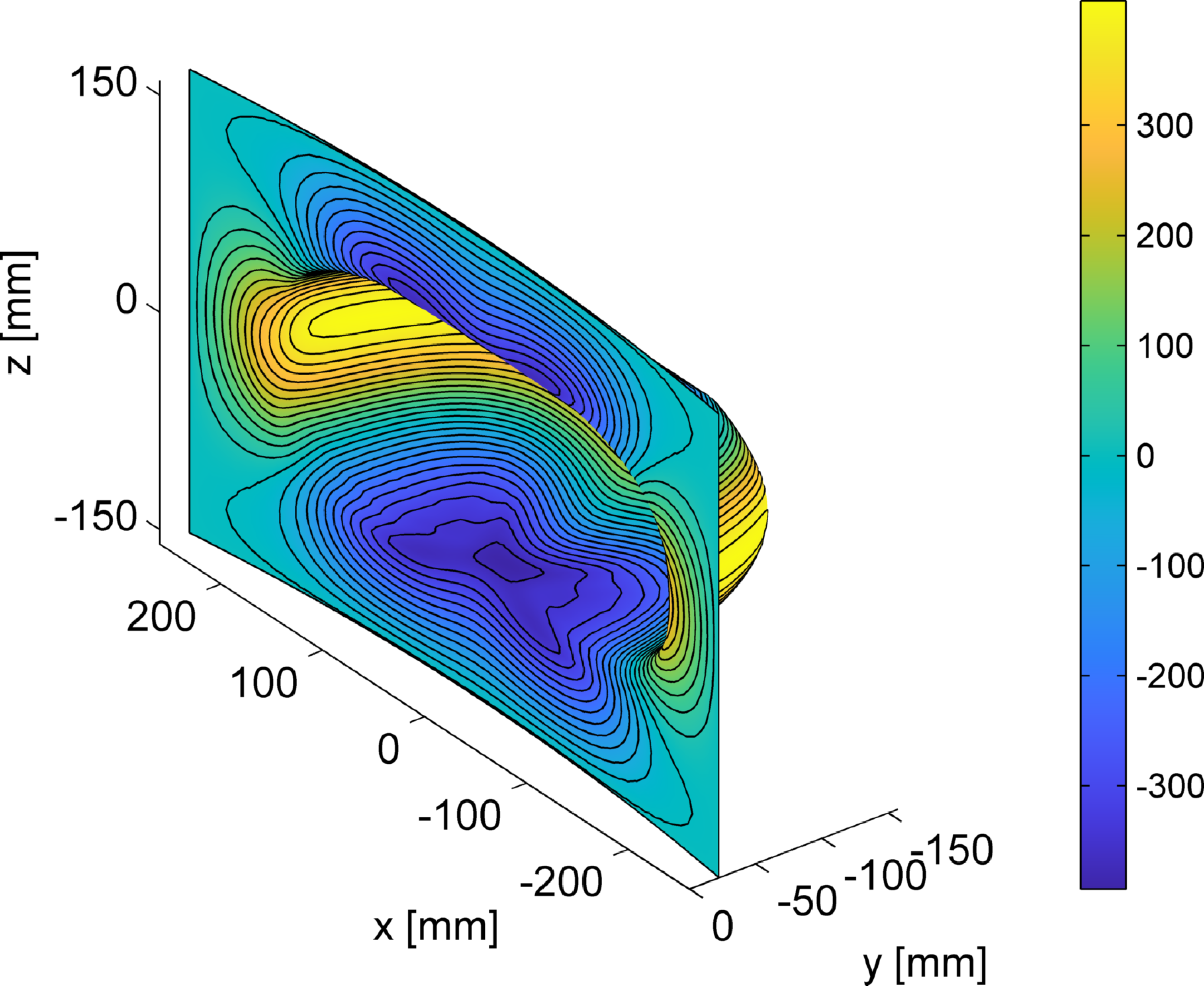}
	} \subfigure[$\alpha_J$ = 0.004875 and $C_y$ = 0.1692]{ \label{fig:SFMD2p5_fig8d} \includegraphics[width=0.475\textwidth]{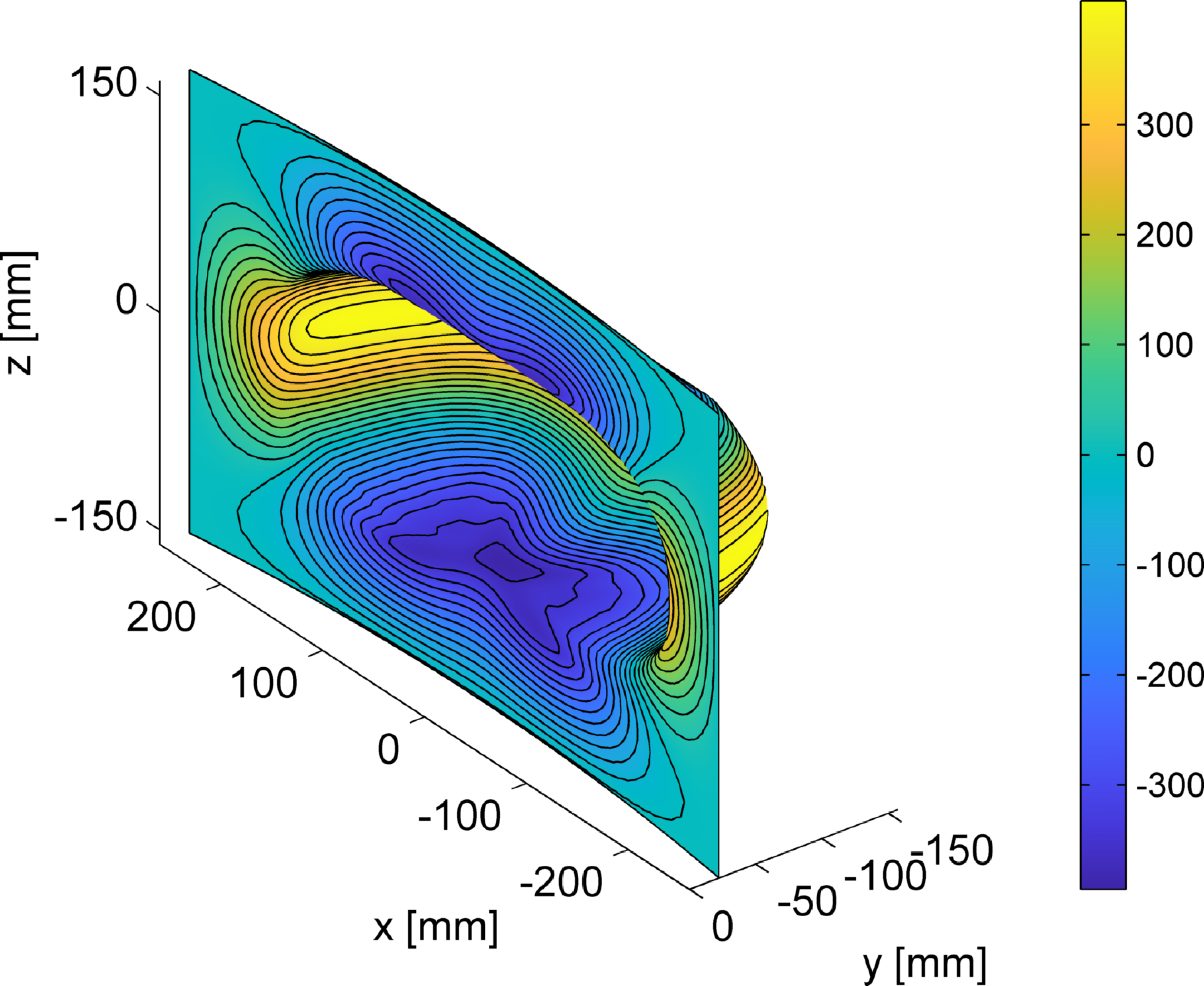}
	}\caption{Spatial distribution of resulting stream function $\psi$ [A] for varying parameters $\alpha_J$ and $C_y$ at $C_s = $ 0.06357. The continuous black contour lines (isolines) define the optimized coil winding layouts mapped onto the 3D surface geometry.}
	\label{fig:SFMD2p5_fig8} 
\end{figure}

\begin{figure}[htbp]
	\centering \subfigure[$\alpha_J$ = 0.00549 and $C_y$ = 0.1191]{ \label{fig:SFMD5_fig9a} \includegraphics[width=0.475\textwidth]{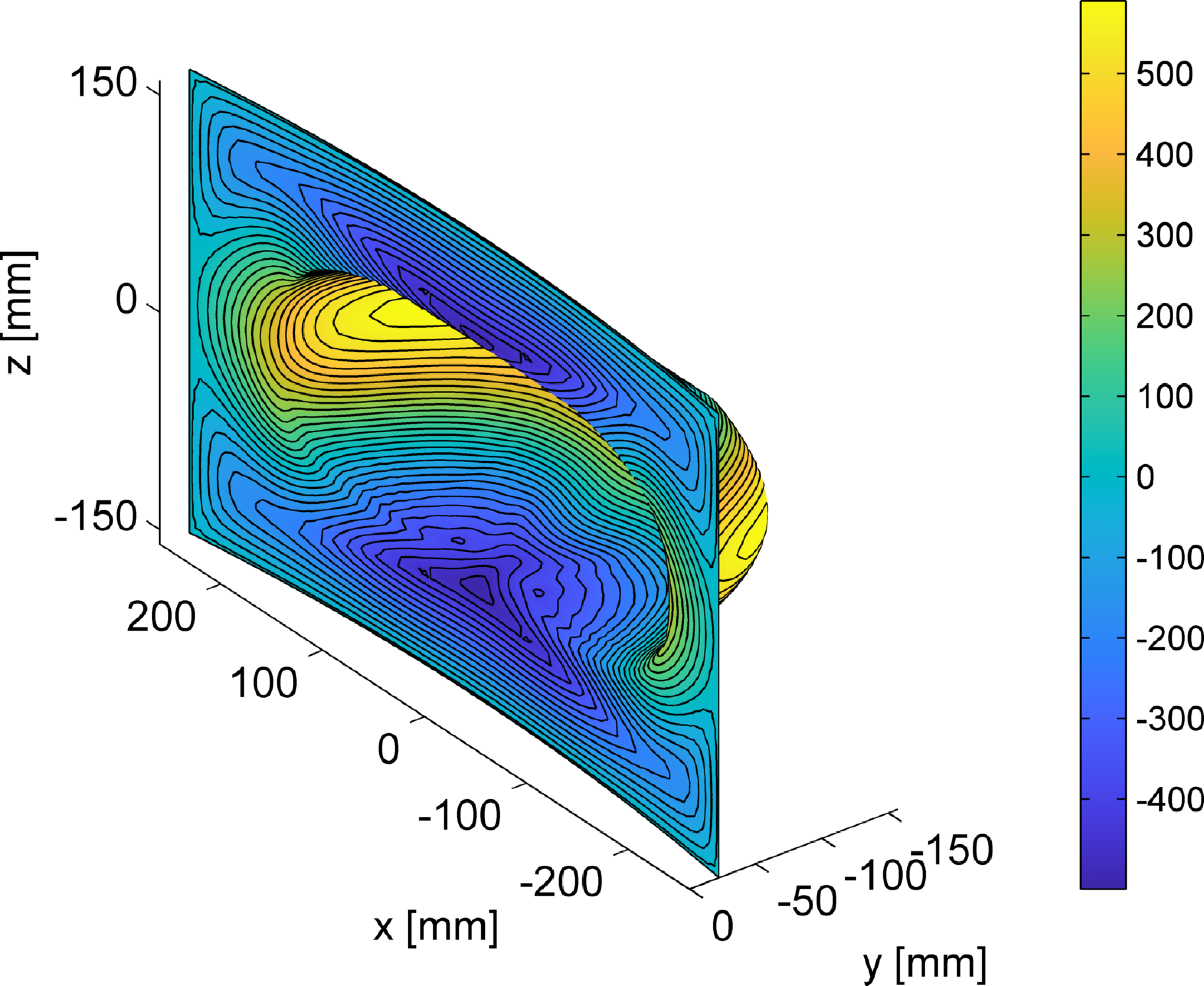}
	} \subfigure[$\alpha_J$ = 0.00126 and $C_y$ = 0.16]{ \label{fig:SFMD5_fig9b} \includegraphics[width=0.475\textwidth]{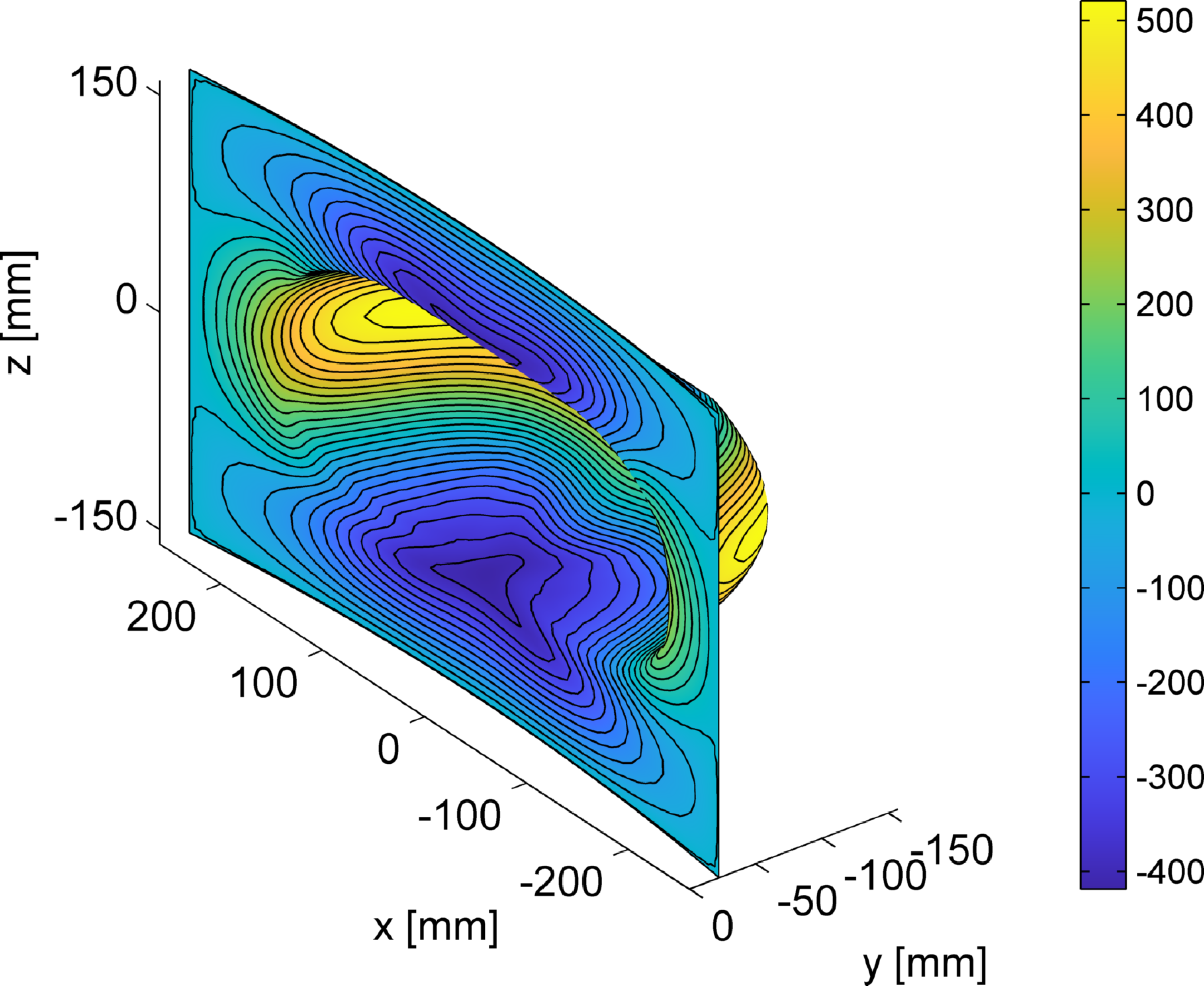}
	} \subfigure[$\alpha_J$ = 0.007125 and $C_y$ = 0.16]{ \label{fig:SFMD5_fig9c} \includegraphics[width=0.475\textwidth]{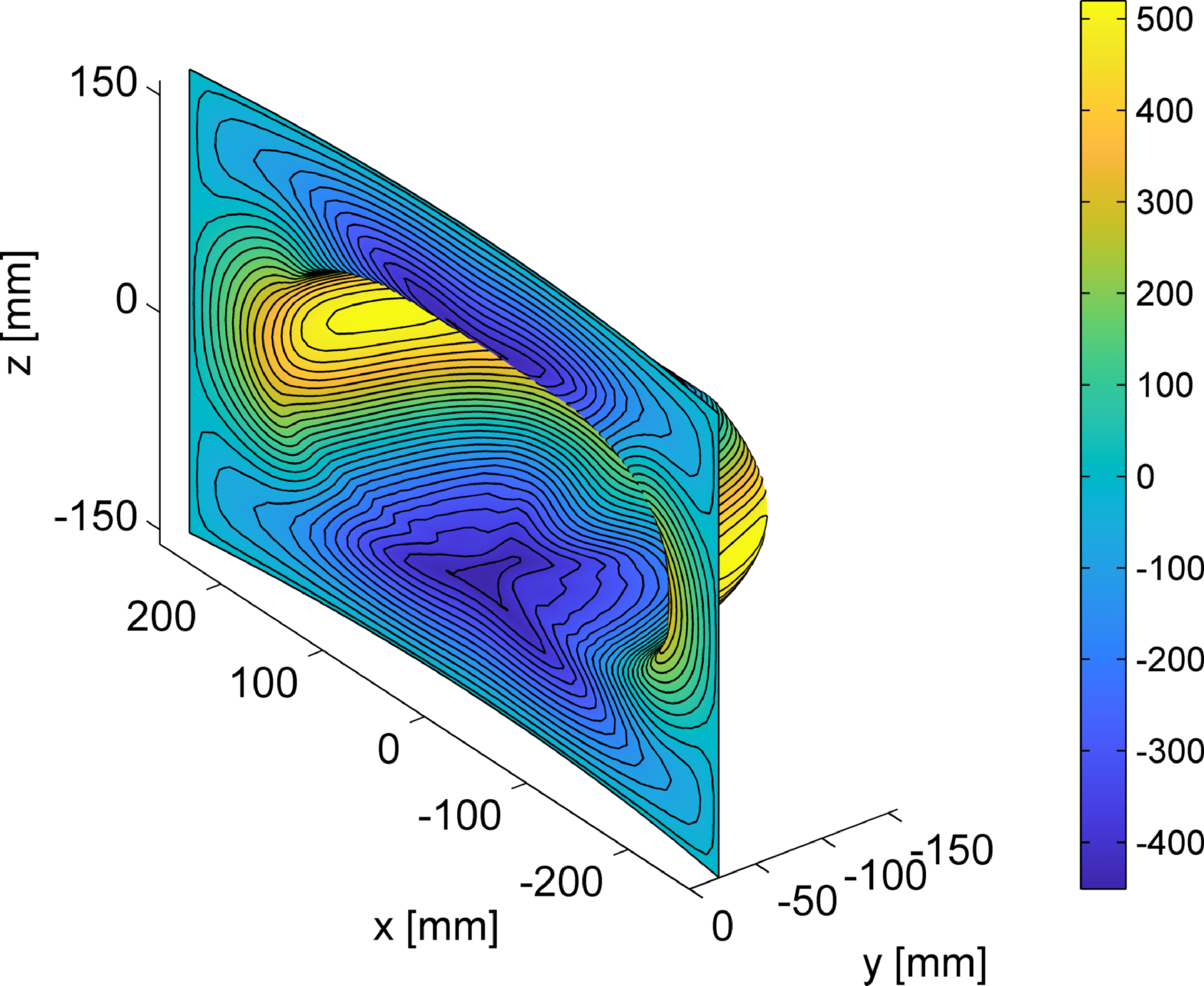}
	} \subfigure[$\alpha_J$ = 0.0044375 and $C_y$ = 0.1817]{ \label{fig:SFMD5_fig9d} \includegraphics[width=0.475\textwidth]{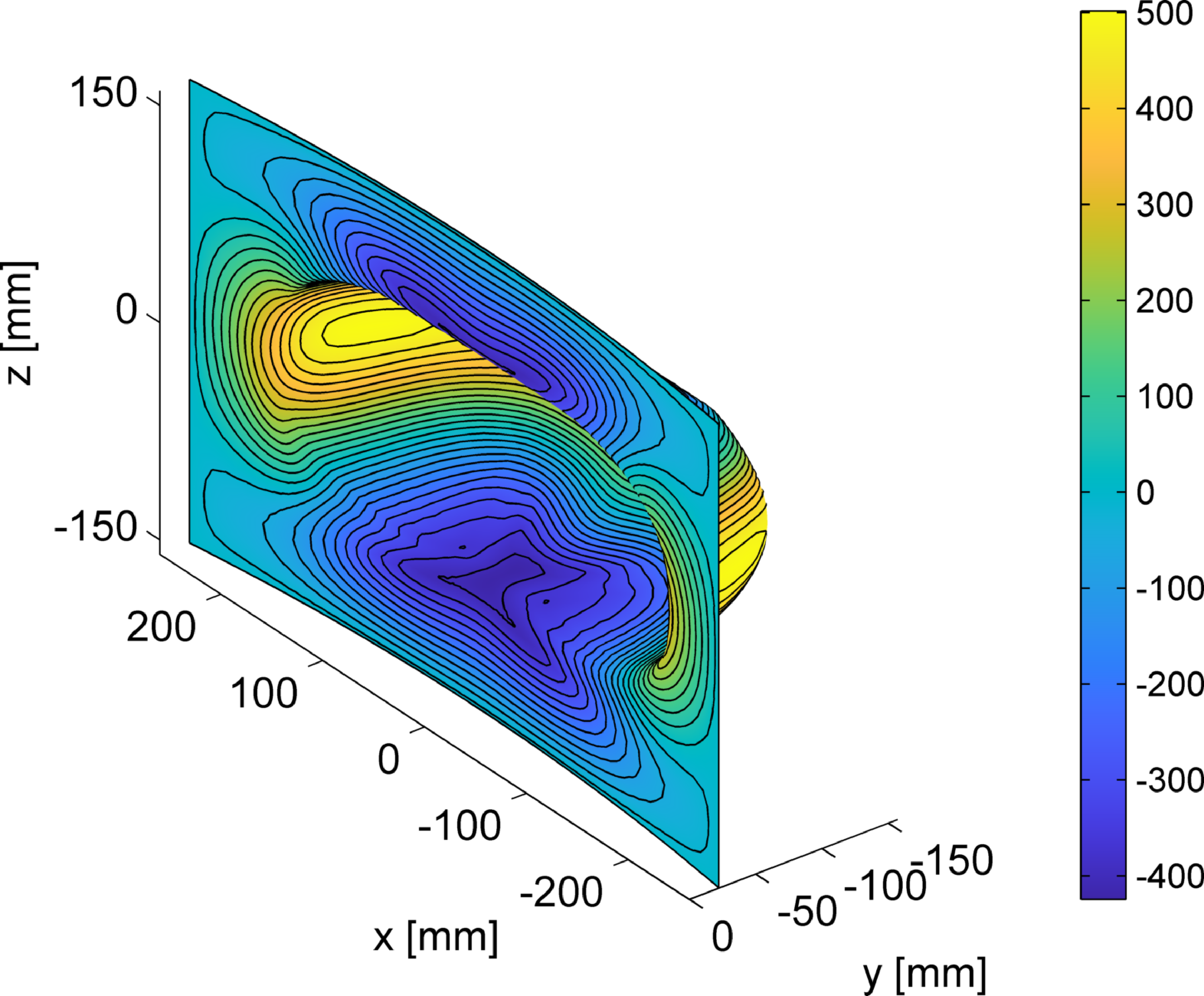}
	}\caption{Spatial distribution of resulting stream function $\psi$ [A] for varying parameters $\alpha_J$ and $C_y$ at $C_s = $ 0.03178. The continuous black contour lines (isolines) define the optimized coil winding layouts.}
	\label{fig:SFMD5_fig9} 
\end{figure}

Figures \ref{fig:EtaD1_fig4}, \ref{fig:EtaD2p5_fig5} and \ref{fig:EtaD5_fig6} show the distribution of coil efficiency $\eta$ over the ROI for the tuning parameters of $C_y$ and $\alpha_J$ presented in Tables \ref{table:NBCD_Non-LinearCoils_CCD1} ($C_s = 0.1589$), \ref{table:NBCD_Non-LinearCoils_CCD2p5} ($C_s = 0.06357$) and \ref{table:NBCD_Non-LinearCoils_CCD5} ($C_s = 0.03178$), respectively. As seen, the coil efficiency $\eta$ for $C_s = 0.03178$ has better gradient uniformity than those for both $C_s = 0.1589$ and $C_s = 0.06357$. Figures \ref{fig:SFMD1_fig7}, \ref{fig:SFMD2p5_fig8} and \ref{fig:SFMD5_fig9} show the corresponding coil layouts, which can assist gradient coil engineers to judge the complexity of coil fabrication.

\subsection{Prototype and field map measurement}
\label{subsec:Experiments}
\begin{figure}[htbp]
	\centering \subfigure[simulated field $B^s_z$]{ \label{fig:simBz_fig10a} \includegraphics[width=0.5\textwidth]{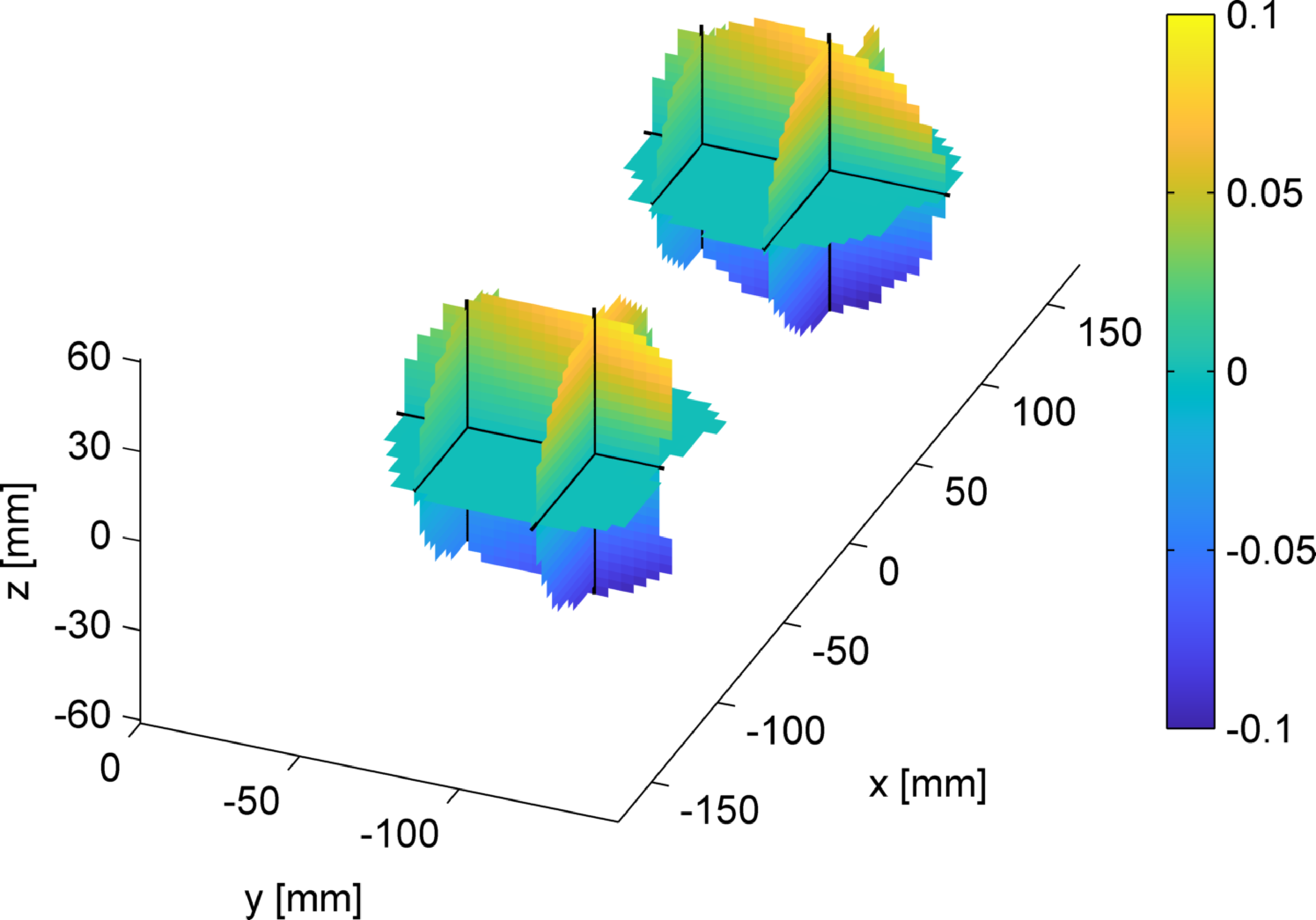}
	} \subfigure[measured field $B^m_z$]{ \label{fig:measuredBz_fig10c} \includegraphics[width=0.5\textwidth]{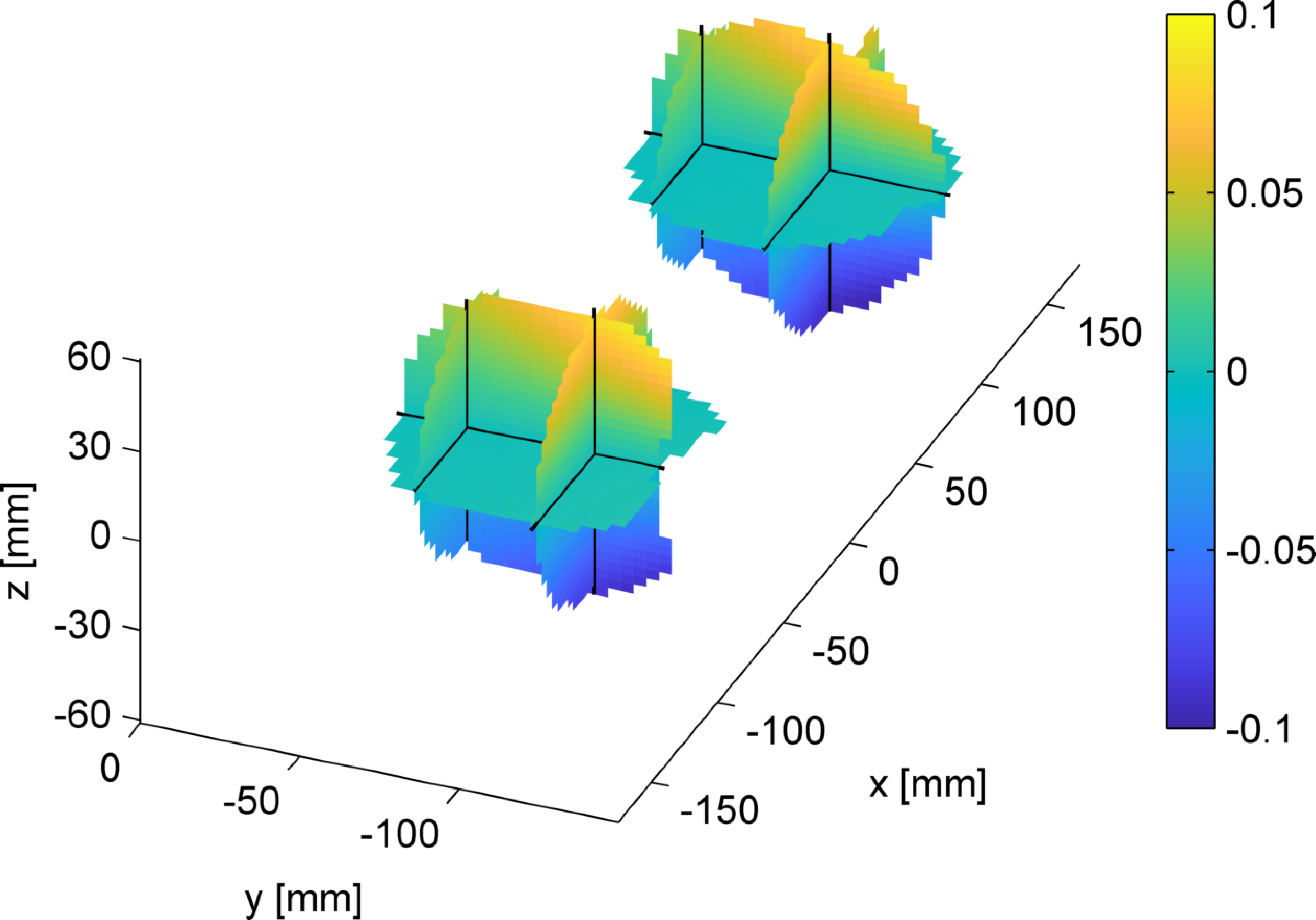}
	} \subfigure[$\left|B^m_z-B^s_z\right|$/$\max(\left|B^s_z\right|)$]{ \label{fig:errBz_fig10e} \includegraphics[width=0.5\textwidth]{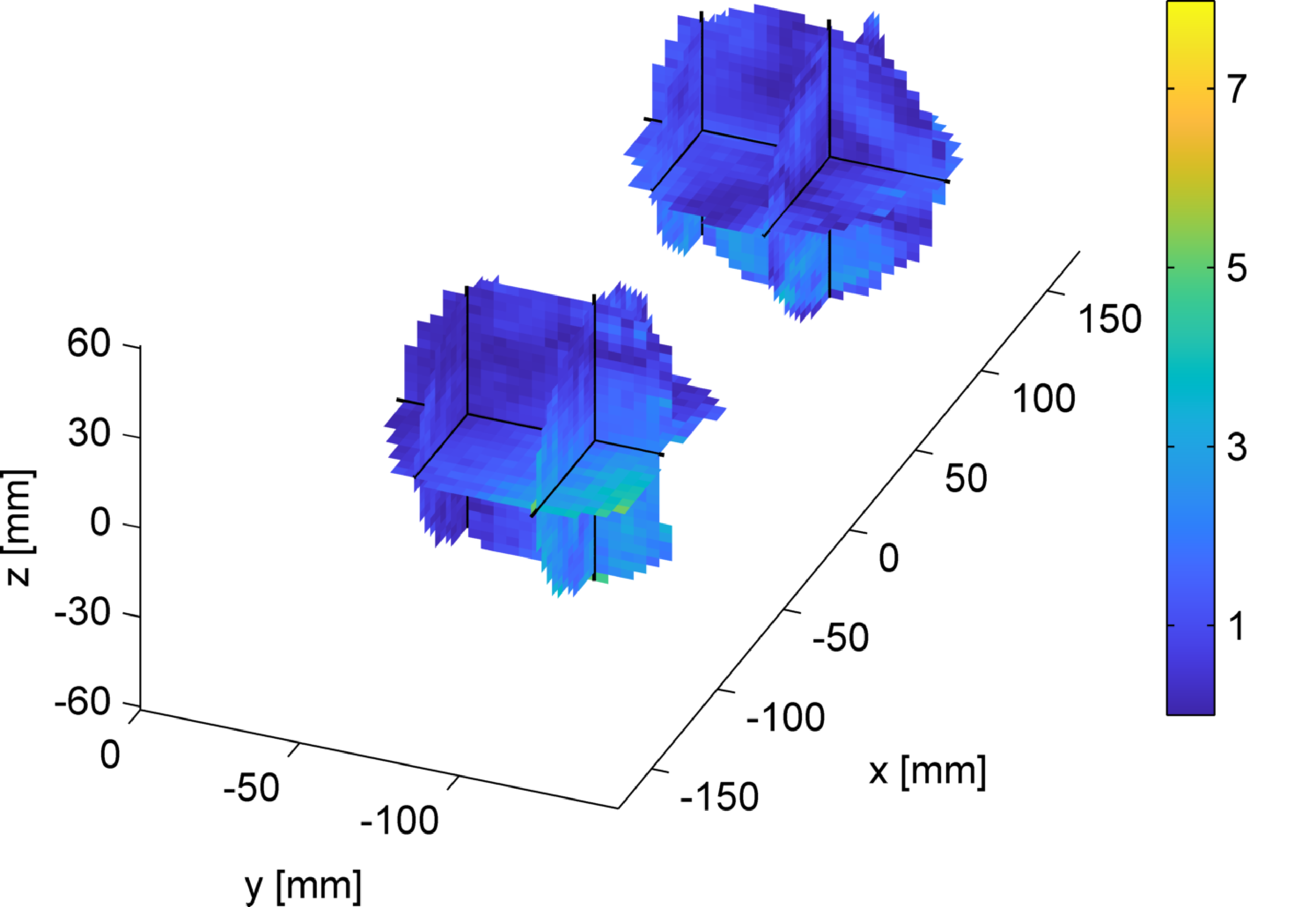}
	}\caption{Comparison of simulated (a) and measured (b) fields. The simulated field $B^s_z$ is calculated using the coil layout shown in Fig. \ref{fig:SFMD5_fig9}.b, and the measured field $B^m_z$ is obtained from the fabricated prototype. The relative errors (c) between the two fields are also depicted.}
	\label{fig:ErrBzEta_fig10} 
\end{figure}

To reduce the difficulty of in-house coil fabrication, the coil designed with $C_s = 0.03178$, $C_y=0.16$ and $\alpha=1.26e-3$ was realized experimentally. Fig. \ref{fig:ErrBzEta_fig10} depicts the simulated magnetic field $B^s_z$ and measured magnetic field $B^m_z$ generated by the built coil, respectively. As can be seen, the measured results match with the simulated fields well. The relative errors of the magnetic field (Fig. \ref{fig:ErrBzEta_fig10}.c) are less than 7.98\% over the whole phantom. These deviations may arise from manufacturing imperfections. In general, the results demonstrate the validity of our design method.

\section{Discussion}
\label{sec.Discussion}
This study presents a novel optimization strategy for controlling the variation of gradient magnitude along the $y$-axis in nonlinear breast gradient coils. The introduction of a width-based figure of merit allows the performance of nonlinear coil designs to be evaluated in a way that accounts for both coil efficiency and engineering constraints, such as minimum wire width. In addition, the coefficient of variation of coil efficiency $\eta$ is employed as a metric to assess the spatial uniformity of gradient strength inside coronal slices or over the ROI. The effectiveness of the proposed framework was demonstrated through a series of numerical examples and validated with experimental results, highlighting its potential for designing high-performance, localized gradient systems for diffusion-weighted breast imaging.

The use of the coefficient of variation of the coil efficiency as a performance indicator provides a direct and spatially resolved measure of uniformity. Notably, as shown in Figure \ref{fig:CVAllCurvesSlices_fig3}, the designs optimized with  $C_s = 0.03178$ achieve intra-slice uniformity levels that are comparable to, or even better than, those of the conventional linear $G_z$ coil ($CV_{ROI}=0.1022$, Table \ref{table:Comp_Linear_NonLinearCoils}), while still offering the spatial selectivity advantages of nonlinear encoding. These results underscore the utility of the proposed formulation for balancing efficiency, manufacturability, and spatial uniformity in the design of next-generation nonlinear breast gradient coils.

Although increasing $C_y$ generally increases gradient variation along the $y$-axis, Figure~\ref{fig:CVAllCurvesSlices_fig3} indicates that its influence on $CV_s$ is not uniform across all coronal slices. This observation may be attributed to the definition of the per-slice constraint~(\ref{equ:NBCD_CoronalSliceConstraints}), which is defined over the entire ROI rather than on a slice-by-slice basis. To better control the spatial uniformity of gradient variation within individual slices, an improved formulation could incorporate additional constraints on selected representative coronal slices. This direction will be explored in future work.

Figure~\ref{fig:SFMD5_fig9}.b illustrates the coil layout of the constructed prototype. Notably, a significant number of windings are positioned along the rim, close to the chest region. As a result, the coil may generate elevated electric fields in thoracic tissues, potentially leading to peripheral nerve stimulation or cardiac stimulation. To evaluate the safety of the prototype for in vivo use, the effects of PNS and the induced electric field in the heart region will be simulated  using the method described in~\cite{JiaLittin2015a} in the near future. Moreover, experimental PNS assessments will also be conducted to evaluate subject safety in future work.

The present designs utilize an unshielded topology for two primary reasons. First, unshielded configurations maximize the achievable gradient strength for a defined minimum wire width, generally offering more coil efficiency compared to shielded variants. Second, as the local coil is positioned relatively far away from the cryostat, only minor eddy-current induction is expected. If eddy-current effects prove non-negligible in diffusion-weighted imaging, two mitigation strategies remain available for future exploration. These include redesigning the system with active shielding or implementing compensation techniques via linear gradient coils, as detailed in~\cite{van_der_velden_novel_2017}.

An initial prototype based on the proposed approach has been fabricated in-house~\cite{ArendsLittin2025}. Figure \ref{fig:ErrBzEta_fig10} show agreement between simulated and measured magnetic fields, confirming the validity of the design process. As this report focuses on the theoretical framework, other characterizations of the prototype’s physical properties including thermal performance, eddy current behavior, PNS thresholds, acoustics and imaging - are currently in progress and will be detailed in a subsequent publication.

\section{Conclusion}
\label{sec:Conclusion}
This study introduced and validated a new optimization method for designing non-linear breast gradient coils for diffusion weighted MRI in breast cancer screening. By adding a y-axis constraint and a width based figure of merit we were able to improve coil efficiency and spatial uniformity near the posterior breast region which is the most challenging area to image. Experimental results show the feasibility and accuracy of these designs and their potential for clinical use. Future work will look into safety and refine the coil performance, especially manufacturing and physiological limitations.


\section{Data availability}
The data that support the findings of this study are not openly available due to intellectual property restrictions, but are available from the corresponding author upon reasonable request. Data are located in controlled access data storage at University medical center Freiburg.

\section*{Acknowledgments}
This work was supported by the German Research Foundation (DFG) Project number 468440804 "High-Power Diffusion Probe for Human Breast MRI – Phase 2" and the Eurostars 3 Project number 3325 "FEM-SCAN: Fast and Efficient MRI Scanning for breast cancer detection".

\section*{Ethics declarations}
Funding: This study was funded by the German Research Foundation (DFG) Project number 468440804 "High-Power Diffusion Probe for Human Breast MRI – Phase 2" and the Eurostars 3 Project number 3325 "FEM-SCAN: Fast and Efficient MRI Scanning for breast cancer detection".

Conflict of Interest: All authors declare that they have no conflict of interest.

Ethical approval: All procedures performed in studies involving human participants were in accordance with the ethical standards of the institutional and/or national research committee and with the 1964 Helsinki declaration and its later amendments or comparable ethical standards.

Informed consent: Informed consent was obtained from all individual participants included in the study.

\bibliography{NBCD}

@article{molendowska_diffusion_2024,
	title = {Diffusion {MRI} in prostate cancer with ultra-strong whole-body gradients},
	volume = {37},
	doi = {10.1002/nbm.5229},
	pages = {e5229},
	number = {12},
	journaltitle = {{NMR} in Biomedicine},
	author = {Molendowska, Malwina and Palombo, Marco and Foley, Kieran G. and Narahari, Krishna and Fasano, Fabrizio and Jones, Derek K. and Alexander, Daniel C. and Panagiotaki, Eleftheria and Tax, Chantal M. W.},
	year = {2024}
}

@article{lemdiasov_stream_2005,
	title = {A stream function method for gradient coil design},
	volume = {26B},
	doi = {10.1002/cmr.b.20040},
	pages = {67-80},
	number = {1},
	journal = {Concepts in Magnetic Resonance Part B: Magnetic Resonance Engineering},
	shortjournal = {Concepts Magn. Reson.},
	author = {Lemdiasov, Rostislav A. and Ludwig, Reinhold},
	year = {2005}
}

@INPROCEEDINGS{JiaKorvink2011a,
  author = {Feng Jia and Zhenyu Liu and Jan G. Korvink},
  title = {A novel coil design method for manufacturable configurations at optimal
	performance},
  booktitle = {Proceedings of the ISMRM 20th Scientific Meeting and Exhibition},
  year = {2011},
  pages = {3780},
  address = {Montreal}
}

@INPROCEEDINGS{ArendsLittin2025,
	author = {Gerrit C. Arends and Edwin Versteeg and Feng Jia and Rokus Valentijn and Emma D.M. Cooijmans and Dennis J.W. Klomp and Maxim Zaitsev and Chantal M.W. Tax and Sebastian Littin},
	title = {Bilateral breast gradient insert prototype for strong diffusion encoding at 3{T}},
	booktitle = {Proceedings of 2025 ISMRM Annual Meeting and Exhibition, Honolulu, Hawaii},
	year = {2025},
	pages = {1339},
	address = {Online}
}

@ARTICLE{JiaLittin2015a,
	author = {Feng Jia and Axel Vom Endt and Philipp Amrein and Maximilian Frederik Russe and Heiko Rohdjess and Martino Leghissa and Maxim Zaitsev and Sebastian Littin},
	title = {Initial assessment of PNS safety for interventionalists during image-guided procedures},
	journal = {Magnetic Resonance Materials in Physics, Biology and Medicine},
	year = {2025},
	volume = {38},
	pages = {239-251},
	doi = {10.1007/s10334-025-01228-4}
}

@ARTICLE{peeren_stream_2003,
  author = {Peeren, G. N.},
  title = {Stream function approach for determining optimal surface currents},
  journal = {Journal of Computational Physics},
  year = {2003},
  volume = {191},
  pages = {305--321},
  number = {1},
  month = oct,
  doi = {10.1016/S0021-9991(03)00320-6},
  issn = {0021-9991},
  urldate = {2014-07-02}
}

@ARTICLE{PooleCrozier2012a,
  author = {Michael S. Poole and Peter T. While and Hector Sanchez Lopez and
	Michael Ng and Stuart Crozier},
  title = {Minimax current density coil design},
  journal = {Journal of Physics D: Applied Physics},
  year = {2010},
  volume = {43},
  pages = {095001},
  number = {9},
  doi = {10.1088/0022-3727/43/9/095001},
  urldate = {2014-09-05}
}

@ARTICLE{AnnaZaitsev2013a,
  author = {Anna Masako Welz and Chris Cocosco and Andrew Dewdney and Daniel
	Gallichan and Feng Jia and Heinrich Lehr and Zhenyu Liu and Hans
	Post and Hartmut Schmidt and Gerrit Schultz and Frederik Testud and
	Hans Weber and Walter Witschey and Jan Korvink and J\"{u}rgen Hennig
	and Maxim Zaitsev},
  title = {Development and characterisation of an unshielded PatLoc gradient
	coil for human head imaging},
  journal = {Concepts in Magnetic Resonance Part B: Magnetic Resonance Engineering},
  year = {2013},
  volume = {43B},
  pages = {111-125},
  doi = {10.1002/cmr.b.21244}
}

@thesis{van_der_velden_novel_2017,
	location = {Utrecht, Netherlands},
	title = {Novel techniques for 7 tesla breast {MRI}},
	url = {http://dspace.library.uu.nl/handle/1874/351284},
	institution = {Utrecht University},
	type = {Dissertation},
	author = {van der Velden, T. A.},
	urldate = {2017-11-02},
	year = {2017}
}

@article{jia_design_2017,
	title = {Design of a shielded coil element of a matrix gradient coil},
	volume = {281},
	doi = {10.1016/j.jmr.2017.06.006},
	journal = {Journal of Magnetic Resonance},
	author = {Jia, Feng and Littin, Sebastian and Layton, Kelvin J. and Kroboth, Stefan and Yu, Huijun and Zaitsev, Maxim},
	year = {2017},
	pages = {217-228}
}

@article{turner_minimum_1988,
	title = {Minimum inductance coils},
	volume = {21},
	doi = {10.1088/0022-3735/21/10/008},
	pages = {948-952},
	number = {10},
	journaltitle = {Journal of Physics E: Scientific Instruments},
	shortjournal = {J. Phys. E: Sci. Instrum.},
	author = {Turner, R.},
	year = {1988},
}

@article{sanchez_simple_2007,
	title = {A Simple Relationship for High Efficiency–Gradient Uniformity Tradeoff in Multilayer Asymmetric Gradient Coils for Magnetic Resonance Imaging},
	volume = {43},
	doi = {10.1109/TMAG.2006.887177},
	pages = {523-532},
	number = {2},
	journal = {{IEEE} Transactions on Magnetics},
	shortjournal = {{IEEE} Trans. Magn.},
	author = {Sanchez, H. and Liu, F. and Trakic, A. and Crozier, S.},
	year = {2007},
}

@article{jones_microstructural_2018,
	title = {Microstructural imaging of the human brain with a ‘super-scanner’: 10 key advantages of ultra-strong gradients for diffusion {MRI}},
	volume = {182},
	doi = {10.1016/j.neuroimage.2018.05.047},
	pages = {8-38},
	journal = {{NeuroImage}},
	author = {Jones, D. K. and Alexander, D. C. and Bowtell, R. and Cercignani, M. and Dell'Acqua, F. and {McHugh}, D. J. and Miller, K. L. and Palombo, M. and Parker, G. J. M. and Rudrapatna, U. S. and Tax, C. M. W.},
	year = {2018}
}

@article{iima_road_2023,
	title = {The road to breast cancer screening with diffusion {MRI}},
	volume = {13},
	doi = {https://doi.org/10.3389/fonc.2023.993540},
	journaltitle = {Frontiers in Oncology},
	author = {Iima, Mami and Le Bihan, Denis},
	pages = {993540},
	year = {2023},
}

@article{baltzer_diffusion-weighted_2020,
	title = {Diffusion-weighted imaging of the breast—a consensus and mission statement from the {EUSOBI} International Breast Diffusion-Weighted Imaging working group},
	volume = {30},
	doi = {10.1007/s00330-019-06510-3},
	pages = {1436-1450},
	number = {3},
	journal = {European Radiology},
	shortjournal = {Eur Radiol},
	author = {Baltzer, Pascal and Mann, Ritse M. and Iima, Mami and Sigmund, Eric E. and Clauser, Paola and Gilbert, Fiona J. and Martincich, Laura and Partridge, Savannah C. and Patterson, Andrew and Pinker, Katja and Thibault, Fabienne and Camps-Herrero, Julia and Le Bihan, Denis and {On behalf of the EUSOBI international Breast Diffusion-Weighted Imaging working group}},
	year = {2020},
	langid = {english},
}

@article{baxter_meta-analysis_2019,
	title = {A Meta-analysis of the Diagnostic Performance of Diffusion {MRI} for Breast Lesion Characterization},
	volume = {291},
	doi = {10.1148/radiol.2019182510},
	pages = {632-641},
	number = {3},
	journal = {Radiology},
	author = {Baxter, Gabrielle C. and Graves, Martin J. and Gilbert, Fiona J. and Patterson, Andrew J.},
	year = {2019},
}

@article{partridge_diffusion-weighted_2017,
	title = {Diffusion-weighted breast {MRI}: Clinical applications and emerging techniques},
	volume = {45},
	doi = {https://doi.org/10.1002/jmri.25479},
	pages = {337-355},
	number = {2},
	journal= {Journal of Magnetic Resonance Imaging},
	author = {Partridge, Savannah C. and Nissan, Noam and Rahbar, Habib and Kitsch, Averi E. and Sigmund, Eric E.},
	year = {2017},
	langid = {english},
}

@article{setsompop_pushing_2013,
	title = {Pushing the limits of in vivo diffusion {MRI} for the Human Connectome Project},
	volume = {80},
	doi = {10.1016/j.neuroimage.2013.05.078},
	pages = {220-233},
	journal = {{NeuroImage}},
	author = {Setsompop, K. and Kimmlingen, R. and Eberlein, E. and Witzel, T. and Cohen-Adad, J. and {McNab}, J. A. and Keil, B. and Tisdall, M. D. and Hoecht, P. and Dietz, P. and Cauley, S. F. and Tountcheva, V. and Matschl, V. and Lenz, V. H. and Heberlein, K. and Potthast, A. and Thein, H. and Van Horn, J. and Toga, A. and Schmitt, F. and Lehne, D. and Rosen, B. R. and Wedeen, V. and Wald, L. L.},
	year = {2013}
}

@article{laun_nmr-based_2012,
	title = {{NMR}-based diffusion pore imaging},
	volume = {86},
	doi = {10.1103/PhysRevE.86.021906},
	pages = {021906},
	number = {2},
	journal = {Physical Review E},
	author = {Laun, Frederik Bernd and Kuder, Tristan Anselm and Wetscherek, Andreas and Stieltjes, Bram and Semmler, Wolfhard},
	year = {2012}
}

@article{lasic_apparent_2016,
	title = {Apparent exchange rate for breast cancer characterization},
	volume = {29},
	doi = {10.1002/nbm.3504},
	pages = {631-639},
	number = {5},
	journal = {{NMR} in Biomedicine},
	author = {Lasi{$\check{c}$}, Samo and Oredsson, Stina and Partridge, Savannah C. and Saal, Lao H. and Topgaard, Daniel and Nilsson, Markus and Bryskhe, Karin},
	year = {2016}
}

@article{zhang_peripheral_2003,
	title = {Peripheral nerve stimulation properties of head and body gradient coils of various sizes},
	volume = {50},
	doi = {10.1002/mrm.10508},
	pages = {50-58},
	number = {1},
	journal = {Magnetic Resonance in Medicine},
	author = {Zhang, Beibei and Yen, Yi-Fen and Chronik, Blaine A. and {McKinnon}, Graeme C. and Schaefer, Daniel J. and Rutt, Brian K.},
	year = {2003}
}

@article{tan_peripheral_2020,
	title = {Peripheral nerve stimulation limits of a high amplitude and slew rate magnetic field gradient coil for neuroimaging},
	volume = {83},
	doi = {10.1002/mrm.27909},
	pages = {352-366},
	number = {1},
	journal = {Magnetic Resonance in Medicine},
	author = {Tan, Ek T. and Hua, Yihe and Fiveland, Eric W. and Vermilyea, Mark E. and Piel, Joseph E. and Park, Keith J. and Ho, Vincent B. and Foo, Thomas K. F.},
	year = {2020},
}

@article{jia_design_2021,
	title = {Design of a high-performance non-linear gradient coil for diffusion weighted {MRI} of the breast},
	volume = {331},
	doi = {10.1016/j.jmr.2021.107052},
	pages = {107052},
	journal = {Journal of Magnetic Resonance},
	year = {2021},
	shortjournal = {Journal of Magnetic Resonance},
	author = {Jia, Feng and Littin, Sebastian and Amrein, Philipp and Yu, Huijun and Magill, Arthur W. and Kuder, Tristan A. and Bickelhaupt, Sebastian and Laun, Frederik and Ladd, Mark E. and Zaitsev, Maxim},
}

@article{amrein_coilgen_2022,
	title = {CoilGen: Open-source {MR} coil layout generator},
	volume = {88},
	doi = {10.1002/mrm.29294},
	pages = {1465-1479},
	journal = {Magnetic Resonance in Medicine},
	year = {2022},
	shortjournal = {Magnetic Resonance in Medicine},
	author = {Amrein, Philipp and Jia, Feng and Zaitsev, Maxim and Littin, Sebastian},
}

@article{ludwig_diffusion_2022,
	title = {Diffusion pore imaging in the presence of extraporal water},
	volume = {339},
	doi = {10.1016/j.jmr.2022.107219},
	pages = {107219},
	journal = {Journal of Magnetic Resonance},
	author = {Ludwig, Dominik and Laun, Frederik Bernd and Klika, Karel D. and Rauch, Julian and Ladd, Mark Edward and Bachert, Peter and Kuder, Tristan Anselm},
	year = {2022}
}

@INPROCEEDINGS{littin_single_2020,
	author = {Littin, Sebastian and Jia, Feng and Amrein, Philipp and Yu, Huijun and Magill, Arthur W. and Kuder, Tristan A. and Ladd, Mark E. and Laun, Frederik and Bickelhaupt, Sebastian and Zaitsev, Maxim},
	title = {Single channel non-linear breast gradient coil for diffusion encoding},
	booktitle = {Proceedings of the 2020 ISMRM Virtual Conference and Exhibition},
	year = {2020},
	pages = {1134}
}

@INPROCEEDINGS{littin_Approaching_2021,
	author = {Littin, Sebastian and Kuder, Tristan A. and Jia, Feng and Magill, Arthur W. and Amrein, Philipp and Laun, Frederik and Bickelhaupt, Sebastian and Klein, Valerie and Ladd, Mark E. and Zaitsev, Maxim},
	title = {Approaching order of magnitude increase of gradient strength: Non-linear breast gradient coil for diffusion encoding},
	booktitle = {Proceedings of the 2021 ISMRM Virtual Conference and Exhibition},
	year = {2021},
	pages = {3096}
}

@INPROCEEDINGS{Arends_Littin_2025,
	author = {Gerrit C. Arends and Edwin Versteeg and Feng Jia and Rokus Valentijn and Emma D.M. Cooijmans and Dennis J.W. Klomp and Maxim Zaitsev and Chantal M.W. Tax and Sebastian Littin},
	title = {Bilateral breast gradient insert prototype for strong diffusion encoding at 3{T}},
	booktitle = {Proceedings of 2025 ISMRM-ISMRT Annual Meeting and Exhibition, Honolulu, Hawaii},
	year = {2025},
	pages = {1339}
}

@INPROCEEDINGS{Arends_Tax_2026,
	author = {Gerrit C. Arends and Edwin Versteeg and Feng Jia and Dennis J.W. Klomp and Maxim Zaitsev and Sebastian Littin and Chantal M.W. Tax},
	title = {Towards high-performance breast diffusion MRI: evaluation of temperature, nerve stimulation, and diffusion measurements},
	booktitle = {Proceedings of 2026 ISMRM-ISMRT Annual Meeting and Exhibition, Cape Town},
	year = {2026},
	pages = {1766}
}
\bibliographystyle{elsarticle-num}

\newpage{}


\begin{table}[htb]
	\centering
	\begin{tabular}{c|c|c|c}
		\hline Properties  & linear $G_z$ coil & old non-linear coil & new non-linear coil \\ 
		\hline 	$C_s$ & 0.1589 & 0.1589 & 0.1589 \\
		$C_y$ & 0.1191 & - & 0.1191 \\
		$D_y/k/C_g$ & 0.1191 & 0 & 0.1191 \\
		$\alpha_J$ & - & 0 & 0 \\
		Number of contours & 26 & 26 & 26 \\
		Current $I$ [A] & 65.61 & 23.55 & 27.88 \\
		$\eta:=|\nabla{B_z}|/I$ [mT/m/A] & [0.4497,1.338] & [0.1392, 5.367] & [0.366,3.207] \\
		$\bar{\eta}=$mean($\eta$) [mT/m/A] & 0.7826 & 2.1806 & 1.842 \\  
		$\sigma_{\eta}:=$std($\eta$) [mT/m/A] & 0.08 & 1.2269 & 0.666 \\
		$CV_{ROI}:=\sigma_{\eta}$/$\bar{\eta}$ & 0.1022 & 0.5626 & 0.3616 \\   
		$\beta_J$ [$\mu$T/A] & 1.402 & 5.3 & 5.954 \\
		$\beta_P$ [mT/(m$\cdot$A$\cdot\Omega^{1/2}$)] & 4.2 & 11.6 & 10.2 \\
		$\beta_W$ [T/(m$\cdot$A$\cdot H^{1/2}$)] & 0.0967 & 0.2565 & 0.2186 \\
		Coil inductance [$\mu$H] & 130.98 & 144.55 & 141.94 \\
		Coil resistance [m$\Omega$] & 34.75 & 35.2 & 32.57 \\
		$\max (|M_x|,|M_y|,|M_z|)$  [N$\cdot$m] & 1.0e-4 & 1.0e-4 & 1.0e-4 \\
		$w_{\min}$ [mm] & 1.79 & 2.431 & 3.233 \\
		\hline 
	\end{tabular} 
	\caption{Performance comparison among the linear gradient coil, the old and new non-linear gradient coils with the same $C_s$. The old non-linear coil was designed without incorporating the new constraint~(\ref{equ:NBCD_YAxisConstraints}), as described in the previous work~\cite{jia_design_2021}. In contrast, the new non-linear coil was obtained by including the $y$-axis gradient variation constraint~(\ref{equ:NBCD_YAxisConstraints}) in the optimization process.
	} \vspace{2mm}
	\label{table:Comp_Linear_NonLinearCoils}
\end{table}

\begin{table}[htb]
	\centering
	\begin{tabular}{c|cccc}
		\hline Properties  & \multicolumn{4}{c}{Non-linear coils when $C_s$ = 0.1589} \\ 
		\hline 	Case & 1 & 2 & 3 & 4 \\
		$C_y$ & 0.1191 & 0.1361 & 0.1532 & 0.16 \\
		$\alpha_J$ & 6.375e-3 & 6.6875e-3 & 7.4375e-3 & 3.094e-3 \\
		Number of contours & 32 & 30 & 28 & 26   \\
		Current $I$ [A]& 25.83 & 25.18 & 24.97 & 25.42  \\
		$\eta:=|\nabla{B_z}|/I$ [mT/m/A] & [0.47,3.384] & [0.406,3.542] & [0.3434,3.919] & [0.292,3.992] \\
		$\bar{\eta}=$mean($\eta$) [mT/m/A] & 1.9884 & 2.039 & 2.057 & 2.02 \\  
		$\sigma_{\eta}:=$std($\eta$) [mT/m/A] & 0.6502 & 0.72 & 0.793 & 0.8247 \\ 
		$CV_{ROI}:=\sigma_{\eta}$/$\bar{\eta}$ & 0.327 & 0.353 & 0.3856 & 0.4081 \\  
		$\beta_J$ [$\mu$T/A] & 6.9611 & 7.1384 & 7.2026 & 7.077\\
		$\beta_P$ [mT/(m$\cdot$A$\cdot\Omega^{1/2}$)] & 9.71 & 10.17 & 10.44 & 10.78 \\
		$\beta_W$ [T/(m$\cdot$A$\cdot H^{1/2}$)] & 0.2084 & 0.2187 & 0.225 & 0.232\\
		Coil inductance [$\mu$H] & 182.1 & 173.87 & 167 & 151.35 \\
		Coil resistance [m$\Omega$] & 41.93 & 40.2 & 38.85 & 35.11 \\
 		$\max (|M_x|,|M_y|,|M_z|)$  [N$\cdot$m] & 1.0e-4 & 1.0e-4 & 1.0e-4 & 1.0e-4\\
		$w_{\min}$ [mm] & 3.5 & 3.5  & 3.5 & 3.5 \\
		\hline 
	\end{tabular} 
	\caption{Performance comparison among non-linear gradient coils designed with $C_s$ = 0.1589.} \vspace{2mm}
	\label{table:NBCD_Non-LinearCoils_CCD1}
\end{table}

\begin{table}[htb]
	\centering
	\begin{tabular}{c|cccc}
		\hline Properties  & \multicolumn{4}{c}{Non-linear coils when $C_s$ = 0.06357} \\ 
		\hline 		Case & 1 & 2 & 3 & 4 \\
		$C_y$ & 0.1191 & 0.1316 & 0.16 & 0.1692 \\
		$\alpha_J$ & 1.875e-3 & 4.875e-3 & 4.625e-3 & 4.875e-3 \\
		Number of contours & 28 & 28 & 26 & 26   \\
		Current $I$ [A]& 30.97 & 30.41 & 31.04 & 31.06  \\
		$\eta:=|\nabla{B_z}|/I$ [mT/m/A] & [0.651,2.832] & [0.625,3.149] & [0.521,3.616] & [0.519,3.634] \\
		$\bar{\eta}=$mean($\eta$) [mT/m/A] & 1.658 & 1.689 & 1.6542 & 1.6535 \\  
		$\sigma_{\eta}:=$std($\eta$) [mT/m/A] & 0.503 & 0.566 & 0.6712 & 0.6743 \\
		$CV_{ROI}:=\sigma_{\eta}$/$\bar{\eta}$ & 0.3034 & 0.3352 & 0.4057 & 0.4078 \\   
		$\beta_J$ [$\mu$T/A] & 5.81 & 5.91 & 5.7908 & 5.7927\\
		$\beta_P$ [mT/(m$\cdot$A$\cdot\Omega^{1/2}$)] & 8.61 & 8.7 & 9.037 & 9.033 \\
		$\beta_W$ [T/(m$\cdot$A$\cdot H^{1/2}$)] & 0.179 & 0.182 & 0.19 & 0.1902\\
		Coil inductance [$\mu$H] & 171.73 & 171.73 & 151.37 & 151.21 \\
		Coil resistance [m$\Omega$] & 37.17 & 37.67 & 33.51 & 33.51 \\
		$\max (|M_x|,|M_y|,|M_z|)$  [N$\cdot$m] & 1.0e-4 & 1.0e-4 & 1.0e-4 & 1.0e-4\\
		$w_{\min}$ [mm] & 3.5 & 3.5  & 3.5 & 3.5 \\
		\hline 
	\end{tabular} 
	\caption{Performance comparison among non-linear gradient coils designed with $C_s$ = 0.06357.} \vspace{2mm}
	\label{table:NBCD_Non-LinearCoils_CCD2p5}
\end{table}


\begin{table}[htb]
	\centering
	\begin{tabular}{c|ccccc}
		\hline Properties  & \multicolumn{4}{c}{Non-linear coils when $C_s$ = 0.03178} \\ 
		\hline 		Case & 1 & 2 & 3 & 4 & 5\\
		$C_y$ & 0.1191 & 0.1348 & 0.16 & 0.16 & 0.1817 \\
		$\alpha_J$ & 5.49e-3 & 4.4625e-3 & 1.26e-3 & 7.125e-3 & 4.4375e-3 \\
		Number of contours & 32 & 30 & 26 & 28 & 26   \\
		Current $I$ [A]& 34.44 & 34.51 & 36.14 & 34.67 & 35.62 \\
		$\eta:=|\nabla{B_z}|/I$ [mT/m/A] & [0.729,2.635] & [0.671,2.854] & [0.5616,3.057] & [0.593,3.219] & [0.523 3.433]\\
		$\bar{\eta}=$mean($\eta$) [mT/m/A] & 1.491 & 1.488 & 1.421 & 1.481 & 1.442\\  
		$\sigma_{\eta}:=$std($\eta$) [mT/m/A] & 0.453 & 0.511 & 0.576 & 0.599 & 0.652\\  
		$CV_{ROI}:=\sigma_{\eta}$/$\bar{\eta}$ & 0.3037 & 0.3436 & 0.4055 & 0.4042 & 0.4523\\ 
		$\beta_J$ [$\mu$T/A] & 5.219 & 5.21 & 4.976 & 5.187 & 5.045\\
		$\beta_P$ [mT/(m$\cdot$A$\cdot\Omega^{1/2}$)] & 7.12 & 7.51 & 8.166 & 7.86 & 8.156\\
		$\beta_W$ [T/(m$\cdot$A$\cdot H^{1/2}$)] & 0.152 & 0.159 & 0.171 & 0.166 & 0.172\\
		Coil inductance [$\mu$H] & 192.39 & 174.62 & 138.4 & 158.27 & 140.19\\
		Coil resistance [m$\Omega$] &43.87 & 39.21 & 30.28 & 35.52 & 31.24\\
		$\max (|M_x|,|M_y|,|M_z|)$  [N$\cdot$m] & 1.0e-4 & 1.0e-4 & 1.0e-4 & 1.0e-4 & 1e-4\\
		$w_{\min}$ [mm] & 3.5 & 3.5  & 3.5 & 3.5 & 3.5\\
		\hline 
	\end{tabular} 
	\caption{Performance comparison among non-linear gradient coils designed with $C_s$ = 0.03178.} \vspace{2mm}
	\label{table:NBCD_Non-LinearCoils_CCD5}
\end{table}

\newpage{}

\end{document}